\PassOptionsToPackage{pdftex}{graphicx}
\documentclass{article}
\usepackage{arxiv}
\usepackage[utf8]{inputenc}
\usepackage{listings}
\usepackage[dvipsnames]{xcolor}
\usepackage[utf8]{inputenc}
\usepackage{todonotes}
\usepackage{url}
\usepackage{listings}
\usepackage{nameref}

\usepackage[justification=justified,singlelinecheck=false]{caption}
\usepackage{tikz}

\usepackage{enumitem}

\usepackage{wrapfig}
\usepackage{lipsum}

\usepackage{soul}

\newcommand{\hlcolor}{Yellow!35}
\newcommand{\hlcolorTwo}{LimeGreen!35}
\makeatletter
\newenvironment{btHighlight}[1][]
{\begingroup\tikzset{bt@Highlight@par/.style={#1}}\begin{lrbox}{\@tempboxa}}
{\end{lrbox}\bt@HL@box[bt@Highlight@par]{\@tempboxa}\endgroup}

\newcommand\btHL[1][]{%
  \begin{btHighlight}[#1]\bgroup\aftergroup\bt@HL@endenv%
}
\def\bt@HL@endenv{%
  \end{btHighlight}%
  \egroup
}
\newcommand{\bt@HL@box}[2][]{%
  \tikz[#1]{%
    \pgfpathrectangle{\pgfpoint{1pt}{0pt}}{\pgfpoint{\wd #2}{\ht #2}}%
    \pgfusepath{use as bounding box}%
    \node[anchor=base west, fill=\hlcolor,outer sep=0pt,inner xsep=1pt, inner ysep=0pt, rounded corners=2pt, minimum height=\ht\strutbox+2pt,#1]{\raisebox{1pt}{\strut}\strut\usebox{#2}};
  }%
}

\newenvironment{btHighlightTwo}[1][]
{\begingroup\tikzset{bt@HighlightTwo@par/.style={#1}}\begin{lrbox}{\@tempboxa}}
{\end{lrbox}\bt@HLTwo@box[bt@HighlightTwo@par]{\@tempboxa}\endgroup}

\newcommand\btHLTwo[1][]{%
  \begin{btHighlightTwo}[#1]\bgroup\aftergroup\bt@HLTwo@endenv%
}
\def\bt@HLTwo@endenv{%
  \end{btHighlightTwo}%
  \egroup
}
\newcommand{\bt@HLTwo@box}[2][]{%
  \tikz[#1]{%
    \pgfpathrectangle{\pgfpoint{1pt}{0pt}}{\pgfpoint{\wd #2}{\ht #2}}%
    \pgfusepath{use as bounding box}%
    \node[anchor=base west, fill=\hlcolorTwo,outer sep=0pt,inner xsep=1pt, inner ysep=0pt, rounded corners=2pt, minimum height=\ht\strutbox+2pt,#1]{\raisebox{1pt}{\strut}\strut\usebox{#2}};
  }%
}

\usepackage{inconsolata}
\usepackage{listings}
\usepackage{etoolbox}
\newtoggle{InString}{}
\togglefalse{InString}
\newcommand*{\ColorIfNotInString}[1]{\iftoggle{InString}{#1}{\color{blue}#1}}%
\newcommand*{\ProcessQuote}[1]{#1\iftoggle{InString}{\global\togglefalse{InString}}{\global\toggletrue{InString}}}%
\newcommand{\indentrule}{\color{code_indent}\vrule\hspace{2pt}}
\definecolor{code_indent}{HTML}{CCCCCC}

\newenvironment{figureAsListing}
    {
    \addtocounter{figure}{-1}
    \refstepcounter{lstlisting}
     
    \begin{figure}[!htbp]
        
        \centering
    }
    { 
        \end{figure} 
    }

\newenvironment{figureAsListingWide}
    {
    \addtocounter{figure}{-1}
    \refstepcounter{lstlisting}
     
    \begin{figure*}[!htbp]
        
        \centering
    }
    { 
        \end{figure*} 
    }

\usepackage{multicol}

\newboolean{showcomments}
\setboolean{showcomments}{false}
\ifthenelse{\boolean{showcomments}}
{ \newcommand{\mynote}[3]{
    \fbox{\bfseries\sffamily\scriptsize#1}
{\small$\blacktriangleright$\textsf{\emph{\color{#3}{#2}}}$\blacktriangleleft$}}
}{
\newcommand{\mynote}[3]{}}

\newcommand*\justify{%
  \fontdimen2\font=0.4em
  \fontdimen3\font=0.2em
  \fontdimen4\font=0.1em
  \fontdimen7\font=0.1em
  \hyphenchar\font=`\-
}
\newcommand{\sqs}{\texttt{\justify CancellableQueueSynchronizer}}

\title{
CQS: A Formally-Verified Framework for Fair and Abortable Synchronization
}

\author{
  Nikita Koval \\
  JetBrains Research\\
  \texttt{nikita.koval@jetbrains.com}
  \And
  Dmitry Khalanskiy\\
  JetBrains\\
  \texttt{dmitry.khalanskiy@jetbrains.com} \\
  \And
  Dan Alistarh\\
  IST Austria\\
  \texttt{dan.alistarh@ist.ac.at} \\
}

\begin{document}


\maketitle

\begin{abstract}
Writing concurrent code that is both correct and efficient is notoriously difficult. Thus, programmers often prefer to use synchronization abstractions, which render  code simpler and easier to reason about. Despite a wealth of work on this topic, there is still a gap between the rich semantics provided by  synchronization abstractions in modern programming languages---specifically, \emph{fair} FIFO ordering of synchronization requests and support for \emph{abortable} operations---and frameworks for implementing it correctly and efficiently. 
Supporting such semantics is critical given the rising popularity of constructs for asynchronous programming, such as coroutines, which abort frequently and are cheaper to suspend and resume compared to native threads.

This paper introduces a new framework called \sqs{} (CQS), 
which enables simple yet efficient implementations of a wide range of fair and abortable synchronization primitives: mutexes, semaphores, barriers, count-down latches, and blocking pools. 
Our main contribution is algorithmic, as implementing both fairness and abortability efficiently at this level of generality is non-trivial. 
Importantly, all our algorithms, including the CQS framework and the primitives built on top of it, come with \emph{formal proofs} in the Iris framework for Coq for many of their properties. These proofs are modular, so it is easy to show correctness for new primitives implemented on top of CQS. 
From a practical perspective, implementation of CQS for native threads on the JVM improves throughput by up to two orders of magnitude over Java's \texttt{AbstractQueuedSynchronizer}, the only practical abstraction offering similar semantics. 
Further, we successfully integrated CQS as a core component of the popular Kotlin Coroutines library, validating the framework's practical impact and expressiveness in a real-world environment. 
In sum, \sqs{} is the first framework to combine expressiveness with formal guarantees and solid practical performance. Our approach should be extensible to other languages and families of synchronization primitives.
\end{abstract}

\section{Introduction}\label{section:intro}


Providing the ``right'' set of programming abstractions to enable efficient and correct concurrent code is a question as old as the field of concurrency~\cite{dijkstra2001solution, knuth1966additional}.  
One of the most basic primitives is the \emph{mutex}, which allows access to the critical section to at most one thread, via \texttt{lock()} and \texttt{unlock()} invocations. 
Standard libraries, e.g., the Java concurrency library~\cite{lea2005java},  
provide more general primitives, such as the \emph{semaphore}, which allows at most a fixed number of threads to be in the critical section simultaneously, the \emph{barrier}, which allows a set of threads to wait for each other at a common program point, 
and the \emph{count-down latch}, which allows threads to wait until a given set of operations is completed.

Although basic versions of the above primitives exist in most specialized libraries, 
programmers often require \emph{stronger semantics} from synchronization abstractions, which are supported by modern programming languages such as Java, C\#, Go, Kotlin and Scala. 
One particularly desirable property is \emph{fairness}~\cite{izraelevitz2017generality}, by which the order of critical section traversals should respect the FIFO order of arrivals, to avoid starvation. 
A second key property is \emph{abortability}, which enables a thread to \emph{cancel} its request, due to a time-out or user-specified behavior. 
Abortability and scalability are especially important in the context of \emph{coroutines}~\cite{kahn1976coroutines, kotlincoroutines}, the number of which can be in the millions simultaneously, and which can be frequently cancelled. 
Coroutines are also significantly cheaper to suspend and resume compared to native threads:
internally, they are scheduled on a thread pool, so when a coroutine is suspended, the corresponding thread immediately picks up another one and executes it instead, so the native thread never blocks. 
%
In such a setting, efficient cancellation is vital, while fairness becomes less expensive and more natural. Coroutines are now a key component of modern programming languages such as Java, C++, Go, Scala, and Kotlin. 

Implementing fairness and cancellation efficiently in a concurrent context is known to be very challenging, and there is a long line of work proposing highly non-trivial designs~\cite{jayanti2003adaptive, lee2010fast, danek2010closing, alon2018deterministic, GW17, pareek2012rmr}. 
Modern languages and libraries typically either restrict the generality of the semantics, providing more efficient \emph{unfair} synchronization, or implement complex constructs, which may lead to correctness and performance issues. 
This paper addresses the question of implementing \emph{fair} and \emph{abortable} synchronization primitives in a way that is general, efficient, and easy to reason. 

For intuition, we begin with the observation that most of the synchronization operations we focus on are inherently \emph{blocking}: 
threads attempt to acquire a shared resource or synchronize, and may have to wait for the resource to become available. 
For example, the mutex \texttt{lock()} operation either acquires the lock instantly or adds the currently running thread to a queue of waiting operations and suspends. The \texttt{unlock()} invocation resumes the first waiting \texttt{lock()} request, handing the lock over.
To our knowledge, the only practical abstraction to implement such synchronization primitives in full generality is the \texttt{AbstractQueuedSynchronizer} in Java~\cite{lea2005java}, which maintains a FIFO queue of suspended requests in a way reminiscent of the state-of-the-art CLH mutex~\cite{CLH}. While the \texttt{AbstractQueuedSynchronizer} has been extremely influential, its design does not scale to high contention. 
Our goal is to provide a design that is general enough to support a wide range of abstractions, but also efficient enough to support modern usage scenarios.

\paragraph{Our Contribution.}
We introduce a new framework called \sqs{} (CQS), which enables simple and efficient implementations of a wide range of \emph{fair} and \emph{abortable} synchronization and communication primitives, such as mutexes, semaphores, barriers, count-down latches, and blocking pools.
We show that CQS can implement this wide range of synchronization primitives, for which we provide both formal proofs and extensive experimental validation, showing significant practical improvements over the state of the art. 
%
%
%

Conceptually, the goal of the \sqs{} framework is to efficiently maintain a FIFO queue of waiting threads, corresponding to operations to be completed. To this end, CQS provides two main operations:
    (1) \texttt{suspend()}, which adds the current thread as a waiter into the queue and suspends, and 
    (2) \texttt{resume(result)}, which tries to retrieve and resume the first waiter with the specified~result.
One key advantage of the CQS semantics is that it allows operations to invoke \texttt{resume(..)} before \texttt{suspend()}: we actively use this property for implementation simplicity and better performance. 
Despite the relative simplicity of the \sqs{} API, it allows us to support a rich set of synchronization and communication primitives. 

The data structure behind \sqs{} uses techniques from modern concurrent queue implementations~\cite{LCRQ,YM16}, leveraging the \texttt{Fetch-and-Add} instruction for better scalability. In brief, our solution is based on a logically-infinite array, equipped with two position counters, indexing \texttt{suspend()} and \texttt{resume(..)} operations, respectively. 
Each operation starts by incrementing its counter via \texttt{Fetch-and-Add}, thus reserving the cell in the first-come-first-served order. The rest of the synchronization is performed in the cell: \texttt{suspend()} stores the current thread, while \texttt{resume(..)} wakes up the suspended thread.

The main novelty behind \sqs{} is the efficient built-in support for \emph{aborting/cancelling} operations. We emulate the infinite array with a linked list of fixed-sized \emph{cell segments}, so each cell stores a waiting operation. On thread cancellation, the cell state should be reclaimed to avoid memory leaks, but also segments full of cancelled requests should be physically removed from the linked list. 
One naive way to implement such functionality is by linearly searching the waiting queue for the corresponding segment, which is then unlinked~\cite{lea2005java}. However, this would have a worst-case linear time in the queue size, making frequent cancellations lead to significant overheads. While for threads this approach is sufficient (since their number is small and they rarely abort), for coroutines, of which millions can exist at the same time and which cancel frequently, significant complexity improvements are required.
We propose a more efficient design, where segments form a concurrent doubly-linked list, enabling constant time removals via careful pointer manipulations. This also allows us to support different cancellation modes: in case of \emph{simple} cancellation, \texttt{resume(..)} is allowed to fail if the waiter located in the corresponding cell was cancelled, whereas \emph{smart} cancellation provides a mechanism to efficiently skip a sequence of aborted requests but requires complex mechanisms to ensure that some thread is always resumed. 

\paragraph{Formal Proofs in Iris/Coq.}
The complexity of the resulting CQS implementation renders manual correctness proofs quite challenging and error-prone. 
We provide \emph{modular formal proofs} for all the presented primitives in Coq~\cite{krebbers2017interactive} using the Iris separation logic~\cite{jung2018iris}. 
We formally specify the CQS operations and then demonstrate that they obey the semantics corresponding to the primitives we consider. 
One property we do not show formally is the FIFO order, which is notoriously difficult to approach in Iris but can be demonstrated through classical proofs. 
We emphasize the complexity of our formalization task, as only a few similar real-world implementations are formally verified~\cite{krogh2016verifying, krishna2020verifying, vindum2021contextual, jung2017rustbelt, chajed2021gojournal}.
Our proofs for CQS span approximately 8000 lines of Coq code, often requiring non-trivial reasoning. 
Yet, the proofs are \emph{modular}, so they can be employed as the basis for proving new synchronization primitives implemented on top of CQS, reducing formalization effort. 
Specifically, proving each higher-order CQS-based primitive presented in the paper on this basis takes only around 500 lines.

\paragraph{Evaluation.}
We integrated the CQS framework as part of the standard Kotlin Coroutines library and used it to implement several fundamental synchronization primitives. 
To validate performance, we implemented \sqs{} on the JVM for native threads and compared it against the state-of-the-art \texttt{AbstractQueuedSynchronizer} framework in Java~\cite{lea2005java}, which aims to solve the same problem, and, to our knowledge, is the only practical abstraction that provides similarly general semantics.
We present different versions of mutex and semaphore, barrier and count-down-latch primitives, and two versions of blocking pools. Our algorithms outperform existing  implementations in almost all scenarios and are sometimes faster by orders of magnitude.

\noindent
In particular, our semaphore implementation outperforms the standard Java solution, which is implemented via \texttt{AbstractQueuedSynchronizer}~\cite{lea2005java}, up to \texttt{4x} in the uncontended case where the number of threads does not exceed the number of permits, and up to \texttt{90x} in a highly-contended scenario. 
%
For the count-down-latch implementation, our solution shows up to \texttt{7x} speedup compared to the Java library, while the barrier synchronization is faster by up to \texttt{4x}.
%
For blocking pools, which share a limited set of resources, our approach is faster than the Java library implementation by up to \texttt{150x}.
In some cases, the \emph{fair} synchronization primitives we present even outperform the \emph{unfair} variants in the Java standard library.
Finally, results show that the cancellation support of CQS is more efficient than the one of the \texttt{AbstractQueuedSynchronizer} framework.
Our analysis shows that these improvements come mainly because from the superior scalability of our design.

\section{Basic CQS Algorithm}\label{sec:basic}
In this section, we describe the key ideas behind the \sqs{} algorithm in an iterative fashion, using a simple non-abortable mutex construct as an example.
We then focus on the complexities of supporting cancellation/abortability in the next section. 

\paragraph{Thread Management.} We manipulate threads to suspend and resume operations. While our main application is coroutines, we will use  threads for illustration, as they may be more familiar to the readers. Listing~\ref{listing:threadsapi} presents the API we use in the paper. We emphasize that our approach can be directly adapted to any concurrency model, such as coroutines, futures, or continuations.
\footnote{Many languages support asynchronous programming either explicitly via \texttt{Future}-s, or implicitly via the \texttt{async/await} construct that internally manipulates continuation objects.}
Our implementations for Java native threads and Kotlin coroutines (Section~\ref{sec:experiments}) support this claim.


\setlength{\columnsep}{1.7em}
\begin{wrapfigure}[8]{r}{0.545\textwidth}
\vspace{-1.2em}
\begin{lstlisting}[xleftmargin=0em, label={listing:threadsapi},
caption={
Thread management API.
}]
interface Thread {
 fun park(onCancel: lambda () -> Unit#\footnotemark{}#): Any
 fun unpark(result: Any): Bool
 fun cancel() // invoked by user
}
fun currentThread(): Thread
\end{lstlisting}

\end{wrapfigure}

\footnotetext{\texttt{Unit} is a type with only one value: the \texttt{Unit} object. This type corresponds to the \texttt{void} type in Java.}

Our API assumes that the currently-running thread can be obtained by calling \texttt{currentThread()}, and suspended by invoking \texttt{park(..)}. 
While suspended, the thread can be aborted via \texttt{cancel()} call, becoming unable to resume. In that case, the \texttt{onCancel} lambda provided in \texttt{park(..)} is executed. If a thread is cancelled in an active state, the cancellation takes effect with the following \texttt{park(..)} invocation.

To resume a thread, the \texttt{unpark(result)} function should be called. It returns \texttt{true} if the resumption succeeds, so the corresponding \texttt{park(..)} invocation completes with the specified result. Otherwise, if the thread is already cancelled, \texttt{unpark(result)} returns \texttt{false}. Notably, \texttt{unpark(result)} can be called before \texttt{park(..)} -- in this case, the following \texttt{park(..)} invocation immediately completes without suspension, returning the provided result.

\paragraph{Environment.}
For simplicity, we assume the sequentially-consistent memory model, which matches our implementation, as all real-world weak memory models provide sequential consistency for data-race-free programs. In addition to plain reads and writes, we use atomic \texttt{Compare-and-Swap$\:$(CAS)}, \texttt{Get-and-Set}, and \texttt{Fetch-and-Add$\:$(FAA)} instructions, which are available in all modern programming languages.
We also assume that the runtime environment supports garbage collection (GC).
Reclamation techniques such as hazard pointers~\cite{michael2004hazard} or hazard eras~\cite{ramalhete2017brief} can be used in environments without GC.

\paragraph{High-Level Algorithm Overview.}
At the logical level, the \sqs{} maintains a first-in-first-out (FIFO) queue of waiting requests and provides two main functions:
\begin{itemize}[nosep]
    \item \texttt{suspend(): T}, which adds the current thread as a waiter into the queue and suspends, and
    \item \texttt{resume(result: T): Bool}, which tries to retrieve and resume the next waiter, passing the specified value of type \texttt{T}.
\end{itemize}

A key advantage is that the framework allows to invoke \texttt{resume(..)} before \texttt{suspend()} as long as it is known that \texttt{suspend()} will happen eventually, so synchronization primitive implementations can allow such races. In Section~\ref{sec:primitives}, we present several algorithms that leverage this property for better performance and simplicity.


\begin{figureAsListingWide}
\begin{minipage}[t]{0.5\textwidth}
\begin{lstlisting}[]
val cells = InfiniteArray() #\label{line:sqs_ho_cells}#
var suspendIdx: Int#64# = 0 #\label{line:sqs_ho_idx_start}# 
var resumeIdx:  Int#64# = 0 #\label{line:sqs_ho_idx_end}#

fun suspend(): T {
  i := FAA(&suspendIdx, +1) #\label{line:sqs_ho_susp_inc}#
  // Try to suspend in #\color{Mahogany}cells[i]#.
  t := currentThread() #\label{line:sqs_ho_susp_createreq}#
  if CAS(&cells[i], null, t): #\label{line:sqs_ho_susp_cas}#
  #\indentrule#  return park() // enqueued, suspend#\label{line:sqs_ho_susp_park}#
  // Read the result and finish.
  result := cells[i]; cells[i] = TAKEN #\label{line:sqs_ho_susp_retrieve}#
  return result #\label{line:sqs_ho_susp_imm}#
}
\end{lstlisting}
\end{minipage}
\hfill
\begin{minipage}[t]{0.48\textwidth}
\begin{lstlisting}[firstnumber=15]
fun resume(result: T) {
  i := FAA(&resumeIdx, +1) #\label{line:sqs_ho_resume_inc}#
  t := cells[i]
  if t == null: // is the cell empty?#\label{line:sqs_ho_resume_is_empty}#
  #\indentrule#  // `suspend()` is coming, try to 
  #\indentrule#  // install the result and finish.
  #\indentrule#  if CAS(&cells[i], null, result): #\label{line:sqs_ho_resume_elim_cas}#
  #\indentrule#  #\indentrule#  return 
  #\indentrule#  // The cell stores a thread.
  #\indentrule#  t = cells[i]  #\label{line:sqs_ho_resume_reread}#
  // Resume the waiting request.
  cells[i] = RESUMED #\label{line:sqs_ho_resume_clean}#
  t.unpark(result) // t is Thread #\label{line:sqs_ho_resume_complete}#
}
\end{lstlisting}
\end{minipage}
\vspace{-1em}
\caption{High-level CQS implementation on top of an infinite array without abortability support.
}
\vspace{-0.5em}
\label{listing:sqs_highlevel}
\end{figureAsListingWide}

A useful mental image of CQS is that of an infinite array supplied with two counters: one that references the cell in which the new waiter should be enqueued as part of the next \texttt{suspend()} call, and one that references the next cell for \texttt{resume(..)}. The intuition is that \texttt{suspend()} atomically increments its counter via \texttt{Fetch-and-Add}, stores the currently running thread in the corresponding cell, and suspends. Likewise, \texttt{resume(..)} increments its counter, visits the corresponding cell, and resumes the stored thread with the specified value. However, if \texttt{resume(..)} comes before \texttt{suspend()}, it simply places the value in the cell and finishes {---} \texttt{suspend()} grabs the value later and completes without an actual suspension.
\footnote{The \texttt{suspend()} and \texttt{resume(..)} race behavior is similar to the thread parking mechanism in both our API and Java, where \texttt{unpark(..)} followed by \texttt{park()} results in the latter operation returning immediately.} 

Listing~\ref{listing:sqs_highlevel} provides a high-level pseudocode for this simplified \sqs{}, without abortability support. An infinite array \texttt{cells} (line~\ref{line:sqs_ho_cells}) stores waiting threads and values inserted by racing resumptions.
Counters \texttt{suspendIdx} and \texttt{resumeIdx} (lines~\ref{line:sqs_ho_idx_start}--\ref{line:sqs_ho_idx_end}) reference cells for the next \texttt{suspend()} and \texttt{resume(..)} operations.

When \texttt{suspend()} starts, it first gets its index and increments the counter atomically via \texttt{Fetch-And-Add (FAA)}, which returns the value right before the increment (line~\ref{line:sqs_ho_susp_inc}). Next, it obtains the currently running thread to be inserted into the cell (line~\ref{line:sqs_ho_susp_createreq}) and tries to do so via \texttt{Compare-And-Swap (CAS)} (line~\ref{line:sqs_ho_susp_cas}). If this \texttt{CAS} succeeds, the operation parks the thread, finishing when resumed (line~\ref{line:sqs_ho_susp_park}). Otherwise, a concurrent \texttt{resume(..)} has already visited the cell {---} thus, \texttt{suspend()} extracts the placed value, cleans the cell by placing a special \texttt{TAKEN} token (line~\ref{line:sqs_ho_susp_retrieve}), and returns the extracted value (line~\ref{line:sqs_ho_susp_imm}). Note that in the mutex implementation, we always pass \texttt{Unit} through CQS; other data structures, such as blocking pools discussed in Section~\ref{subsec:pools}, may pass different values.

\begin{figureAsListingWide}
\begin{minipage}[t]{0.58\textwidth}
\begin{lstlisting}[]
val cqs = CQS<Unit>()#\label{line:mutex_basic_sqs}#
var state: Int = 1 // "unlocked" intially#\label{line:mutex_basic_state}#
fun lock() {
  s := FAA(&state, -1) #\label{line:mutex_basic_dec}#
  if s > 0: return // was the lock acquired? #\label{line:mutex_basic_lock_locked}#
  cqs.suspend() // suspend otherwise #\label{line:mutex_basic_lock_susp}#
}
\end{lstlisting}
\end{minipage}
\hfill
\begin{minipage}[t]{0.38\textwidth}
\begin{lstlisting}[firstnumber=8]
fun unlock() {
  s := FAA(&state, +1) #\label{line:mutex_basic_unlock_inc}#
  // Resume the first waiting
  // request if there is one.
  if s < 0: cqs.resume(Unit) #\label{line:mutex_basic_unlock_resume}#
}
\end{lstlisting}
\end{minipage}
\vspace{-1em}
\caption{
Basic mutex algorithm without abortability support using the CQS framework.
}
\vspace{-1em}
\label{listing:mutex_basic}
\end{figureAsListingWide}

\noindent
Symmetrically, \texttt{resume(..)} increments \texttt{resumeIdx} first (line~\ref{line:sqs_ho_resume_inc}). It then checks whether the cell is empty (line~\ref{line:sqs_ho_resume_is_empty}), in which case it tries to place the resumption value directly into the cell (line~\ref{line:sqs_ho_resume_elim_cas}). If the attempt fails, a waiter is already stored in the cell, so the algorithm re-reads it (line~\ref{line:sqs_ho_resume_reread}). 
After the waiter is extracted, the operation stores a special \texttt{RESUMED} token in the cell to avoid memory leaks and resumes the extracted thread (lines~\ref{line:sqs_ho_resume_clean}--\ref{line:sqs_ho_resume_complete}).

\paragraph{Mutex on Top of CQS.}
To illustrate how primitives should use CQS, consider the simple mutex implementation from Listing~\ref{listing:mutex_basic}.  
The rough idea is to maintain a \texttt{state} field (line~\ref{line:mutex_basic_state}) that stores $1$ if the mutex is unlocked, and $w \leq 0$ if the mutex is locked. In the latter case, the negated value of $w$ is the number of waiters on this mutex.

Initially, the mutex is unlocked and its \texttt{state} equals $1$. 
When a \texttt{lock()} operation arrives, it atomically decrements the state, setting it to $0$ (line~\ref{line:mutex_basic_dec}), so the logical state becomes ``locked''. 
Since the previous logical state was ``unlocked'', the operation completes immediately (line~\ref{line:mutex_basic_lock_locked}).
However, if another \texttt{lock()} arrives after that, it changes \texttt{state} to $-1$, keeping the logical state as ``locked'' and incrementing the 
number of waiters. Since the mutex was already locked, this invocation suspends via CQS (line~\ref{line:mutex_basic_lock_susp}).
Likewise, \texttt{unlock()} increments \texttt{state}, either making the mutex ``unlocked'' if the counter was $0$, or decrementing the number of waiters (line~\ref{line:mutex_basic_unlock_inc}). In the latter case, \texttt{unlock()} resumes the first waiter via CQS (line~\ref{line:mutex_basic_unlock_resume}). 
It is worth emphasizing that \texttt{lock()} and \texttt{unlock()} contain only five lines of easy-to-follow code in total. 

\paragraph{Non-Blocking Operations.} 
Synchronization primitives typically provide non-blocking variants of operations, such as the \texttt{tryLock()} sibling of \texttt{Mutex.lock()}, which succeed only when the operation does not require suspension. 
However, supporting them becomes non-trivial when \texttt{resume(..)} comes before \texttt{suspend()}, so the data (e.g., the lock permit) is stored in CQS and cannot be extracted without suspension; thus, the non-blocking sibling cannot access it.

To solve the problem, we introduce a special \emph{synchronous} resumption mode, so that \texttt{resume(..)} always makes a rendezvous with \texttt{suspend()} and does not leave the value in CQS, failing when this rendezvous cannot happen in bounded time.
We view this as an extension to \sqs{} and present it in Appendix~\ref{appendix:sync_async}.

\begin{wrapfigure}[6]{r}{0.5\textwidth}
    \vspace{-1.6em}
    \centering
    \includegraphics[width=0.5\textwidth]{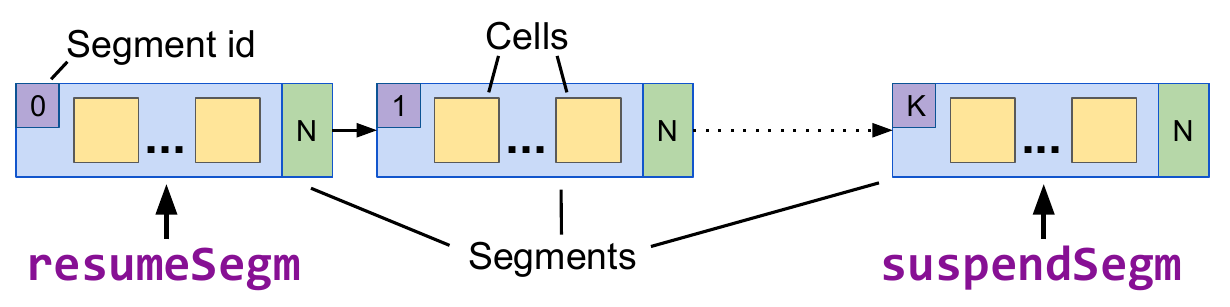}
    \vspace{-2.1em}
    \caption{An infinite array as a linked list of cell segments.}
    \label{fig:segments}
\end{wrapfigure}

\paragraph{Infinite Array Implementation.}
The last building block of the basic CQS implementation is the emulation of an infinite array. Since all cells are processed in sequential order, the algorithm only requires having access to the cells between \texttt{resumeIdx} and \texttt{suspendIdx} and does not need to store an infinite number of cells.
We follow the approach behind the channels implementation in Kotlin~\cite{channels_kotlin}, maintaining a linked list of cell segments, each containing a fixed number of cells, as illustrated in Figure~\ref{fig:segments}.

Each segment has a unique id and can be seen as a node in a Michael-Scott queue~\cite{michael1996simple}. Following this structure, we maintain only those cells that are in the current active range (between \texttt{resumeIdx} and \texttt{suspendIdx}) and access them similarly to an array. Specifically, we change the current working segment after completing operations equal to the number of cells in each segment.

Despite conceptual simplicity, the implementation of this structure is non-trivial, as shown in~\cite{channels_kotlin}. We discuss the implementation and the required changes to the CQS algorithm in Appendix~\ref{appendix:infarr}. 

\section{Cancellation Support}\label{sec:abortability}
In this section, we extend the basic construct above with cancellation support.
We assume that threads can be aborted via \texttt{Thread.cancel()} call, which bounds the following \texttt{unpark(..)} to fail. Additionally, the \texttt{onCancel} \emph{cancellation handler} provided in the \texttt{park(..)} call\footnote{Since in practice we manipulate threads or coroutines, cancellation should be handled via an existing mechanism. In Java, for example, aborted threads throw \texttt{InterruptedException}, which can be caught and processed by the user. Moreover, some coroutines libraries, such as Kotlin Coroutines~\cite{kotlincoroutines}, already support an API similar to the one we use.} is invoked when the thread aborts. We will use this functionality later in this section.


We support two cancellation modes: \emph{simple} and \emph{smart}. Intuitively, the difference is that in the \emph{simple} cancellation mode, \texttt{resume(..)} fails if the thread in the corresponding cell has been cancelled, whereas the \emph{smart} cancellation enables efficient skipping a sequence of aborted requests. 

\subsection{Simple Cancellation}\label{subsection:simple_cancellation}
\begin{wrapfigure}[15]{r}{0.4\textwidth}
\vspace{-1em}
    \vspace{-0.2em}
    \centering
    \includegraphics[width=0.4\textwidth]{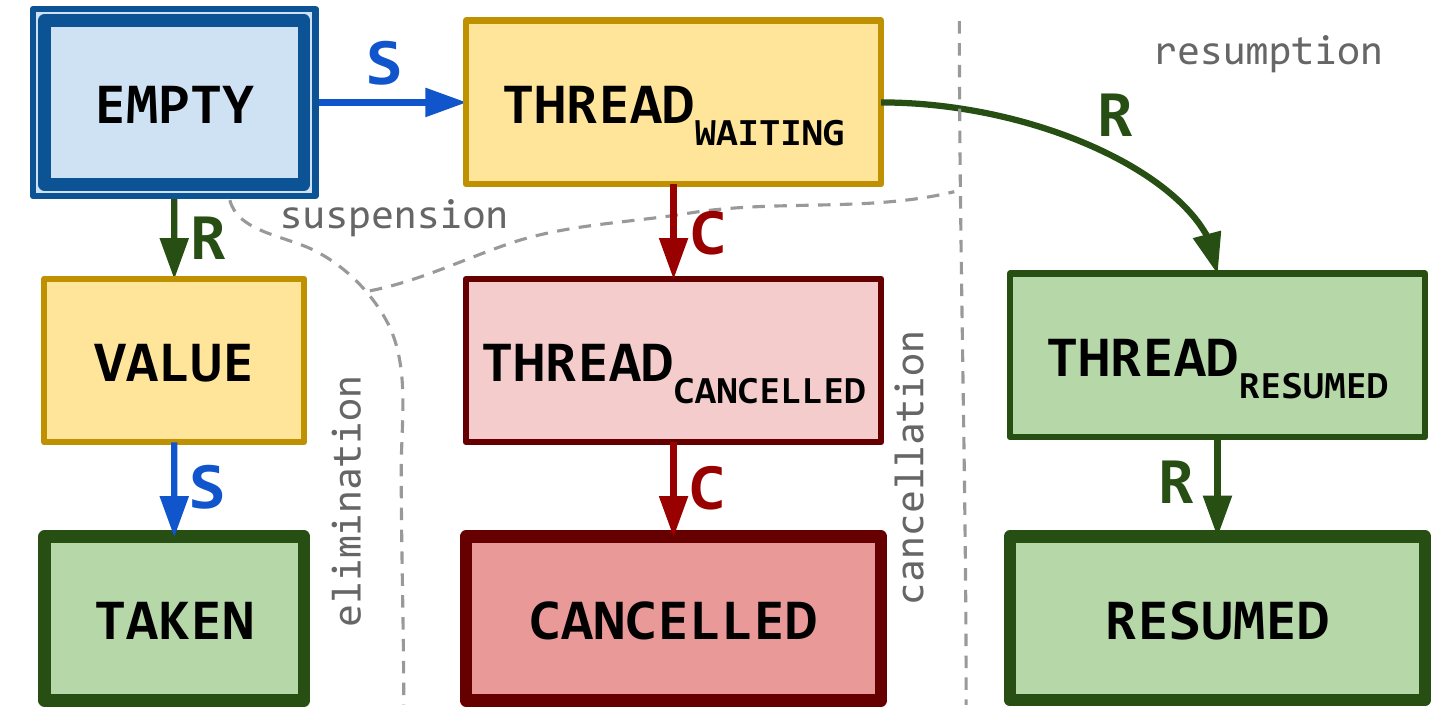}
    \vspace{-1.6em}
    \caption{Cell life-cycle with \emph{simple} cancellation. When a thread becomes cancelled by a successful \texttt{Thread.cancel()} invocation (tagged with ``\texttt{C}''), its content changes to \texttt{CANCELLED} to avoid memory leaks, and the corresponding \texttt{resume(..)} fails. The edges marked with \texttt{S} and \texttt{R} correspond to the transitions by \texttt{suspend()} and \texttt{resume(..)}.}
    \label{fig:cell_simple_cancellation}
\end{wrapfigure}

The \emph{simple} cancellation mode is relatively straightforward~{---} when a waiter becomes cancelled, the \texttt{resume(..)} operation that processes this cell is bound to fail. 
Thus, the code for \texttt{resume(..)} in Listing~\ref{listing:sqs_highlevel} should return \texttt{true} if \texttt{t.unpark(result)} at line~\ref{line:sqs_ho_resume_complete} succeeds, and \texttt{false} on failure, indicating that the thread has already been aborted. Figure~\ref{fig:cell_simple_cancellation} shows the corresponding cell life-cycle.

An important technical detail is that aborted threads should be physically removed from the waiting queue to allow the garbage collector to reclaim the related memory. Thus, we specify a \emph{cancellation handler} that replaces the aborted \texttt{Thread} with a special \texttt{CANCELLED} marker, according to the diagram in Figure~\ref{fig:cell_simple_cancellation}.
In addition, we must remove segments full of cancelled cells from the linked list to avoid memory leaks; we discuss the corresponding part of the algorithm in Appendix~\ref{appendix:infarr}.

\paragraph{Mutex with Simple Cancellation.} Please recall the mutex algorithm presented in Listing~\ref{listing:mutex_basic}. 
With simple cancellation, \texttt{resume(..)} fails when the resuming thread is already aborted ($\mathtt{THREAD}_\mathtt{CANCELLED}$ or \texttt{CANCELLED} state). In this case, the corresponding \texttt{lock()} request is no longer valid, and the \texttt{unlock()} invocation, which performs \texttt{resume(..)} on this cell, incremented the counter (which must have been decremented by a \texttt{lock()} operation earlier). Thus, the balance is met, and \texttt{unlock()} should restart.


\paragraph{Limitations.}
One issue with the cancellation logic above is that it requires the \texttt{resume(..)} operation to process all the cancelled cells. 
Consider $N$ \texttt{lock()} operations which execute and then immediately abort {---} the following call to \texttt{unlock()} increments \texttt{state} and unsuccessfully invokes \texttt{resume(..)} exactly $N$ times. This leads to $\Theta(N)$ complexity, which is, nevertheless, amortized by \texttt{Thread.cancel()} invocations. 
Ideally, however, \texttt{unlock()} should require $O(1)$ time under no contention and should not ``pay'' for the cancelled requests.

\noindent
Another problem is that it is sometimes infeasible to wait until a \textit{resume(..)} operation observes that the waiter is cancelled. We often wish to immediately learn about a waiter being cancelled and change the state correspondingly. 
As an example, consider a readers-writer lock and the following execution: (1) a reader comes and takes a lock, (2) a writer arrives and suspends, (3) then, another reader arrives and also suspends, because it should take a lock after the suspended writer. After that, (4) the suspended writer becomes cancelled, so the second reader should be resumed and take the lock. However, with simple cancellation, the effect of cancellation is postponed until another operation tries to resume the cancelled waiter, so the reader does not wake up. Making cancellations take effect immediately is critical in this context.

\subsection{Smart Cancellation}
A better option would be to skip cancelled waiters in \texttt{resume(..)} and install a cancellation handler that de-registers the operation when it  aborts. For mutex, this could be incrementing the \texttt{state} field. 
However, a naive approach where \texttt{resume(..)} simply skips aborted threads would be incorrect.

\begin{wrapfigure}[10]{r}{0.55\textwidth}
    \centering
    \vspace{-2em} 
    \includegraphics[width=0.55\textwidth]{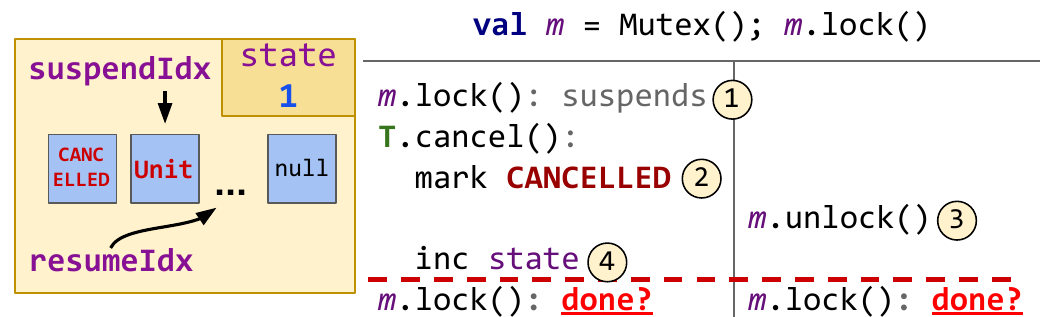}
    \vspace{-1.4em}
    \caption{An incorrect execution of mutex with naive cancellation strategy, where the cancellation handler increments \texttt{state} back, while \texttt{resume(..)} simply skips cancelled cells.}
    \label{fig:mutex_smart_cancellation_problem}
\end{wrapfigure}

\paragraph{The Problem.} 
Figure~\ref{fig:mutex_smart_cancellation_problem} illustrates a potential problematic execution with such a mutex. Assume it is initially locked and two threads start. The first thread invokes \texttt{lock()}, placing itself in the CQS, and immediately aborts; however, the \texttt{state} is not incremented back yet. After that, the second thread calls \texttt{unlock()}, which increments the \texttt{state} counter (so it becomes $0$) and intends to wake up a waiting \texttt{lock()} operation. The corresponding \texttt{resume(..)} sees the first cell in \texttt{CANCELLED} state and places its value in the next empty cell. The execution switches back to the first thread, and the cancellation handler of the aborted \texttt{lock()} increases the counter to $1$. The resulting state is shown in the figure. Finally, two \texttt{lock()} calls by both threads are performed (they are under the dashed red line). One of them decrements \texttt{state} to $0$ and enters the critical section; the other suspends via CQS and, observing the value in the cell, also proceeds to enter the critical section, thus breaking the mutex semantics.

\paragraph{The \texttt{REFUSE} State.}
Notice that the naive version above would work fine in cases where the cancellation handler does not change the mutex state from ``locked'' to ``unlocked'' (thus, \texttt{state} stays non-positive). The problem occurs when the ``last'' waiter becomes cancelled, and a concurrent \texttt{resume(..)} tries to complete it. In this case, \texttt{resume(..)} must be informed that there is no longer any waiter in the \sqs{} that could receive the value.

\begin{wrapfigure}[6]{r}{0.4\textwidth}
    \vspace{-0.9em}
    \centering
    \includegraphics[width=0.25\textwidth]{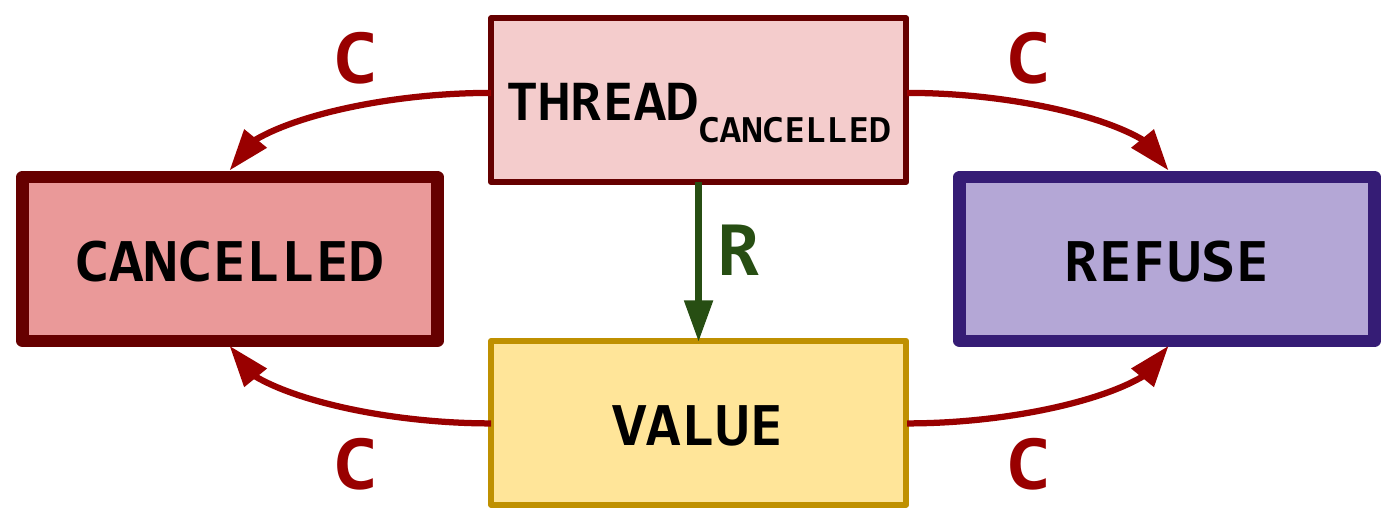}
    \vspace{-0.3em}
    \caption{Cell life-cycle for \emph{smart} cancellation, with a special \texttt{REFUSE} state. The suspension, elimination, and resumption parts stay unchanged (see Figure~\ref{fig:cell_simple_cancellation}).}
    \label{fig:cell_smart_cancellation}
\end{wrapfigure}

To signal this, a new \texttt{REFUSE} state is added to the cell life-cycle; see Figure~\ref{fig:cell_smart_cancellation} on the right for the updated cancellation part. This state signals that an operation attempted to abort, but determined that there is an upcoming \texttt{resume(..)} and the aborted waiter was the last one in the CQS. Thus, the \texttt{resume(..)} that inevitably visits the cell should be refused by CQS and will no longer attempt to pass the value to any waiter.



\begin{wrapfigure}[9]{r}{0.45\textwidth}
\vspace{-1.85em}
\begin{lstlisting}[xleftmargin=0em, caption={Smart cancellation API.},label={listing:smart_cancellation_api}]
// Invoked on cancellation and returns 
// `true` if the cell should enter to 
// CANCELLED state and `false` when it
// should  transition to REFUSE.
fun onCancellation(): Bool
// Defines how to process a refused 
// resume(..) with the specified value
fun completeRefusedResume(value: T)
\end{lstlisting}
\end{wrapfigure}

\paragraph{Smart Cancellation API.}
Users who develop primitives on top of CQS with smart cancellation should implement \texttt{onCancellation()} and \texttt{completeRefusedResume(value)} functions, whose semantics are described in Listing~\ref{listing:smart_cancellation_api}.
When a waiter is cancelled (the cell state changes to $\mathtt{THREAD}_\mathtt{CANCELLED}$), the cancellation handler invokes \texttt{onCancellation()}, which tries to logically remove the waiter from the data structure. If the \texttt{resume(..)} operation that sees this cell can safely skip it and still match with another non-cancelled \texttt{suspend()}, the operation returns \texttt{true}, and the cell state becomes \texttt{CANCELLED}. Otherwise, when the cell state becomes \texttt{REFUSE}, and the corresponding \texttt{resume(..)} should be refused, \texttt{onCancellation()} should return \texttt{false}. This way, the refused \texttt{resume(..)} invokes \texttt{completeRefusedResume(...)} to complete the operation. 

Note that the behavior of \texttt{resume(..)} depends on whether the aborted thread moves the cell to \texttt{CANCELLED} or \texttt{REFUSE} state. However, if \texttt{resume(..)} observes $\mathtt{THREAD}_\mathtt{CANCELLED}$ state (the cell stores a \texttt{Thread} instance while \texttt{unpark(..)} fails), the expected behavior can not yet be predicted. 
We resolve this race by \emph{delegating} the rest of the current \texttt{resume(..)} to the cancellation handler, replacing the thread instance with the resumption value {---} see the corresponding transition from $\mathtt{THREAD}_\mathtt{CANCELLED}$ to $\mathtt{VALUE}$ in Figure~\ref{fig:cell_smart_cancellation}. After that, when the cancellation handler changes the cell's state to \texttt{CANCELLED} or \texttt{REFUSE}, it receives the value and completes the resumption correspondingly. In this case, the value passed to \texttt{resume(..)} can be out of the data structure for a while but is guaranteed to be processed eventually.
Note that \texttt{resume(..)} never fails when using the smart cancellation mode.

\paragraph{Mutex with Smart Cancellation.} Consider the mutex example again. Listing~\ref{listing:mutex_smart_cancellation} presents the \texttt{onCancellation()} and \texttt{completeRefusedResume(..)} implementations for the basic algorithm from Listing~\ref{listing:mutex_basic}; the rest stays the same. 


\setlength{\columnsep}{1.9em}
\begin{wrapfigure}[17]{r}{0.46\textwidth}
\vspace{-1.7em}
\begin{lstlisting}[xleftmargin=0em,
label={listing:mutex_smart_cancellation},
caption={Cancellation handling in the \emph{smart} mode for the basic mutex algorithm from Listing~\ref{listing:mutex_basic}.}
]
fun onCancellation(): Bool {
 // Increment the counter back. 
 s := FAA(&state, +1) #\label{line:mutex:onCancellation:inc}#
 // #\color{Mahogany} s < 0:# still in the "locked" state;
 // #\color{Mahogany} s = 0: the# mutex has become "unlocked", 
 //       refuse the upcoming #\color{Mahogany}resume(..)#.
 return s < 0
}
fun completeRefusedResume(permit: Unit) {
 // Do nothing, the mutex has already  
 // been moved to the "unlocked" state.
}
\end{lstlisting}
\end{wrapfigure}

When a \texttt{lock()} request aborts, the \texttt{onCancellation()} operation increments \texttt{state} (thus, decrementing the number of waiters). 
However, when the increment changes \texttt{state} to $1$ (``unlocked''), the operation must return \texttt{false} to refuse the upcoming \texttt{resume(..)}. After that, the \texttt{resume(..)} that comes to the cell sees it in \texttt{REFUSE} state and invokes \texttt{completeRefusedResume(..)}. For the mutex, the lock is already successfully returned at the moment of incrementing \texttt{state} in \texttt{onCancellation()}, so this function does nothing. However, when CQS is used to transfer elements (see blocking pools in Subsection~\ref{subsec:pools} as an example), the refused element should be returned back to the data structure via \texttt{completeRefusedResume(..)}.


\clearpage

\begin{figureAsListingWide}
\begin{minipage}[t]{0.5\textwidth}
\begin{lstlisting}[]
fun resume(value: T): Bool {
 i := FAA(&resumeIdx, +1)  #\label{line:resumeAndCancel:resume:inc}#
 while (true): // modify the cell state#\label{line:resumeAndCancel:resume:modifyCell:start}#
 #\indentrule# cur := cells[i]  #\label{line:resumeAndCancel:resume:curState}#
 #\indentrule# when {
 #\indentrule# #\indentrule#cur == null: #\label{line:resumeAndCancel:resume:empty}#
 #\indentrule# #\indentrule#  if CAS(&cells[i], null, value): #\label{line:resumeAndCancel:resume:tryElimination0}#
 #\indentrule# #\indentrule#  #\indentrule#  return true #\label{line:resumeAndCancel:resume:tryElimination1}#
 #\indentrule# #\indentrule#cur is Thread: #\label{line:resumeAndCancel:resume:isRequest}#
 #\indentrule# #\indentrule#  if cur.unpark(value): #\label{line:resumeAndCancel:resume:tryComplete}#
 #\indentrule# #\indentrule#  #\indentrule#  cells[i] = RESUMED #\label{line:resumeAndCancel:resume:resumed0}#
 #\indentrule# #\indentrule#  #\indentrule#  return true #\label{line:resumeAndCancel:resume:resumed1}#
 #\indentrule# #\indentrule#  // The thread is cancelled.
 #\indentrule# #\indentrule#  if cancellationMode == SIMPLE: #\label{line:resumeAndCancel:resume:simpleFail0}#
 #\indentrule# #\indentrule#  #\indentrule#  return false #\label{line:resumeAndCancel:resume:simpleFail1}#
 #\indentrule# #\indentrule#  // In smart cancellation, delegate 
 #\indentrule# #\indentrule#  // this resume(..) completion  
 #\indentrule# #\indentrule#  // to the cancellation handler.
 #\indentrule# #\indentrule#  if CAS(&cells[i], cur, value): #\label{line:resumeAndCancel:resume:putValue0}#
 #\indentrule# #\indentrule#  #\indentrule#  return true #\label{line:resumeAndCancel:resume:putValue1}#
 #\indentrule# #\indentrule#cur == CANCELLED: #\label{line:resumeAndCancel:resume:isCancelled}#
 #\indentrule# #\indentrule#  // Fail with simple cancellation.
 #\indentrule# #\indentrule#  if cancellationMode == SIMPLE: #\label{line:resumeAndCancel:resume:isCancelled_failSimple0}#
 #\indentrule# #\indentrule#  #\indentrule#  return false #\label{line:resumeAndCancel:resume:isCancelled_failSimple1}#
 #\indentrule# #\indentrule#  // Skip the cell in the smart mode
 #\indentrule# #\indentrule#  return resume(value) #\label{line:resumeAndCancel:resume:isCancelled_skipSmart}#
 #\indentrule# #\indentrule#cur == REFUSE: #\label{line:resumeAndCancel:resume:isRefused}#
 #\indentrule# #\indentrule#  ##@completeRefusedResume(value)@#\label{line:resumeAndCancel:resume:completeRefusedResume}#
 #\indentrule# #\indentrule#  return true #\label{line:resumeAndCancel:resume:finishRefuse}# #\label{line:resumeAndCancel:resume:modifyCell:end}#
 #\indentrule# }
}
\end{lstlisting}
\end{minipage}
\hfill
\begin{minipage}[t]{0.49\textwidth}
\begin{lstlisting}[firstnumber=32]
fun cancellationHandler(s: Segment, 
                        i: Int) {
 // Which cancellation mode do we use?
 if cancellationMode == SIMPLE: #\label{line:resumeAndCancel:handler:checkSimple}#
 #\indentrule#  // Mark the cell state to  
 #\indentrule#  // CANCELLED and finish.
 #\indentrule#  s[i] = CANCELLED #\label{line:resumeAndCancel:handler:simple_processCell0}#
 #\indentrule#  ##^s.onCancelledCell()^ #\label{line:resumeAndCancel:handler:simple_processCell1}# 
 #\indentrule#  return #\label{line:resumeAndCancel:handler:simpleEnd}#
 // Smart cancellation mode is used.
 markCancelled := @onCancellation()@ #\label{line:resumeAndCancel:handler:onCancellation}#
 if markCancelled: 
 #\indentrule#  // Mark the cell as CANCELLED.
 #\indentrule#  old := GetAndSet(&s[i], CANCELLED)#\label{line:resumeAndCancel:handler:markCancelled}#
 #\indentrule#  // Did it store an aborted thread?
 #\indentrule#  if old is Thread: #\label{line:resumeAndCancel:handler:isRequest}#
 #\indentrule#  #\indentrule#  ##^s.onCancelledCell()^ #\label{line:resumeAndCancel:handler:onCancelledCellSmart}#
 #\indentrule#  else: // old is a value of type T
 #\indentrule#  #\indentrule#  // A concurrent resume(..) has 
 #\indentrule#  #\indentrule#  // delegated its completion.
 #\indentrule#  #\indentrule#  resume(old) #\label{line:resumeAndCancel:handler:resume}#
 else: #\label{line:resumeAndCancel:handler:onCancellationFalse}#
 #\indentrule#  // Move the cell state to REFUSE.
 #\indentrule#  old := GetAndSet(&s[i], REFUSE) #\label{line:resumeAndCancel:handler:markRefused}#
 #\indentrule#  // Did it store an aborted thread?
 #\indentrule#  if old is Thread: return #\label{line:resumeAndCancel:handler:successfullyRefused}#
 #\indentrule#  // A concurrent resume(..) has 
 #\indentrule#  // delegated its completion;
 #\indentrule#  // old is a value of type T.
 #\indentrule#  ##@completeRefusedResume(old)@ #\label{line:resumeAndCancel:handler:completeRefusedResume}#
}
\end{lstlisting}
\end{minipage}
\vspace{-1.3em}
\caption{Pseudocode for \texttt{resume(..)} that supports all cancellation modes and the corresponding cancellation handler. The \texttt{suspend()} implementation stays the same.
The user-specified operations are highlighted in yellow.
The \texttt{onCancelledCell()} operation, highlighted with green, informs the segment about a new cancelled cell {---} we have to remove segments full of cancelled cells to avoid memory leaks; the details are discussed in Appendix~\ref{appendix:infarr}.
}
\label{listing:resume_and_cancellation}
\end{figureAsListingWide}

\paragraph{The \texttt{resume(..)} Operation.}
Listing~\ref{listing:resume_and_cancellation} presents a pseudocode for \texttt{resume(..)} that supports all cancellation modes and for the cancellation handler {---} the function that is invoked when \texttt{Thread} becomes cancelled; it is set in the \texttt{park(..)} invocation (see Listing~\ref{listing:threadsapi}). For simplicity, we assume that CQS uses an infinite array in \texttt{resume(..)}; the changes required for support of cancellation in its emulation are discussed in Appendix~\ref{appendix:infarr}. 

Like before, \texttt{resume(..)} increments \texttt{resumeIdx} first (line \ref{line:resumeAndCancel:resume:inc}). After that, the corresponding cell should be modified {---} this logic is wrapped with a \texttt{while(true)} loop (lines~\ref{line:resumeAndCancel:resume:modifyCell:start}--\ref{line:resumeAndCancel:resume:modifyCell:end}); the current cell state is obtained in the beginning of it (line~\ref{line:resumeAndCancel:resume:curState}). 
When the cell is empty (line~\ref{line:resumeAndCancel:resume:empty}), \texttt{resume(..)} tries to set the resumption value to the cell (lines~\ref{line:resumeAndCancel:resume:tryElimination0}--\ref{line:resumeAndCancel:resume:tryElimination1}). If the corresponding \texttt{CAS} succeeds, this \texttt{resume(..)} finishes immediately. If the \texttt{CAS} fails, the cell modification procedure restarts. 


When the cell stores a suspended thread (line~\ref{line:resumeAndCancel:resume:isRequest}), \texttt{resume(..)} tries to complete it (line~\ref{line:resumeAndCancel:resume:tryComplete}). If successful, the cell value is cleared for garbage collection, and the operation finishes (lines~\ref{line:resumeAndCancel:resume:resumed0}--\ref{line:resumeAndCancel:resume:resumed1}). Otherwise, the thread has been cancelled. In the simple cancellation mode, \texttt{resume(..)} simply fails (lines~\ref{line:resumeAndCancel:resume:simpleFail0}--\ref{line:resumeAndCancel:resume:simpleFail1}). 
With the smart cancellation, \texttt{resume(..)} tries to replace the cancelled waiter with the resumption value, thus, delegating its completion to the cancellation handler, and finishes on success (line~\ref{line:resumeAndCancel:resume:putValue0}--\ref{line:resumeAndCancel:resume:putValue1}). On failure, one of the branches below will be entered.


\noindent
When the cell is in \texttt{CANCELLED} state (line~\ref{line:resumeAndCancel:resume:isCancelled}), \texttt{resume(..)} either fails in the simple cancellation mode (lines~\ref{line:resumeAndCancel:resume:isCancelled_failSimple0}--\ref{line:resumeAndCancel:resume:isCancelled_failSimple1}), or skips this cell in the smart one, invoking \texttt{resume(..)} one more time (line~\ref{line:resumeAndCancel:resume:isCancelled_skipSmart}). In Appendix~\ref{appendix:infarr}, we describe how to skip a sequence of \texttt{CANCELLED} cells in $O(1)$ under no contention, with the infinite array implemented as a linked list of segments.

In the remaining case, when the cell is in the \texttt{REFUSE} state (line~\ref{line:resumeAndCancel:resume:isRefused}), this \texttt{resume(..)} should be refused, and \texttt{completeRefusedResume(..)} is called (line~\ref{line:resumeAndCancel:resume:completeRefusedResume}). After that, the operation successfully finishes (line~\ref{line:resumeAndCancel:resume:finishRefuse}).


\paragraph{The Cancellation Handler.}
The cancellation handler can be specified as a parameter of the \texttt{park(..)} call (see Listing~\ref{listing:threadsapi}) and is invoked when the thread becomes aborted. Here, the \texttt{cancellationHandler(..)} function accepts the segment and the location index of the cell inside it {---} we know them at the point of invoking \texttt{park(..)}, so the handler has access to the cell and can update its state to \texttt{CANCELLED} or \texttt{REFUSE}.

In the first case, when the simple cancellation mode is used (lines~\ref{line:resumeAndCancel:handler:checkSimple}--\ref{line:resumeAndCancel:handler:simpleEnd}), the cell state is always updated to \texttt{CANCELLED} and a special \texttt{onCancelledCell()} function is invoked on the segment (lines~\ref{line:resumeAndCancel:handler:simple_processCell0}--\ref{line:resumeAndCancel:handler:simple_processCell1}). This \texttt{onCancelledCell()} function signals that one more cell in this segment was cancelled and removes the segment if all the cells become cancelled (see Appendix~\ref{appendix:infarr} for details).

With smart cancellation, \texttt{onCancellation()} is invoked first (line~\ref{line:resumeAndCancel:handler:onCancellation}). If it succeeds (returns \texttt{true}), then the cell state can be moved to \texttt{CANCELLED}. However, a concurrent \texttt{resume(..)} may come and replace the aborted thread with its resumption value, see the cell state diagram in Figure~\ref{fig:cell_smart_cancellation}. Therefore, we put the \texttt{CANCELLED} token via an atomic \texttt{GetAndSet} operation, which returns the previous cell state (line~\ref{line:resumeAndCancel:handler:markCancelled}).
If a thread instance was stored in the cell (line~\ref{line:resumeAndCancel:handler:isRequest}), \texttt{resume(..)} has not come there: the handler signals about a new cancelled cell, removing the segment if needed (line~\ref{line:resumeAndCancel:handler:onCancelledCellSmart}), and finishes. Otherwise, if a resumption value was stored in the cell, the cancellation handler completes the corresponding resumption by invoking \texttt{resume(..)} with this value (line~\ref{line:resumeAndCancel:handler:resume}).

In case \texttt{onCancellation()} returns \texttt{false} (line~\ref{line:resumeAndCancel:handler:onCancellationFalse}), the matching \texttt{resume(..)} should be refused. Thus, the cell state moves to \texttt{REFUSE} via an atomic \texttt{GetAndSet} (line~\ref{line:resumeAndCancel:handler:markRefused}). If the cell stored the cancelled thread, the handler finishes (line~\ref{line:resumeAndCancel:handler:successfullyRefused}). Otherwise, a concurrent \texttt{resume(..)} has replaced it with the resumption value {---} we complete it with \texttt{completeRefusedResume(..)} (line~\ref{line:resumeAndCancel:handler:completeRefusedResume}).
\section{Synchronization Primitives on Top of CQS}\label{sec:primitives}
To show the expressiveness of the \sqs{} framework, we present several algorithms developed on top of it. Starting with the barrier, we present a new count-down-latch algorithm, then several semaphore algorithms, and finish with blocking pools. 


\subsection{Barrier}
A simple but popular synchronization abstraction is the \emph{barrier}, which allows a set of parallel threads wait for each other at a common program point, via a provided \texttt{arrive()} operation.

\setlength{\columnsep}{1.7em}
\begin{wrapfigure}[9]{r}{0.52\textwidth}
\vspace{-1.65em}
\begin{lstlisting}[xleftmargin=0em,
  label={listing:barrier},
  caption={Barrier algorithm via CQS. 
  %Like the standard Java implementation, it does not support cancellation as it would require a practically unavailable ability to resume multiple waiters atomically.
  }
]
val cqs = CQS<Unit>()
var remaining: Int = parties #\label{line:barrier_remaining}# 

fun arrive() {
  r := FAA(&remaining, -1) #\label{line:barrier_faa}#
  if r > 1: return cqs.suspend() #\label{line:barrier_susp}#
  repeat(parties - 1) { cqs.resume(Unit) }#\label{line:barrier_resume}#
}
\end{lstlisting}
\end{wrapfigure}

\paragraph{Algorithm.}
Listing~\ref{listing:barrier} on the right presents the algorithm on top of CQS. The implementation is straightforward: it maintains a counter of the parties who arrived (line~\ref{line:barrier_remaining}) and increments it in the beginning of the \texttt{arrive()} operation (line~\ref{line:barrier_faa}). All but the last \texttt{arrive()} invocations suspend (line~\ref{line:barrier_susp}), while the latter one resumes all those who previously arrived (line~\ref{line:barrier_resume}).

\noindent
Once the last thread arrives, all the waiters should be resumed. However, if any of these waiters becomes cancelled, the barrier contract is violated {---} fewer waiters will be successfully resumed and overcome the barrier. Unfortunately, solving this problem would require an ability to \emph{atomically} resume a set of waiters (so either all the waiters are resumed or none), but no real system provides such a primitive.
Thus, similarly to the implementation in Java, we do not support cancellation. However, instead of breaking the barrier when a thread is cancelled, we ignore cancellation. The intuition behind this design is that even if a waiter has been cancelled, it has successfully reached the barrier point and should not block the other parties from continuing.



\subsection{Count-Down-Latch}
The next synchronization primitive we consider is the \emph{count-down-latch}, which allows waiting until the specified number of operations are completed.
It is initialized with a given count, and each \texttt{countDown()} invocation decrements the number of operations yet to be completed. Meanwhile, the \texttt{await()} operation suspends until the count reaches zero.

\paragraph{Basic Algorithm.}
The pseudocode of our count-down-latch implementation is presented in Listing~\ref{listing:cdl}. 
Essentially, the latch maintains two counters: \texttt{count}, representing the number of remaining operations (line~\ref{line:cdl_count}), and \texttt{waiters}, which stores the number of pending \texttt{await()}-s (line~\ref{line:cdl_waiters}). 

\begin{figureAsListingWide}
\vspace{-0.6em}
\begin{minipage}[t]{0.47\textwidth}
\begin{lstlisting} 
val cqs = CQS<Unit>(
 cancellationMode = SMART #\label{line:cdl_sqs_2}#
)
// initialized by user
var count: Int = initCount #\label{line:cdl_count}# 
// the number of waiters 
var waiters: Int = 0 #\label{line:cdl_waiters}# 

fun countDown() {
 r := FAA(&count, -1) #\label{line:cdl_cd_inc}#
 // Has the counter reached zero?
 if r <= 1: resumeWaiters()  #\label{line:cdl_cd_rw}#
}

fun await() {
 if count <= 0: return #\label{line:cdl_a_check}#
 w := FAA(&waiters,#\thinspace#+1) #\label{line:cdl_a_winc}#
 // Is DONE_BIT set?
 if w#\thinspace#&#\thinspace#DONE_BIT != 0: return #\label{line:cdl_a_bitcheck}#
 // Suspend until count reaches zero
 cqs.suspend() #\label{line:cdl_a_susp}#
}
\end{lstlisting}
\end{minipage}
\hfill
\begin{minipage}[t]{0.52\textwidth}
\begin{lstlisting}[firstnumber=23]
fun resumeWaiters() = while(true) {
 w := waiters
 // Is DONE_BIT set?
 if w#\thinspace#&#\thinspace#DONE_BIT != 0: return #\label{line:cdl_rw_bitcheck}#
 // Set DONE_BIT and resume waiters.
 if CAS(&waiters, w, w#\thinspace#|#\thinspace#DONE_BIT): #\label{line:cdl_rw_setbit}#
 #\indentrule#  repeat(w) { cqs.resume(Unit) }#\label{line:cdl_rw_resume}#
 #\indentrule#  return
}

fun onCancellation(): Bool { #\label{line:cdl_onc_start}#
 w := FAA(&waiters, -1) #\label{line:cdl_onc_decw}#
 // Move the cell to CANCELLED if the  
 // bit is unset; otherwise, to REFUSE.
 return w#\thinspace#&#\thinspace#DONE_BIT == 0 #\label{line:cdl_onc_bitcheck}#
} #\label{line:cdl_onc_end}#

fun completeRefusedResume(token: Unit) {#\label{line:crr0}#
 // Ignore cancelled await()-s.
} #\label{line:crr1}#

const DONE_BIT = 1 << 31
\end{lstlisting}
\end{minipage}
\vspace{-1em}
\caption{Count-down-latch implementation on top of CQS with smart cancellation. 
When manipulating with \texttt{DONE\_BIT}, we use bitwise "and", "or", and "left shift" operators, denoted as \textbf{\texttt{\&}}, \textbf{\texttt{|}}, and \textbf{\texttt{<}}\textbf{\texttt{<}}, respectively.}
\label{listing:cdl}
\vspace{-0.5em}
\end{figureAsListingWide}

The \texttt{countDown()} function is straightforward: it decrements the number of remaining operations (line~\ref{line:cdl_cd_inc}), resuming the waiting \texttt{await()}-s if the count reached zero (line~\ref{line:cdl_cd_rw}).\footnote{We allow the number of \texttt{countDown()} calls to be greater than the initially specified one. However, we could check in \texttt{countDown()} that the counter is still non-negative after the decrement, throwing an exception otherwise.}
Meanwhile, \texttt{await()} checks whether the counter of remaining operations has already reached zero, immediately completing in this case (line~\ref{line:cdl_a_check}). 
If \texttt{await()} observes that \texttt{count} is positive, it increments the number of waiters (line~\ref{line:cdl_a_winc}) and suspends (line~\ref{line:cdl_a_susp}). 

\noindent
Given that \texttt{resumeWaiters()}, which is invoked by the last \texttt{countDown()}, can be executed concurrently with \texttt{await()}, they should synchronize. For this purpose, \texttt{resumeWaiters()} sets the \texttt{DONE\_BIT} in the \texttt{waiters} counter (line~\ref{line:cdl_rw_setbit}), forbidding further suspensions and showing that this count-down-latch has reached zero. Thereby, \texttt{await()} checks for this \texttt{DONE\_BIT} before suspension and completes immediately if the bit is set (line~\ref{line:cdl_a_bitcheck}).



\paragraph{Cancellation.} 
The simplest way to support cancellation is to do nothing: the algorithm already works with the simple cancellation mode, where \texttt{resume(..)}-s silently fail on cancelled \texttt{await()} requests (line~\ref{line:cdl_rw_resume}). 
This strategy results in resuming cancelled waiters, which makes \texttt{resumeWaiters()} work in a linear time on the total number of \texttt{await()} invocations, including the aborted~ones.

Smart cancellation, on the other hand, makes it possible to optimize \texttt{resumeWaiters()} so that the number of steps is bounded by the number of non-cancelled \texttt{await()}-s. The \texttt{onCancellation()} function 
is invoked when a waiter becomes cancelled. It attempts to decrement the number of waiters (line~\ref{line:cdl_onc_decw}), making \texttt{resume(..)} skip the corresponding cell in the CQS. However, if the \texttt{DONE\_BIT} is already set at the moment of the decrement, a concurrent \texttt{resumeWaiters()} is going to resume this cancelled waiter. The corresponding \texttt{resume(..)} call should be ignored, so \texttt{onCancellation()} returns \texttt{false}, while \texttt{completeRefusedResume(..)} does nothing (lines~\ref{line:crr0}--\ref{line:crr1}).

\subsection{Semaphores}\label{subsec:sema}
The barrier and count-down latch algorithms described above do not actually require waiting requests to be resumed in FIFO order. However, this property is critical for some primitives such as the mutex or the semaphore. While the mutex allows at most one thread to be in the critical section protected by \texttt{lock()} and \texttt{unlock()} invocations, the semaphore is a generalization of mutex that allows the specified number of threads to be the critical section simultaneously by taking a permit via \texttt{acquire()} and returning it back via \texttt{release()}. 

In fact, the semaphore algorithm is almost the same as the one for the mutex, presented under the CQS framework presentation in Listing~\ref{listing:mutex_basic} (the basic version) and Listing~~\ref{listing:mutex_smart_cancellation} (the cancellation part). The only difference is that the \texttt{state} field is initialized with $K$ instead of $1$, when $K$ is the number of threads allowed to be in the critical section concurrently. We present the implementation details in Appendix~\ref{appendix:primitives}.

\vspace{-0.2em}
\subsection{Blocking Pools}\label{subsec:pools}
\vspace{-0.1em}
While the previous algorithms use \sqs{} only for synchronization, it is also possible to develop \emph{communication} primitives on top of it. Here, we discuss two blocking pool implementations. 
When using expensive resources such as database connections or sockets, it is common to reuse them {---} this usually requires an efficient and accessible mechanism. The \emph{blocking pool} abstraction maintains a set of elements that can be retrieved in order to process some operation, after which the element is placed back in the pool:
\begin{itemize}[nosep]
 \item \texttt{take()} either retrieves one of the elements (in an unspecified order), suspending until an element appears if the pool is empty;
 \item \texttt{put(element)} either resumes the first waiting \texttt{take()} operation and passes the element to it, or puts the element into the pool.
\end{itemize}

Intuitively, the blocking pool contract reminds the semaphore one. Similarly to the semaphore, it transfers resources, with the only difference being that the semaphore shares logical non-distinguishable permits while blocking pool works with real elements. The rest is almost the~same. 
In Appendix~\ref{appendix:primitives}, we present two pool implementations: queue-based and stack-based. Intuitively, the queue-based implementation is faster since queues can be built on segments similar to CQS and leverage \texttt{Fetch-And-Add} on the contended path. In contrast, the stack-based pool retrieves the last inserted, thus the ``hottest'' element.
Please note that both algorithms we discuss are \emph{not} linearizable and can retrieve elements out-of-order under some races. However, since pools do not guarantee that the stored elements are ordered, these queue- and stack-based versions should be considered bags with specific heuristics; these semantics matches practical applications.


\section{Correctness and Progress Guarantees}
In this section, we discuss correctness and progress guarantees for both CQS operations and the primitives we built on top of the framework.

\subsection{Formal Proofs of Correctness in Iris/Coq}
Correctness is formally proven in the state-of-the-art concurrent higher-order separation logic Iris~\cite{jung2018iris} using its Coq formalization~\cite{krebbers2017interactive}. Here, we highlight the key ideas behind the proofs and discuss their limitations. \textbf{The source code of the proofs is available on GitHub~\cite{proofRepo}. We complement this with a detailed outline in Appendix~\ref{sec:proofs}.}


\paragraph{Operation Specifications.} 
Iris is a framework designed for reasoning about the safety of concurrent programs, and several non-trivial algorithms have already been formally proved using it~\cite{krogh2016verifying, krishna2020verifying, vindum2021contextual, jung2017rustbelt, chajed2021gojournal, vindum2022mechanized, carbonneaux2022applying}. When constructing formal proofs, one should provide a specification for each of the data structure operations. In the Iris logic, operations manipulate \emph{resources}, which are pieces of knowledge about the system-wide state and can be held by threads or the data structure itself. These resources are \emph{logical} and do not affect the program execution. A specification describes which resources are required for the operation to start and how they change when it finishes.

Consider again the mutex as an example. Intuitively, we parameterize it with a resource $R$, which serves as an exclusive right to be in the critical section and, thus, to invoke \texttt{unlock()}. Initially, this resource $R$ is held by the mutex object. The specification ensures that:
\begin{enumerate}[nosep]
    \item when the \texttt{lock()} operation finishes, the resource $R$ is transferred to the caller thread, and
    \item when \texttt{unlock()} starts, the corresponding thread must provide $R$.
\end{enumerate}
If the resource $R$ is unique, the specification still holds; however, as it cannot be held by multiple threads by construction, the mutual exclusion contract is satisfied. 

All our specifications are defined in a similar manner. For example, to specify the semaphore contract, we simply need to maintain $K$ non-distinguishable copies of $R$; thus, allowing at most $K$ threads to enter the critical section. However, the actual specifications in Coq contain many additional details, mainly due to support for cancellation semantics. Please refer to the proofs outline in Appendix~\ref{sec:proofs} and the source code~\cite{proofRepo} for details.


\vspace{-0.1em}
\paragraph{Modularity.}
Our Iris proofs are \emph{modular}: specifications treat each operation separately and do not concern the state of the system as a whole, locally manipulating logical resources instead.
As a result, the proof of CQS itself spans 8000 lines of Coq; by comparison, the proof of the barrier, including its definition, takes only 400 lines, the semaphore proof requires less than 300 lines, and the proofs for the count-down-latch and blocking pools take up to 700 lines each. Modularity dramatically reduces the effort for someone wishing to formally verify their CQS-based primitive.


\vspace{-0.1em}
\paragraph{Limitations.}
One main limitation is that the existing formal specifications do not highlight the FIFO semantics, allowing the waiting operations to complete in any order. Instead, these specifications verify high-level properties, such as ``at most one thread can be in the critical section'' for the mutex. This limitation stems from the modularity of proofs and the fact that the user code parameterizes the cancellation handler in  CQS. The fairness of end-to-end structures on top of the CQS is easy to see by the analogy with the state-of-the-art linearizable queues~\cite{LCRQ,YM16}, but proofs of such form are not modular. While the modular Iris proofs are powerful enough to show fairness, this requires significant effort even for simple data structures such as the classic Michael-Scott queue~\cite{vindum2021contextual}, and constructing them for non-trivial and, especially, higher-order structures like CQS is currently impractical. Most importantly, a modular proof of fairness of structures on top of the CQS would require placing highly involved contracts on the cancellation handler as well as the uses of \texttt{suspend} operations that may interact with it, making it significantly more difficult to prove the correctness of primitives on top of the CQS for the end user.

Another limitation of the provided Iris specifications is that they do not assert the lack of memory leaks. In particular, they do not prevent us from always storing the whole infinite array. Nevertheless, the lack of memory leaks follows by construction, as we always physically remove segments full of canceled cells. Beyond that, we have thoroughly tested our implementation for the absence of memory leaks via the Lincheck framework~\cite{LINCHECK}, which enables model checking of concurrent algorithms on the JVM.

Finally, we assume a strong sequentially-consistent memory model. We find this assumption reasonable as almost all the operations that manipulate shared data are atomic in the presented algorithms, while considering relaxed memory may significantly increase the proofs complexity~\cite{dang2019rustbelt,kaiser2017strong}. We also rely on the SC-DRF (sequential consistency for data-race-free programs) property of all real-world weak memory models, such as C++11 and JMM, which makes reasoning in the strong memory model sufficient. However, we plan to extend our proofs to support the release-acquire semantics~\cite{kaiser2017strong} and, thus, match the LLVM memory model for languages such as C/C++ and Rust.

\vspace{-0.4em}
\subsection{Progress Guarantees}
\vspace{-0.1em}
Similarly to the dual data structures formalism~\cite{scherer2006scalable}, we reason about progress independently of whether the operation was suspended. When we say that some blocking operation is lock- or wait-free, we mean that it performs all the synchronization with this progress guarantee, either completing immediately or adding itself to the queue of waiters followed by suspension.


Unfortunately, the progress guarantees cannot be mechanized in our Iris proofs. The reason for this is that, at the time of writing, there are two forms of specifying program behavior in Iris. The first way is to use \emph{(partial) weakest preconditions}, which do not ensure that an operation terminates. In fact, an infinite loop satisfies any such specification.
The second less popular form uses the \emph{total weakest precondition}~\cite{jung2018iris}, which requires that every operation must terminate in a bounded number of steps. This type of specification can be used to show wait-freedom of algorithms, but is not applicable to our case, as some of the operations guarantee only lock-freedom. 

We do not consider the lack of formal proof of progress guarantees a major issue. 
Although it is possible to write such proofs (see~\cite{provingLockFreedomEasily} for a comprehensive analysis), we find it much easier to discuss this question separately.
In essence, most of the presented primitives including the CQS framework itself guarantee wait-freedom under no cancellation and at least lock-freedom when requests may abort. We provide a detailed analysis in Appendix~\ref{appendix:progress_guarantee}.


\section{Evaluation}\label{sec:experiments}
Our main practical contribution is integrating CQS, along with the mutex and semaphore implementations, into the standard Kotlin Coroutines library~\cite{kotlincoroutines}. Other presented synchronization~and communication primitives are implemented in tests, enabling their fast development when needed. 

\noindent
To validate performance, we implemented \sqs{} on the JVM and compared it against the state-of-the-art \texttt{AbstractQueuedSynchronizer} framework for implementing synchronization primitives in Java~\cite{lea2005java}. The latter provides similar semantics to CQS, and is the only practical framework that addresses the same general problem. Notably, CQS-based algorithms  are significantly more straightforward to reason.
For fair performance evaluation, we use threads as waiters in CQS; it should benefit the Java implementation, which is well-optimized for this case. 

Our implementations for coroutines in Kotlin and native threads in Java confirm the flexibility of our design, let alone matching the real-world semantics. 




\paragraph{Experimental Setup.} 
Experiments were run on a server with 4 Intel Xeon Gold 6150 (Skylake) sockets; each socket has 18 2.70 GHz cores, each of which multiplexes 2 hardware threads, for a total of 144 hardware threads. We used \texttt{OpenJDK 15}
in all the experiments and the \texttt{Java Microbenchmark Harness (JMH)} library~\cite{jmh} for running benchmarks.
When measuring operations, we also add some uncontended work after each operation invocation {---} the work size is geometrically distributed with a fixed mean, which we vary in benchmarks. In our CQS implementation, we have chosen the segment size of \texttt{64} based on minimal tuning.


\subsection{Barrier}
We compare the CQS-based barrier implementation with the standard one in Java. In addition, we add a baseline counter-based solution, which is organized in the same way as ours, but performs active waiting instead of suspension, spinning in a loop until the \texttt{remaining} counter becomes zero. 

\paragraph{Benchmark.}
Each of the threads performs barrier point synchronizations followed by some uncontended work. This process is repeated a fixed number of times. 
We measure a single synchronization phase, a set of \texttt{arrive()}-s with additional work for each thread. Without any synchronization, the execution time is expected to stay the same independently of the number of threads. 

\begin{figure}[h]
    \vspace{-0.3em}
    \centering
    \includegraphics[width=0.98\textwidth]{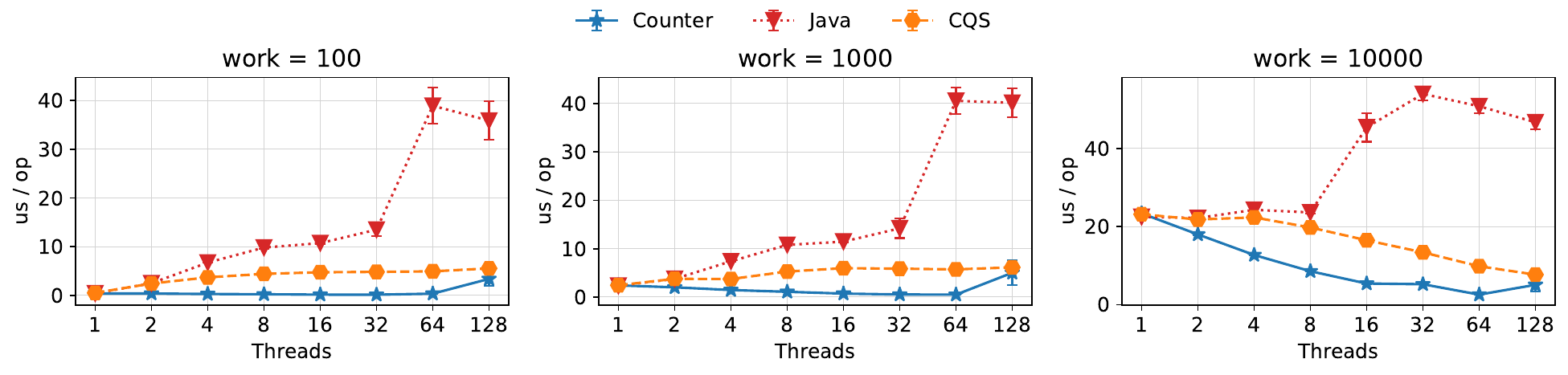}
    \vspace{-1em}
    \caption{Evaluation of \sqs{}-based barrier implementation against the standard one available in Java and a simple counter-based solution with active waiting. The plots show average time per synchronization phase, \textbf{lower is better}.}
    \vspace{-0.3em}
    \label{fig:barrier_experiment}
\end{figure}

\paragraph{Results.}
The experimental results are presented in Figure~\ref{fig:barrier_experiment}. We evaluated all three algorithms on various numbers of threads and with three average work sizes {---} \texttt{100}, \texttt{1000}, and \texttt{10000} uncontended loop iterations on average. The graphs show an average time per operation, so lower is better. 


Since one of the main synchronization costs is thread suspensions and resumptions, the counter-based solution with active waiting is predictably the fastest. Nevertheless, our CQS-based algorithm shows similar performance trends, due to all the operations being based on \texttt{Fetch-And-Add}. In contrast, the solution from the standard concurrency library in Java shows significantly less scalability{---}we find the reason for such performance degradation in using a mutex under the hood; surprisingly, it does not use \texttt{AbstractQueuedSynchronizer} directly.
As a result, we find our simple CQS-based algorithm to provide superior performance.

\subsection{Count-Down-Latch}
Next, we evaluate our count-down latch implementation against the one in Java's concurrency package, which is built on top of the \texttt{AbstractQueuedSynchronizer} framework.

\paragraph{Benchmark.}
We consider a workload with a fixed number of \texttt{countDown()} invocations distributed among threads, each followed by additional uncontended work. Besides, we add a baseline that does not invoke \texttt{countDown()} and only performs the work. Thus, comparing with this baseline we can measure the overhead caused by the count-down-latch synchronization.
%

\begin{figure}[h]
    \centering
    \includegraphics[width=\textwidth]{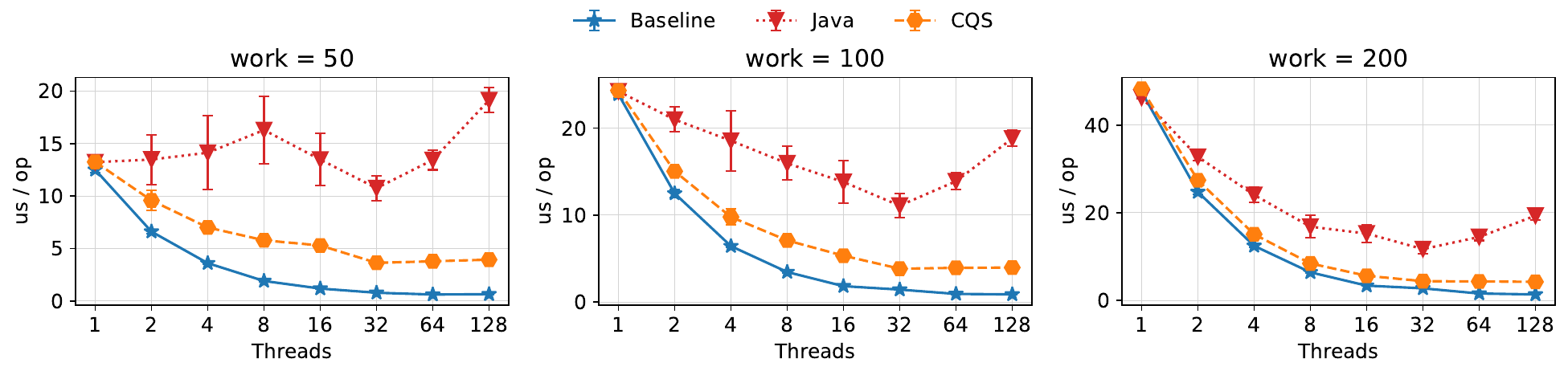}
    \vspace{-1.5em}
    \caption{Evaluation of the CQS-based count-down-latch implementation against the standard one in Java. The baseline illustrates how the operation time would change if \texttt{coundDown()} took no time.  \textbf{Lower is better.}}
    \label{fig:cdl_experiment}
    \vspace{-1em}
\end{figure}

\paragraph{Results.}
Figure~\ref{fig:cdl_experiment} shows the evaluation results with different additional work sizes ($50$ uncontended loop iterations on average on the left, $100$ in the middle, and $200$ on the right). 
It is apparent that the CQS implementation significantly outperforms the standard one from Java, by up to $4\times$. Compared to the baseline, it follows the same trend, providing an extremely small overhead on the right graph, where the work is $200$ uncontended loop cycles.

Similar to our CQS-based algorithm, the implementation in Java maintains a counter of remaining \texttt{countDown()} invocations. However, they update this counter in a \texttt{CAS} loop: the algorithm reads the current counter value and tries to replace it with the reduced by one via \texttt{CAS}, restarting the process on failure. We find this difference the main reason for the superior scalability of our solution. 

\subsection{Mutex and Semaphores}

Since the semaphore is a generalization of the mutex, we equate its implementation with $K=1$  permits as mutual exclusion.
We compare our algorithm against alternatives from the standard Java library, unfair implementations of mutex and semaphore in Java, and the state-of-the-art fair CLH and MCS lock algorithms.
In Section~\ref{sec:basic} we also mention that implementing non-blocking \texttt{Mutex.tryLock()} and \texttt{Semaphore.tryAcquire()} operations would require extending CQS with a special \emph{synchronous} resumption mode, leaving the details to Appendix~\ref{appendix:sync_async}.
We included both semaphore implementations in the experiment to show that the complexity introduced by this \texttt{synchronous} resumption mode does not affect performance.


\paragraph{Benchmark.}
Consider the workload of many operations to be executed by the specified number of threads with the parallelism level restricted via semaphore. Thus, each operation invocation is wrapped with the \texttt{acquire()-release()} pair. When the parallelism level equals \texttt{1}, the semaphore is de facto a mutex, so we can compare our semaphore against other mutex algorithms. As before, the operations are simulated with uncontended geometrically distributed work. In addition, we perform some work before acquiring a permit, thus, simulating a preparation phase for the operation guarded by the semaphore.
We used \texttt{100} uncontended loop iterations on average for both pieces of work; the results for other work sizes do not differ significantly and, therefore, are omitted.

\begin{figure*}
    \centering
    \includegraphics[width=1\textwidth]{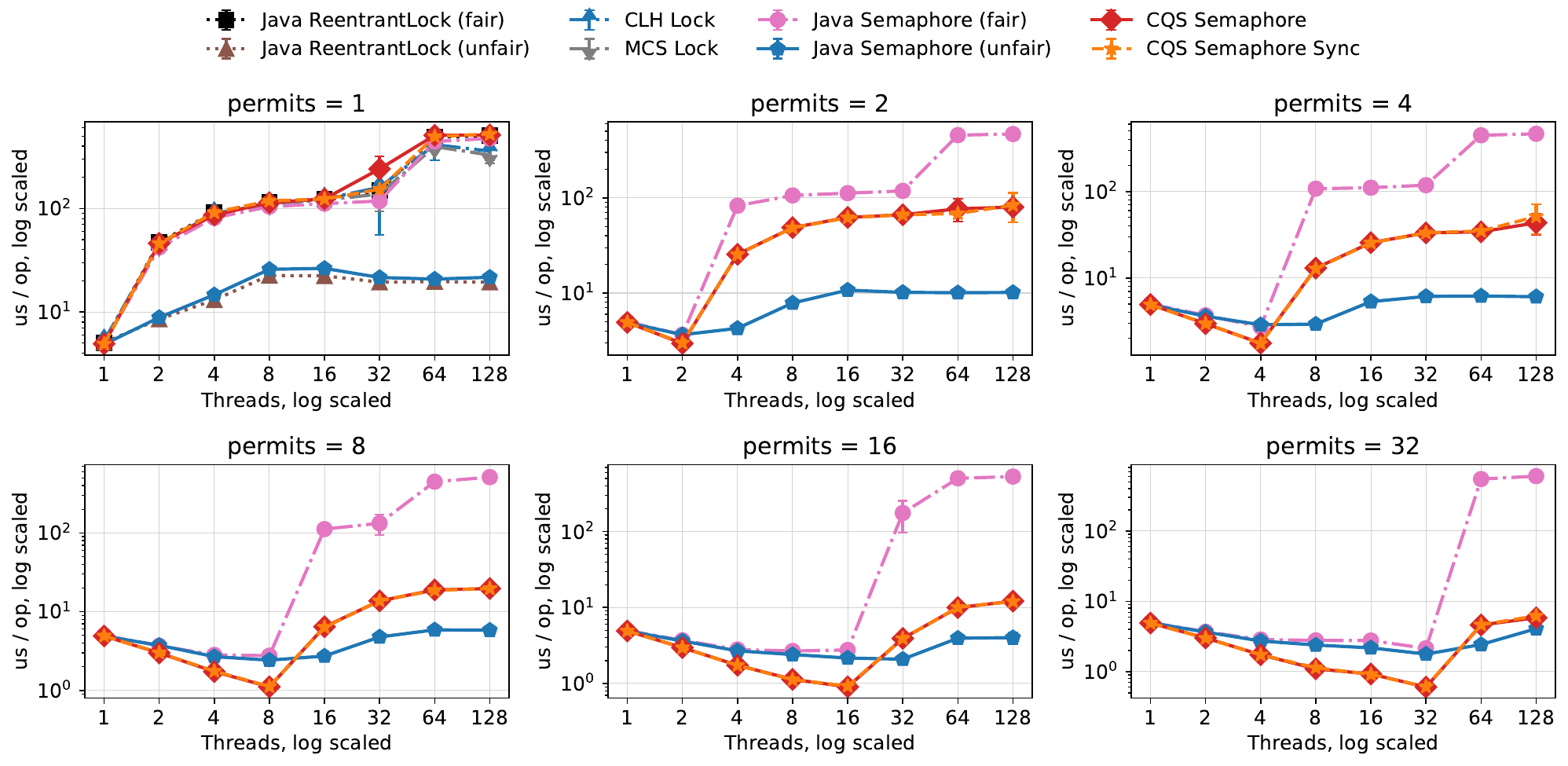}
    \caption{Evaluation of CQS-based semaphore implementations compared to the standard ones in Java, including the unfair variants. In addition, we compare our semaphores against the standard fair and unfair lock implementations in Java and the classic CLH and MCS fair mutexes.
    \textbf{Lower is better.}}
    \label{fig:sema}
    \vspace{-1em}
\end{figure*}

\paragraph{Results.}
The results against both fair and unfair versions of the standard \texttt{ReentrantLock} and \texttt{Semaphore} primitives in Java, as well as against the classic CLH~\cite{CLH} and MCS~\cite{MCS} fair locks, are shown in Figure~\ref{fig:sema}. Our semaphore implementation with the \emph{synchronous} resumption mode in CQS, which enables \texttt{tryAcquire()} implementation, is denoted with the suffix <<\texttt{Sync}>>. 

In the mutex scenario, all the fair algorithms show the same performance, while Java's unfair mutex and semaphore are predictably faster, as unfairness significantly reduces context switches under high contention. 

In the semaphore scenario, our solution outperforms both fair and unfair Java implementations by up to \texttt{4x} when the number of threads does not exceed the number of permits (so suspensions do not happen). The main reason is that our solution leverages \texttt{Fetch-and-Add} to update the number of available permits, which can be negative, indicating the number of waiters. In contrast, the implementation in Java must ensure that this number stays non-negative, so it has to perform this update in a \texttt{CAS} loop, reading the current number of available permits, trying to decrement it via \texttt{CAS} if there is a permit to acquire, and restarting if the \texttt{CAS} fails.

With the increase in the number of threads, our algorithm is almost on par with the unfair version when the number of permits $\geq$\texttt{16}, and significantly outperforms the fair one in all scenarios. 
In particular, our semaphore implementation outperforms the standard fair algorithm in Java by up to \texttt{90x} in a highly-contended scenario. More scalable queue design behind CQS is the key to achieving such an outstanding performance. Notably, the complexity introduced by the synchronous resumption is negligible and does not affect results.

\subsection{Blocking Pools}
We implemented both queue- and stack-based pools and compared them against the existing \texttt{ArrayBlockingQueue} (both fair and unfair) and \texttt{LinkedBlockingQueue} collections from the standard Java library. Notably, they do not leverage the \texttt{AbstractQueuedSynchronizer} framework, as it serves only for synchronization, while CQS enables communication out-of-the-box. Relatively, their solutions provide linearizability, while our pools may be non-linearizable when threads abort. This experiment considers all data structures as solutions for pools of shared resources.

\paragraph{Benchmark.}
We use the same benchmark as for semaphores. In essence, we run many operations on the specified number of threads with a shared pool of elements. Each operation performs some work (\texttt{100} uncontended loop iterations on average in our experiment) first, then takes an element, performs some other work with this element (\texttt{100} more loop iterations on average in our experiment), and returns it to the pool at the end. The results with other work amounts are omitted but were examined and do not differ significantly.

\begin{figure*}[ht!]
    \centering
    \includegraphics[width=1\textwidth]{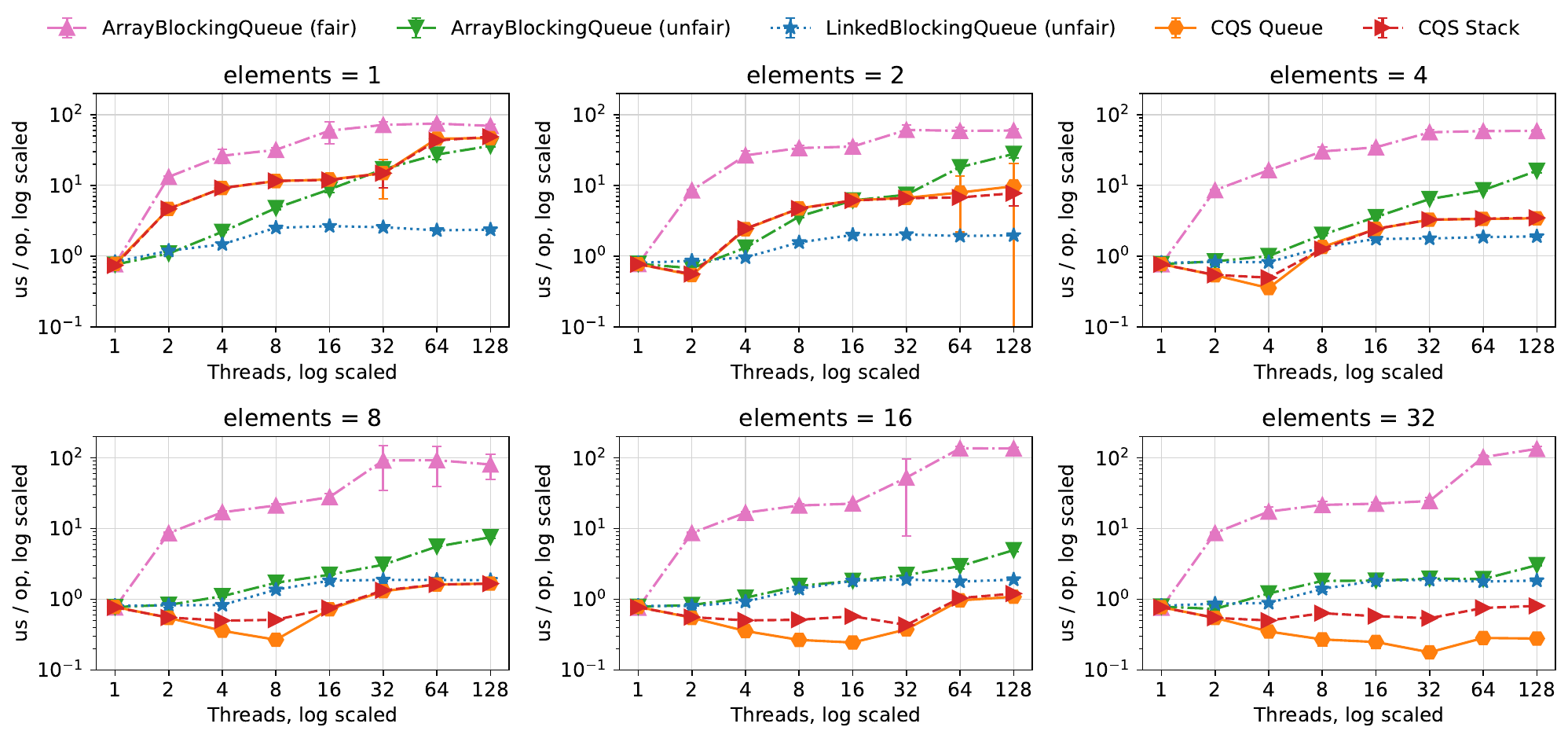}
    \vspace{-1em}
    \caption{Evaluation of the presented queue- and stack-based pool algorithms with various numbers of shared elements against the existing \texttt{ArrayBlockingQueue} (both fair and unfair versions) and \texttt{LinkedBlockingQueue} collections from the standard Java library.
    \textbf{Lower is better.}}
    \label{fig:pools_experiment}
\end{figure*}

\paragraph{Results.}
Figure~\ref{fig:pools_experiment} shows results with different numbers of elements shared in the pool.
First, our queue-based version shows better results on larger numbers of elements, which is expected as the queue perform a \texttt{FAA} on the contended path instead of \texttt{CAS} in the stack-based solution; the latter often fails under high contention, resulting in the operation restart. 
Compared to the fair \texttt{ArrayBlockingQueue}, both of our implementations are more performant by up to \texttt{100}$\times$ times. The synchronization behind \texttt{ArrayBlockingQueue} uses coarse-grained locking, while our solution is non-blocking for storing elements and managing the queue of waiting requests. 

The \emph{unfair} \texttt{LinkedBlockingQueue} is more scalable than the \emph{unfair} version of \texttt{ArrayBlockingQueue}, and they slightly outperform our \emph{fair} implementations on a large number of threads with a small number of shared elements, which is when our solutions suspend a lot. However, both our solutions consistently outperform these unfair primitives by up to \texttt{10}$\times$ times when at least \texttt{8} elements are shared, showing the same or better performance when the number of threads does not exceed the number of elements.



\subsection{Abortability Support}
Up to this point, we have primarily focused on situations where suspended requests do not get aborted. While cancellation performance might not always be crucial, as it typically occurs due to a more resource-intensive coroutine or thread interruption, removing aborted waiters from the queue in constant time remains essential. This is particularly true for coroutines, where thousands may be waiting on a mutex or semaphore. The CQS framework fulfills this need by physically removing aborted threads in $O(1)$ under no contention. In contrast, Java's \texttt{AbstractQueuedSynchronizer} takes linear time in the queue size to remove an interrupted thread.

\begin{wrapfigure}[13]{r}{0.48\textwidth}
    \vspace{-0.7em}
    \centering
    \includegraphics[width=0.48\textwidth]{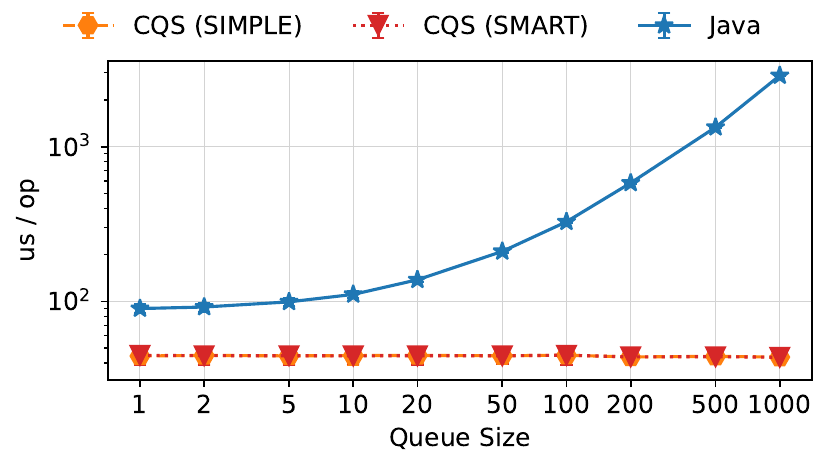}
    \vspace{-2em}
    \caption{Comparison of CQS' and Java's \texttt{AbstractQueuedSynchronizer} cancellation mechanisms. \textbf{Lower is better.}}
    \label{fig:cancellation}
\end{wrapfigure}

To assess the practical impact of constant-time cancellation, we conducted a benchmark comparison between CQS (with both \texttt{SIMPLE} and \texttt{SMART} cancellation modes) and \texttt{AbstractQueuedSynchronizer}. Initially, we populate the synchronization framework with a specified number of suspended threads. After that, we measure the time required to suspend and instantly abort, without parking the native thread. Based on the initial queue size, we anticipate that the performance of Java's solution will degrade, while CQS should consistently deliver the same level of performance. Figure~\ref{fig:cancellation} displays the results, which confirm our hypothesis. In particular, CQS outperforms Java's solution by \texttt{1.9x} on an empty queue, and by \texttt{65x} when the queue contains $1000$ waiters.

Regrettably, it is not feasible to compare the cancellation mechanisms of CQS and \texttt{AbstractQueuedSynchronizer} under concurrent conditions at the time of writing the paper. The reason is that the latter gets into a deadlock in its internal \texttt{cleanQueue()} function.


\section{Related Work}

Our work is part of a wider effort of formalizing and implementing expressive,  safe, and efficient support for asynchronous operations in modern programming languages~\cite{bierman2012pause, okur2014study, ProkopecL18, haller2019reduction, cutner2021safe}. 
In this context, we provide contributions at the level of algorithms, semantics, and formal proofs, with Kotlin/JVM as a practical application. Specifically, we perform one of the first thorough explorations of how fairness and abortability semantics can be efficiently supported at the data structure level, and present one of the first formally-verified such designs for this type of data structure. Importantly, our approach enables high-performance implementations in a range of practical applications, and could serve as a basis for standard library implementations in modern languages. 

We emphasize that few  real-world implementations of similar complexity are formally verified~\cite{krogh2016verifying, krishna2020verifying, vindum2021contextual, jung2017rustbelt, chajed2021gojournal}. 
In line with prior work, we do not formally prove full linearizability, which is notoriously difficult to approach in Iris but can be demonstrated through classical proofs. There are successful linearizability proofs of data structures of comparable complexity using the approach of \emph{contextual refinement}~\cite{vindum2021contextual, vindum2022mechanized} using the ReLoC proof framework. Iris itself permits making specifications \emph{logically atomic}, which can also represent linearizability~\cite{logatomExamples}. We find this approach much less applicable to ensuring linearizability of a \emph{framework}, which depends heavily on the behavior of the code passed to it.

At the algorithmic level, our CQS implementation builds on ideas from both the classic Michael-Scott queue~\cite{michael1996simple} and the highly-efficient LCRQ queue design of Afek and Morrison~\cite{LCRQ}. The latter was also used by Izraelevitz and Scott~\cite{izraelevitz2017generality} to build blocking synchronous queues, and by Koval et al.~\cite{DBLP:conf/europar/KovalAE19, channels_kotlin} to build channels. Relative to these latter modern works,  CQS supports much more general semantics, requiring significant changes to the design, in particular, to support cancellations. 
Specifically, CQS is general enough to provide full support for coroutines, while staying flexible and efficient, whereas these prior design focus on narrower applications, such as blocking queues.  

To our knowledge, the only abstraction that provides similarly-general semantics is the  \texttt{AbstractQueuedSynchronizer} in Java~\cite{lea2005java}, which CQS outperforms by a wide margin due to superior algorithmic design, complemented by formal proofs. 
More precisely, the \texttt{AbstractQueuedSynchronizer} framework combines the classic CLH~\cite{CLH} lock algorithm to maintain the queue of suspended requests with an integer counter, which represents the synchronization primitive state and is updated by \texttt{CAS} operations. 
In contrast, the CQS enables more efficient state updates via \texttt{Fetch-And-Add}-s, also maintaining the queue of waiters with \texttt{FAA}-s on the contended path; thus, providing a more scalable solution.

\section{Discussion}

We have presented a new \sqs{} framework enabling efficient implementations for a whole range of fundamental synchronization primitives in a fair and abortable manner. We observed that the interplay between fairness and cancellation semantics can raise subtle semantic and correctness questions. We found formalization extremely useful when identifying correctness issues in our implementation, notably w.r.t. cancellation semantics. A practical consequence of our work is efficient support for such primitives in the context of Kotlin Coroutines, which we show to generally outperform existing designs offering similar semantics in a wide range of scenarios. 
Specifically, our algorithms on top of CQS outperform existing Java implementations in almost all scenarios and can be faster by orders of magnitude. Surprisingly, the CQS-based primitives frequently surpass even the \emph{unfair} versions of primitives from the standard Java library in our experiments, thanks to the superior scalability of our design.

We believe that CQS could serve as a basis for more complex semantics, designs, and primitives (e.g., fair readers-writer locks and synchronous queues), enabling efficient synchronization not only for Kotlin Coroutines but for other languages and platforms as well, such as C++, Rust, and Go. We plan to investigate this in future work, along with proof extensions to the release-acquire memory model semantics~\cite{kaiser2017strong}.



\bibliography{references}

\clearpage
\appendix

\clearpage
\section{Synchronous Resumption}\label{appendix:sync_async}
In section~\ref{sec:basic} we briefly mentioned that supporting non-blocking variants of blocking operations, such as the \texttt{tryLock()} one in the mutex, would require introducing a special \texttt{synchronous resumption} mode. Indeed, the classic CQS framework does not provide a way to implement such operations correctly. 
Therefore, first, we describe the problem and introduce the \emph{synchronous resumption} mode in the absence of cancellation in Subsection~\ref{subsec:app:sync:basic}. 
After that, we extend it with abortability support in Subsection~\ref{subsec:app:sync:cancellation}.

\subsection{Extension to the Basic CQS}\label{subsec:app:sync:basic}

\begin{wrapfigure}[19]{r}{0.42\textwidth}
\vspace{-1.8em}
\begin{lstlisting}[
caption={The basic mutex algorithm without cancellation support using the CQS framework. This is a copy of the algorithm in Listing~\ref{listing:mutex_basic}.},
label={lst:mutex:basic:copy}
]
val cqs = CQS<Unit>()#\label{line2:mutex_basic_sqs}#
// #\color{Mahogany}1# - unlocked, #\color{Mahogany}$\leq$ 0# - #\color{Mahogany}\## of waiters
var state: Int = 1 // "unlocked" intially#\label{line2:mutex_basic_state}#
fun lock() {
  s := FAA(&state, -1) #\label{line2:mutex_basic_dec}#
  // Is the lock just acquired? 
  if s > 0: return #\label{line2:mutex_basic_lock_locked}#
  cqs.suspend() // suspend otherwise #\label{line2:mutex_basic_lock_susp}#
}
fun unlock() {
  s := FAA(&state, +1) #\label{line2:mutex_basic_unlock_inc}#
  // Resume the first waiting
  // request if there is one.
  if s < 0: cqs.resume(Unit) #\label{line2:mutex_basic_unlock_resume}#
}
\end{lstlisting}
\end{wrapfigure}

\noindent
In Section~\ref{sec:basic}, we presented a mutex algorithm on top of CQS. We copy it in Listing~\ref{lst:mutex:basic:copy} for convenience. 
While this mutex algorithm is simple, extending it to allow a \texttt{tryLock()} operation is not obvious. Specifically, \texttt{tryLock()} should attempt to acquire the lock and return \texttt{true} or \texttt{false} depending on whether it succeeded, attempting to change the logical state from ``unlocked'' to ``locked'' and never manipulating CQS. However, the \texttt{unlock()} implementation in Listing~\ref{lst:mutex:basic:copy} relies on leaving a right to acquire the lock (a ``permit'') as metadata in a CQS cell. A naive implementation of \texttt{tryLock()}, which tries to atomically update \texttt{state} from \texttt{1} (``unlocked'') to \texttt{0} (``locked''), would not observe that the permit in CQS, resulting in an incorrect execution.

\begin{wrapfigure}{r}{0.52\textwidth}
    \centering
    \vspace{-1em}
    \includegraphics[width=0.52\textwidth]{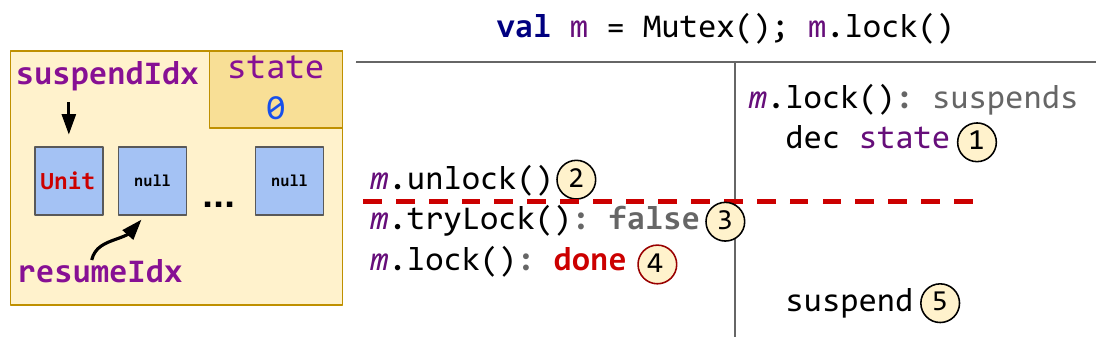}
    \caption{Incorrect behaviour of the mutex from Listing~\ref{lst:mutex:basic:copy} extended with \texttt{tryLock()} that tries to atomically change \texttt{state} from $1$ (unlocked) to $1$ (locked, no waiters).}
    \label{fig:incorrect_try_lock}
\end{wrapfigure}

To illustrate, consider the execution in Figure~\ref{fig:incorrect_try_lock}. First, a new mutex is created, and \texttt{lock()} is invoked. Then, two parallel threads start. The thread on the right also invokes \texttt{lock()}, decrementing \texttt{state} first. Since the mutex is already acquired, it then invokes \texttt{cqs.suspend()}. However, the execution switches to the thread on the left between the \texttt{state} decrement and the \texttt{suspend()} invocation. Then, the \texttt{unlock()} invocation changes \texttt{state} from \texttt{-1} to \texttt{0} and invokes \texttt{cqs.resume(..)}. Since the conjugate \texttt{suspend()} has not been invoked yet, it puts \texttt{Unit} in the first cell and completes. Thus, the mutex is actually in the ``unlocked'' state, while the permit to acquire this mutex is stored not in the \texttt{state} field but in the first cell of the CQS. Therefore, the following invocation of \texttt{tryLock()} fails since \texttt{state} equals \texttt{0}, while the \texttt{lock()} call succeeds {---} it goes to the first cell and completes immediately. 

\paragraph{\texttt{ASYNC} and \texttt{SYNC} Resumption Modes.}
The problem stems from the \texttt{unlock()} behaviour when it intends to pass the lock to a parallel \texttt{lock()} request {---} another \texttt{lock()} operation that happens after this \texttt{unlock()} may illegally acquire the lock from the CQS cell due to a race. 

\begin{wrapfigure}[17]{r}{0.4\textwidth}
  \vspace{-0.9em}
  \begin{center}
    \includegraphics[width=0.4\textwidth]{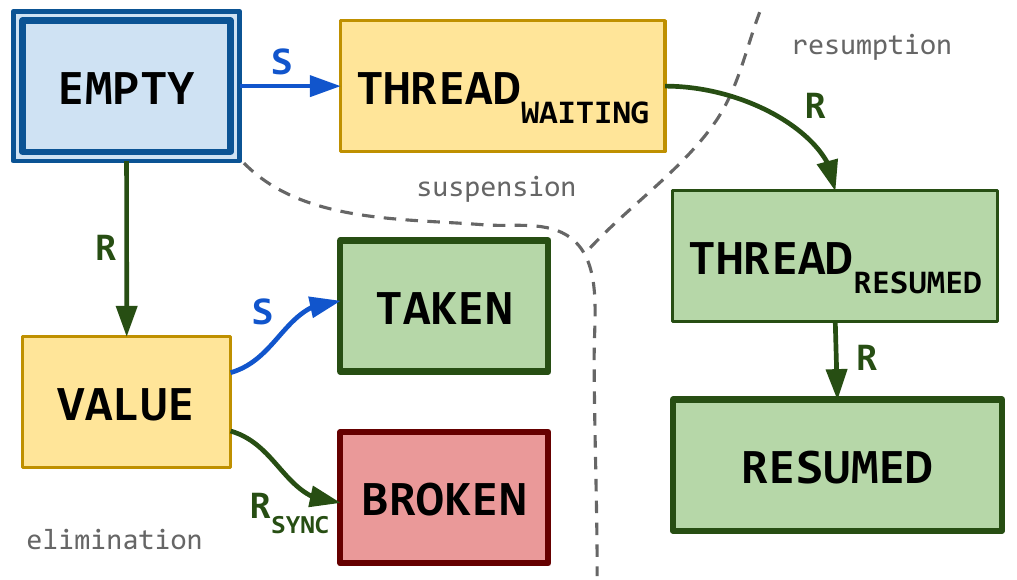}
  \end{center}
  \caption{\sqs{} cell life-cycle with \emph{synchronous resumption mode} (the introduced transition is marked with $\mathtt{R}_\mathtt{SYNC}$) but without cancellation support. The edges marked with \texttt{S} correspond to the transitions by \texttt{suspend()}, and those marked with \texttt{R} represent transitions by \texttt{resume(..)}.}
  \label{fig:cell_no_cancellation}
\end{wrapfigure}

To prevent such a race, we should avoid leaving ``permits'' in cells ``unattended''. As a solution, two resumption modes are proposed {---} \emph{asynchronous} and \emph{synchronous}. The \emph{asynchronous} mode (\texttt{ASYNC} in code) is the standard mode where \texttt{resume(..)} puts the value into an empty cell and completes immediately, thus transferring this value asynchronously. In contrast, the \emph{synchronous} mode (\texttt{SYNC} in code) forces \texttt{resume(..)} to wait until the value is taken, and marks the cell as \texttt{broken} after some bounded time if the value is still not taken by a concurrent \texttt{suspend(..)}. The idea is similar to breaking cells in modern queues~\cite{LCRQ, YM16}. 
In this case, both \texttt{resume(..)} and \texttt{suspend()} operations manipulating this cell fail, so \texttt{suspend(..)} returns \texttt{null} while \texttt{resume(..)} returns \texttt{false}. 
The intuition is that allowing broken cells keeps the balance of paired operations, such as \texttt{lock()} and \texttt{unlock()}, so they should simply restart. 
Figure~\ref{fig:cell_no_cancellation} describes the modified cell life-cycle.

\paragraph{Modifications to \texttt{suspend()} and \texttt{resume(..)}.}
Listing~\ref{listing:sqs_highlevel_sync} shows the updated versions of \texttt{suspend()} and \texttt{resume(..)} operations; the key changes from the basic algorithm in Listing~\ref{listing:sqs_highlevel} in Section~\ref{sec:basic} are highlighted with yellow. The semantics change in that both these operations may fail if the cell is broken; thus, \texttt{suspend()} can return \texttt{null}, while \texttt{resume(..)} returns \texttt{true} on success and \texttt{false} on failure.

\begin{figureAsListingWide}
\begin{minipage}[t]{0.48\textwidth}
\begin{lstlisting}[]
// The resumption mode is specified
// by the user when constructing CQS.
val resumeMode: ASYNC or SYNC

val cells = InfiniteArray() #\label{line2:sqs_ho_cells}#
var suspendIdx: Int#64# = 0 #\label{line2:sqs_ho_idx_start}# 
var resumeIdx:  Int#64# = 0 #\label{line2:sqs_ho_idx_end}#

fun suspend(): T {
 i := FAA(&suspendIdx, +1) #\label{line2:sqs_ho_susp_inc}#
 // Try to suspend in #\color{Mahogany}cells[i]#.
 t := currentThread() #\label{line2:sqs_ho_susp_createreq}#
 if CAS(&cells[i], null, t): #\label{line2:sqs_ho_susp_cas}#
 #\indentrule#  return park() // enqueued, suspend#\label{line2:sqs_ho_susp_park}#
 // Read the result and finish
 // if the cell is not broken.
 ##@result := GetAndSet(&cells[i], TAKEN)@#\label{line2:sqs_ho_susp:getandset}#
 // Was the cell broken?
 ##@if result == BROKEN: return null@  #\label{line2:sqs_ho_susp:broken}#
 // The cell stored a value.
 return result #\label{line2:sqs_ho_susp_imm}#
}
\end{lstlisting}
\end{minipage}
\hfill
\begin{minipage}[t]{0.5\textwidth}
\begin{lstlisting}[firstnumber=23]
fun resume(result: T): Bool {
 i := FAA(&resumeIdx, +1) #\label{line2:sqs_ho_resume_inc}#
 t := cells[i]
 if t == null: // is the cell empty?#\label{line2:sqs_ho_resume_is_empty}#
 #\indentrule#  // `suspend()` is coming, try to 
 #\indentrule#  // install the result and finish.
 #\indentrule#  if CAS(&cells[i], null, result):  #\label{line2:sqs_ho_resume_set_value}#
 #\indentrule#  #\indentrule#  // Finish in ASYNC mode.
 #\indentrule#  #\indentrule#  ##if resumeMode#$\:$#==#$\:$#ASYNC: return true#\label{line2:sqs_ho_resume_async}#
 #\indentrule#  #\indentrule#  // Synchronous resumption. Wait   
 #\indentrule#  #\indentrule#  // until the value is taken.
 #\indentrule#  #\indentrule#  ##@repeat(MAX_SPIN_CYCLES):@ #\label{line2:resumeAndCancel:resume:elimination:loop0}#
 #\indentrule#  #\indentrule#  #\indentrule###@ if cells[i]#$\:$#==#$\:$#TAKEN: return true@ #\label{line2:resumeAndCancel:resume:elimination:loop1}#
 #\indentrule#  #\indentrule#  // The value has not been taken.
 #\indentrule#  #\indentrule#  ##@return !CAS(&cells[i], result,@ #\label{line2:sqs_ho_resume_elim_cas}# #\label{line2:resumeAndCancel:resume:elimination:fail}#
 #\indentrule#  #\indentrule#                         ##@BROKEN)@
 #\indentrule#  // The cell stores a thread.
 #\indentrule#  t = cells[i]  #\label{line2:sqs_ho_resume_reread}#
 // Resume the waiting request.
 cells[i] = RESUMED #\label{line2:clean_cell_resume_fgf}#
 t.unpark(result) // t is Thread #\label{line2:sqs_ho_resume_complete}#
}
\end{lstlisting}
\end{minipage}
\vspace{-1em}
\caption{
High-level CQS implementation without cancellation support but with support of both \emph{asynchronous} and \emph{synchronous} resumption modes. The key changes from the basic algorithm in Listing~\ref{listing:sqs_highlevel} in Section~\ref{sec:basic} are highlighted with yellow.
}
\vspace{-2em}
\label{listing:sqs_highlevel_sync}
\end{figureAsListingWide}

The \texttt{suspend()} operation does not change significantly. As before, it increments \texttt{suspendIdx} first (line~\ref{line2:sqs_ho_susp_inc}), obtains the currently running thread (line~\ref{line2:sqs_ho_susp_createreq}), and tries to install it if the cell is empty (line~\ref{line2:sqs_ho_susp_cas}). However, if the cell is not empty, it could have been broken and, thus, does not necessarily contain a value. Therefore, the algorithm replaces the current cell state with \texttt{TAKEN} via an atomic \texttt{GetAndSet(..)} operation (line~\ref{line2:sqs_ho_susp:getandset}), failing if the cell was in the \texttt{BROKEN} state (line~\ref{line2:sqs_ho_susp:broken}). If the cell did contain a value, it is returned as the result of \texttt{suspend()} (line~\ref{line2:sqs_ho_susp_imm}). 

As for the \texttt{resume(..)} operation, like before, it increments \texttt{resumeIdx} first (line~\ref{line2:sqs_ho_resume_inc}). 
When the cell is empty (line~\ref{line2:sqs_ho_resume_is_empty}), \texttt{resume(..)} tries to set the resumption value to the cell (line~\ref{line2:sqs_ho_resume_set_value}). If the corresponding \texttt{CAS} fails, the cell stores a thread, which is read later at line~\ref{line2:sqs_ho_resume_reread}.
Otherwise, the value is successfully set, and further logic depends on the choice of resumption mode. In the asynchronous mode, \texttt{resume(..)} finishes immediately (line~\ref{line2:sqs_ho_resume_async}). In the synchronous mode, it waits in a bounded loop waiting for the value to be taken and finishes if it happens (lines~\ref{line2:resumeAndCancel:resume:elimination:loop0}--\ref{line2:resumeAndCancel:resume:elimination:loop1}). If the value has not been taken, the operation attempts to break the cell and fail, completing successfully if the value was taken after all and the corresponding \texttt{CAS} fails (line~\ref{line2:resumeAndCancel:resume:elimination:fail}).

\begin{wrapfigure}[11]{r}{0.49\textwidth}
\vspace{-1.8em}
\begin{lstlisting}[
caption={Basic mutex algorithm with \texttt{tryLock()} on top of CQS, without cancellation. 
%The changes from the algorithm in Listing~\ref{lst:mutex:basic:copy} are highlighted with yellow. 
This implementation is also correct with the simple cancellation mode presented in Subsection~\ref{subsection:simple_cancellation}.},
label={listing:mutex_basic_sync2}
]
val cqs = CQS<Unit>(resumeMode = SYNC)#\label{line2:mutex_basic_sync_sqs}#
var state: Int = 0 

fun tryLock(): Bool = CAS(&state, 1, 0)
fun lock(): Unit = while(true) {#\label{line2:mutex_basic_sync_lock_while}#
  s := FAA(&state, -1) 
  if s > 0: return
  ##if cqs.suspend() == Unit :return #\label{line2:mutex_basic_sync_lock_restart}#
}
fun unlock(): Unit = while (true) { #\label{line2:mutex_basic_sync_unlock_while}#
  s := FAA(&state, 1) 
  if s == 0: return
  ##if cqs.resume(Unit): return  #\label{line2:mutex_basic_sync_unlock_restart}#
}
\end{lstlisting}
\end{wrapfigure}

When the cell contains a thread instance, \texttt{resume(..)} cleans the cell (line~\ref{line2:clean_cell_resume_fgf}), resumes the thread (line~\ref{line2:sqs_ho_resume_complete}), and finishes.

\paragraph{Mutex Algorithm with \texttt{tryLock()}.}
Listing~\ref{listing:mutex_basic_sync2} contains a correct mutex implementation extended with the \texttt{tryLock()} operation. As discussed, the synchronous resumption mode is used (line~\ref{line2:mutex_basic_sync_sqs}) and both \texttt{lock()} and \texttt{unlock()} operations are wrapped in an infinite loop (lines~\ref{line2:mutex_basic_sync_lock_while}~and~\ref{line2:mutex_basic_sync_unlock_while}) so that they restart if \texttt{suspend()} and \texttt{resume(..)} fail (lines~\ref{line2:mutex_basic_sync_lock_restart}~and~\ref{line2:mutex_basic_sync_unlock_restart}).
The remainder is the same as in the previous algorithm in Listing~\ref{lst:mutex:basic:copy}.

\subsection{Cancellation Support}\label{subsec:app:sync:cancellation}
To support the \emph{synchronous} resumption mode with cancellation, we need to ensure that \texttt{resume(..)} never leaves the value in CQS without making a rendezvous with \texttt{suspend()}. In the previous Subsection~\ref{subsec:app:sync:basic}, we figured out how the elimination part, when \texttt{resume(..)} comes to the cell before \texttt{suspend()}, should be modified. However, with \emph{smart} cancellation, there is another way to leave the value in CQS {---} when the cell contains an aborted thread, the \texttt{resume(..)} implementation in Listing~\ref{listing:resume_and_cancellation} in Section~\ref{sec:abortability} can delegate its completion by replacing this cancelled thread with the resumption value (lines~\ref{line:resumeAndCancel:resume:putValue0}--\ref{line:resumeAndCancel:resume:putValue1}). With the \emph{synchronous} resumption, we forbid this behaviour and wait in a spin-loop until the state changes to either \texttt{CANCELLED} or \texttt{REFUSE}. 

\paragraph{The \texttt{resume(..)} Modification.}
The resulting pseudocode of the \texttt{resume(..)} operation is presented in Listing~\ref{listing:resume_and_cancellation_sync}. The logic for \texttt{suspend()} stays the same as in Listing~\ref{listing:sqs_highlevel_sync}, while the cancellation handler is identical to the one in Listing~\ref{listing:resume_and_cancellation}.

As usual, the operation starts with incrementing \texttt{resumeIdx} (line~\ref{line3:resumeAndCancel:resume:inc}). After that, if cell is in the empty state, the algorithm tries to perform elimination similarly to Listing~\ref{listing:sqs_highlevel_sync} (lines~\ref{line3:resumeAndCancel:resume:empty}--\ref{line3:resumeAndCancel:resume:resumed1}).
When the cell is in the \texttt{CANCELLED} state, the \texttt{resume(..)} operation either fails in the simple cancellation mode, or skips the cell with smart cancellation (lines~\ref{line3:resumeAndCancel:resume:isCancelled}--\ref{line3:resumeAndCancel:resume:isCancelled_skipSmart}). 
When the cell is in the \texttt{REFUSE} state, the user-specified \texttt{completeRefusedResume(..)} function is called and the operation finishes (lines~\ref{line3:resumeAndCancel:resume:isRefused}--\ref{line3:resumeAndCancel:resume:finishRefuse}).

\begin{wrapfigure}{r}{0.45\textwidth}
\vspace{-2em}
\begin{lstlisting}[
caption={
Pseudo-code for \texttt{resume(..)} that supports all cancellation modes and both \emph{asynchronous} and \emph{synchronous} resumption. The key change from the algorithm with cancellation support in Listing~\ref{listing:resume_and_cancellation} in Section~\ref{sec:abortability} and  Listing~\ref{listing:sqs_highlevel_sync} in Appendix~\ref{subsec:app:sync:basic} is highlighted with yellow.
},
label={listing:resume_and_cancellation_sync}
]
fun resume(result: T): Bool {
 i := FAA(&resumeIdx, +1)  #\label{line3:resumeAndCancel:resume:inc}#
 while (true): // modify the cell #\label{line3:resumeAndCancel:resume:modifyCell:start}#
 #\indentrule# cur := cells[i]  #\label{line3:resumeAndCancel:resume:curState}#
 #\indentrule# when {
 #\indentrule# #\indentrule#cur == null: #\label{line3:resumeAndCancel:resume:empty}#
 #\indentrule# #\indentrule#  // Try to perform elimination 
 #\indentrule# #\indentrule#  // similarly to Listing #\color{Mahogany}\ref{listing:sqs_highlevel_sync}#.
 #\indentrule# #\indentrule#  if CAS(&cells[i], null, result):#\label{line3:resumeAndCancel:resume:tryElimination0}#
 #\indentrule# #\indentrule#  #\indentrule#  return true #\label{line3:resumeAndCancel:resume:tryElimination1}#
 #\indentrule# #\indentrule#cur is Thread: #\label{line3:resumeAndCancel:resume:isRequest}#
 #\indentrule# #\indentrule#  // Try to resume the thread.
 #\indentrule# #\indentrule#  if cur.unpark(result): #\label{line3:resumeAndCancel:resume:tryComplete}#
 #\indentrule# #\indentrule#  #\indentrule#  cells[i] = RESUMED #\label{line3:resumeAndCancel:resume:resumed0}#
 #\indentrule# #\indentrule#  #\indentrule#  return true #\label{line3:resumeAndCancel:resume:resumed1}#
 #\indentrule# #\indentrule#  // The thread is cancelled.
 #\indentrule# #\indentrule#  if cancellationMode == SIMPLE: #\label{line3:resumeAndCancel:resume:simpleFail0}#
 #\indentrule# #\indentrule#  #\indentrule#  return false #\label{line3:resumeAndCancel:resume:simpleFail1}#
 #\indentrule# #\indentrule#  // Smart cancellation is used.
 #\indentrule# #\indentrule#  // With SYNC resumption, wait 
 #\indentrule# #\indentrule#  // until the cell state changes
 #\indentrule# #\indentrule#  // to either CANCELLED or REFUSE
 #\indentrule# #\indentrule#  ##@if resumeMode == SYNC: continue@ #\label{line3resumeAndCancel:resume:continue_sync}#
 #\indentrule# #\indentrule#  // Delegate this resume(..) 
 #\indentrule# #\indentrule#  // completion to the cancel- 
 #\indentrule# #\indentrule#  // lation handler in ASYNC mode.
 #\indentrule# #\indentrule#  if CAS(&cells[i], cur, result): #\label{line3:resumeAndCancel:resume:putValue0}#
 #\indentrule# #\indentrule#  #\indentrule#  return true #\label{line3:resumeAndCancel:resume:putValue1}#
 #\indentrule# #\indentrule#cur == CANCELLED: #\label{line3:resumeAndCancel:resume:isCancelled}#
 #\indentrule# #\indentrule#  // Fail with simple cancellation
 #\indentrule# #\indentrule#  if cancellationMode == SIMPLE: #\label{line3:resumeAndCancel:resume:isCancelled_failSimple0}#
 #\indentrule# #\indentrule#  #\indentrule#  return false #\label{line3:resumeAndCancel:resume:isCancelled_failSimple1}#
 #\indentrule# #\indentrule#  // Skip the cell in SMART mode.
 #\indentrule# #\indentrule#  return resume(result) #\label{line3:resumeAndCancel:resume:isCancelled_skipSmart}#
 #\indentrule# #\indentrule#cur == REFUSE: #\label{line3:resumeAndCancel:resume:isRefused}#
 #\indentrule# #\indentrule#  ##completeRefusedResume(result)#\label{line3:resumeAndCancel:resume:completeRefusedResume}#
 #\indentrule# #\indentrule#  return true #\label{line3:resumeAndCancel:resume:finishRefuse}# #\label{line3:resumeAndCancel:resume:modifyCell:end}#
}
\end{lstlisting}
\end{wrapfigure}

When the cell contains a suspended thread, the algorithm first tries to resume it, finishing on success (lines~\ref{line3:resumeAndCancel:resume:tryComplete}--\ref{line3:resumeAndCancel:resume:resumed1}). Otherwise, this thread is already aborted, and the logic depends on the cancellation mode. 
In simple cancellation, this \texttt{resume(..)} invocation fails (lines~\ref{line3:resumeAndCancel:resume:simpleFail0}--\ref{line3:resumeAndCancel:resume:simpleFail1}).

In contrast, with smart cancellation, the behaviour depends on whether the state changes to \texttt{CANCELLED} or \texttt{REFUSE}. Thus, in order not to leave the value in CQS when the \emph{synchronous} resumption mode is used, we modify the algorithm and wait in a loop until the state changes (line~\ref{line3resumeAndCancel:resume:continue_sync}, highlighted with yellow). Aside from this, with asynchronous resumption, the algorithm stays the same and delegates this resumption completion to the cancellation handler (lines~\ref{line3:resumeAndCancel:resume:putValue0}--\ref{line3:resumeAndCancel:resume:putValue1}).

\paragraph{The Modified Cell Life-Cycle Diagram.}
For readability, we present the full version of the cell life-cycle diagram in Figure~\ref{fig:app:cell_full}, which supports both asynchronous and synchronous resumption as well as simple and smart cancellation modes.

\begin{minipage}{0.48\textwidth}
    \vspace{2em}
    \includegraphics[width=\textwidth]{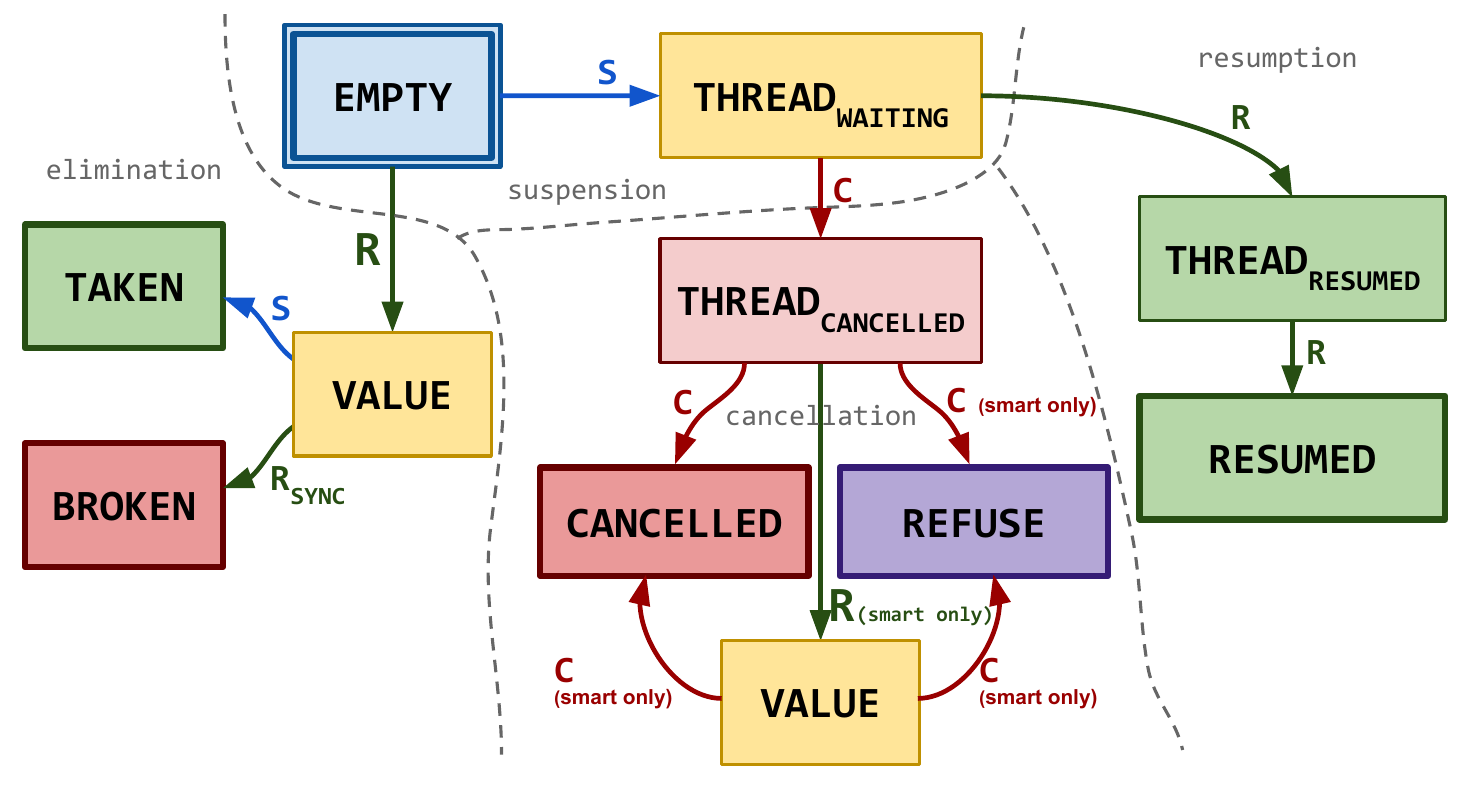}
    \captionof{figure}{The full version of the cell life-cycle diagram supports both asynchronous and synchronous resumption and simple and smart cancellation modes.}
    \label{fig:app:cell_full}
\end{minipage}

\clearpage
\section{Infinite Array Implementation}\label{appendix:infarr}

\begin{wrapfigure}[7]{r}{0.54\textwidth}
    \vspace{-1.8em}
    \centering
    \includegraphics[width=0.54\textwidth]{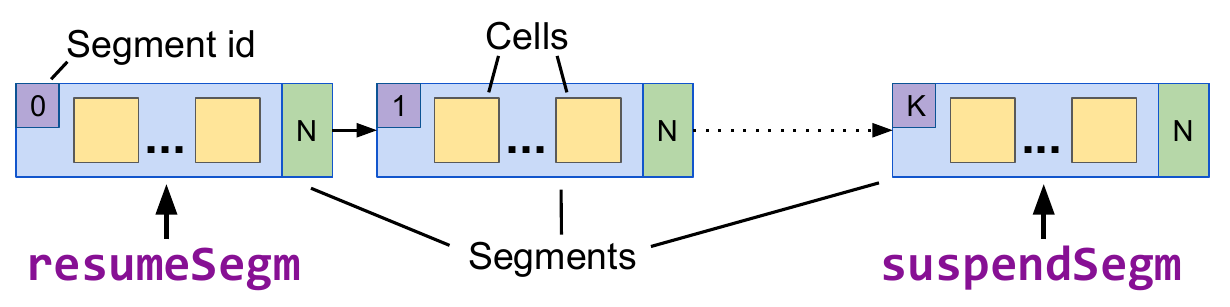}
    \vspace{-1.6em}
    \caption{An infinite array as a linked list of cell segments.}
    \label{fig:segments2}
\end{wrapfigure}

The \sqs{} framework is built on an infinite array, the cells of which are processed in sequential order. To emulate this infinite array, we follow the approach behind the implementation of the channels in Kotlin~\cite{channels_kotlin}, maintaining a linked list of cell segments, each containing a fixed number of cells, as illustrated in Figure~\ref{fig:segments2} (we repeat the illustration of the structure from Section~\ref{sec:basic}).%

Each segment has a unique id and can be seen as a node in a Michael-Scott queue~\cite{michael1996simple}. Following this structure, we maintain only those cells that are in the currently active range (between \texttt{resumeIdx} and \texttt{suspendIdx}) and access them similarly to an array. Specifically, we change the current working segment once every \texttt{SEGM\_SIZE} operations, where \texttt{SEGM\_SIZE} is the number of cells in each segment.

Despite conceptual simplicity, the implementation of this structure is non-trivial, as shown in~\cite{channels_kotlin}. In this section, we discuss the implementation details and the required changes to the CQS algorithm, providing the formal proofs in Appendix~\ref{sec:proofs}. We highlight that the infinite array implementation is not a part of our contribution, but providing all the technical details is necessary to make this paper self-contained and deliver formal proofs.

\subsection{Basics Algorithm without Cancellation}\label{subsec:app:infarr:basic}
Listing~\ref{listing:segments_suspend} presents a pseudo-code for \texttt{suspend()} in terms of these segments; the changes to \texttt{resume(..)} are symmetrical, so they are omitted. 
Instead of maintaining head and tail pointers, as is usually done in concurrent queues~\cite{michael1996simple}, we maintain \texttt{suspendSegm} and \texttt{resumeSegm} pointers related to the last segment used by \texttt{suspend()} and \texttt{resume(..)}, respectively; initially, they reference the same segment of \texttt{id = 0} (lines~\ref{line_apndx:segments_begin0}--\ref{line_apndx:segments_begin1})

Initially, \texttt{suspend()} reads its last used segment  (line~\ref{line_apndx:segments_suspend_suspend_readsegm}) and increments \texttt{suspendIdx} (line~\ref{line_apndx:segments_suspend_suspend_inc}). Then, it locates the required segment by following the chain of \texttt{next} pointers starting from the one already read before the increment, updating \texttt{suspendSegm} if required (line~\ref{line_apndx:segments_suspend_suspend_find}). The \texttt{findAndMoveForwardSusp(..)} implementation is straightforward {---} it finds the required segment first (lines~\ref{line_apndx:segments_suspend_find_findstart}--\ref{line_apndx:segments_suspend_find_findend}), creating new segments if needed (lines~\ref{line_apndx:segments_suspend_find_createsegmstart}--~\ref{line_apndx:segments_suspend_find_createsegmend}), and then updates \texttt{suspendSegm} to the one that was found if it has not been updated to it or one of the later segments yet (lines~\ref{line_apndx:segments_suspend_find_movestart}--\ref{line_apndx:segments_suspend_find_moveend}).

The changes for \texttt{resume(..)} are symmetric and, therefore, are omitted. The only notable difference is that \texttt{findAndMoveForwardResume(..)} should clean the \texttt{prev} pointer to the previous segment in the doubly-linked list structure to ensure that all the processed segments are available for garbage collection. 

\begin{figureAsListingWide}
\begin{minipage}[t]{0.46\textwidth}
\begin{lstlisting}[]
var suspendSegm : Segment #\label{line_apndx:segments_begin0}#
var resumeSegm  : Segment
constructor {
 s := Segment(id = 0)
 suspendSegm = resumeSegm = s
}  #\label{line_apndx:segments_begin1}#

fun suspend(): T {
 // Read the current `suspendSegm`
 // before incrementing the counter.
 s := @suspendSegm@ #\label{line_apndx:segments_suspend_suspend_readsegm}#
 idx := FAA(&suspendIdx, 1) #\label{line_apndx:segments_suspend_suspend_inc}#
 // Find the required segment 
 // and update `suspendSegm`.
 id := idx / SEGM_SIZE
 i := #i \% SEGM\_SIZE#
 s = @findAndMoveForwardSusp(s, id)@ #\label{line_apndx:segments_suspend_suspend_find}#
 // Process the cell #\color{Mahogany}s[i]#.
 t := currentThread() #\label{line_apndx:sqs_ho_susp_createreq}#
 if CAS(&s[i], null, t):  #\label{line_apndx:sqs_ho_susp_cas0}#
 #\indentrule#  return park() #\label{line_apndx:sqs_ho_susp_park}#
 result := s[i]; s[i] = TAKEN
 return result
}
\end{lstlisting}
\end{minipage}
\hfill
\begin{minipage}[t]{0.52\textwidth}
\begin{lstlisting}[firstnumber=25]
fun findAndMoveForwardSusp(start: Segment, 
                      id: Long): Segment {
 cur := start #\label{line_apndx:segments_suspend_find_findstart}#
 // #\color{Mahogany}1.# Find the required segment.
 while cur.id < id: 
 #\indentrule#  if cur.next == null: #\label{line_apndx:segments_suspend_find_createsegmstart}#
 #\indentrule#  #\indentrule#  // Create a new segment if needed.
 #\indentrule#  #\indentrule#  s := Segment(id = cur.id+1)
 #\indentrule#  #\indentrule#  CAS(&cur.next, null, s)#\label{line_apndx:segments_suspend_find_createsegmend}#
 #\indentrule#  cur = cur.next #\label{line_apndx:segments_suspend_find_findend}#
 // #\color{Mahogany}2.# Move `suspendSegm` forward if needed
 while (true):  #\label{line_apndx:segments_suspend_find_movestart}#
 #\indentrule#  s := suspendSegm 
 #\indentrule#  // Already up-to-date?
 #\indentrule#  if s.id #$\geq$# id: break 
 #\indentrule#  // Try to update
 #\indentrule#  if CAS(&suspendSegm, s, cur): break #\label{line_apndx:segments_suspend_find_moveend}#
 // #\color{Mahogany}3.# Return the found segment.
 return cur
}
\end{lstlisting}
\end{minipage}
\vspace{-1em}
\caption{
Implementation of the \texttt{suspend} operation that manipulates a linked list of fixed-size segments. The key change is highlighted {---} the current \texttt{suspendSegm} should be read before the \texttt{suspendIdx} increment, which guarantees that the required segment can be found by following \texttt{next} pointers and all the preceding segments are not needed anymore, so \texttt{suspendSegm} can be safely moved forward. For simplicity, this implementation supports only the asynchronous (default) resumption mode.
}
\label{listing:segments_suspend}
\end{figureAsListingWide}

\subsection{Segment Removal for Cancellation}\label{subsec:app:infarr:cancellation}
Once a cell is in the \texttt{CANCELLED} state, it no longer stores the reference to the cancelled waiter, so the garbage collector is free to collect this reference. However, we also want it to collect the cell itself,  in order to avoid memory leaks.
Since we emulate the infinite array via a concurrent linked list of segments, the segments full of cancelled cells must be physically removed from the list.
This way, we can guarantee that the memory complexity depends only on the number of non-cancelled cells, even in the case where all but one cell in each segment are cancelled, since the segment size is constant. 
The only exception is cells in the \texttt{REFUSE} state, as they are not considered as cancelled, so the corresponding segments cannot be removed. However, the \texttt{REFUSE} state indicates that there is a concurrent \texttt{resume(..)}, which is going to process this cell eventually {---} the number of such \texttt{resume(..)}-s is bounded by the number of threads.
In sum, the total memory complexity should be $O(N + T)$, where $N$ is the number of non-cancelled waiters, and $T$ is the number of threads. (We assume the segment size to be constant; otherwise, the complexity should be multiplied by the segment size.)

Listing~\ref{listing:remove_segment} presents the pseudocode of the segment removal part of the algorithm, including the required changes to the \texttt{resume(..)} and \texttt{findAndMoveForwardResume(..)} functions. The \texttt{suspend()} implementation stays almost the same as in Listing~\ref{listing:segments_suspend} above, with a single addition: a cancellation handler that invokes \texttt{s.onCancelledCell()} should be specified in the \texttt{park(..)} call. The changes to \texttt{findAndMoveForwardSuspend(..)} are symmetric to the ones for \texttt{findAndMoveForwardResume(..)}.

\begin{figureAsListingWide}
\begin{minipage}[t]{0.52\textwidth}
\begin{lstlisting}[
basicstyle=\scriptsize\selectfont\ttfamily
] 
fun resume(value: T): Bool = while(true) { #\label{line_apndx:r:resume0}#
  r := resumeSegm
  i := FAA(&resumeIdx, 1)
  id := i / SEGM_SIZE
  s := findAndMoveForwardResume(r, id)
  // All the previous segments are processed.
  if s.id != id: #\label{line_apndx:r:notFound}#
  #\indentrule#  if cancellationMode == SIMPLE: return false #\label{line_apndx:r:simpleFail}#
  #\indentrule#  // Update resumeIdx to skip the sequence of
  #\indentrule#  // cancelled segments in smart cancellation
  #\indentrule#  CAS(&resumeIdx, i + 1, s.id * SEGM_SIZE) #\label{line_apndx:r:smartSkip0}#
  #\indentrule#  continue #\label{line_apndx:r:smartSkip1}#
  // Process the cell #\color{Mahogany}s[#i % SEGM_SIZE#\color{Mahogany}]#
  ... 
} #\label{line_apndx:r:resume1}#
// Returns the first non-removed segment with
// id equal to or greater than the specified,
// creating new segments if needed.
fun findAndMoveForwardResume(start: Segment, 
                           id: Long): Segment#$\:$#{
  while (true): 
  #\indentrule#  s := findSegm(start, id) #\label{line_apndx:r:findSegm}#
  #\indentrule#  // Try to update `resumeSegm`, and 
  #\indentrule#  // restart if the found segment is
  #\indentrule#  // removed and is not the tail one.
  #\indentrule#  if moveForwardResume(s): break #\label{line_apndx:r:move}#
  #\indentrule#  ##@s.prev = null@; return s #\label{line_apndx:r:cleanPrevXXXXXX}#
}

fun findSegm(start:#$\:$#Segment,#$\:$#id:#$\:$#Long):#$\:$#Segment#$\:$#{
  cur := start
  while cur.id < id || ##@cur.removed()@: #\label{line_apndx:r:highlighted}#
  #\indentrule#  if cur.next == null: 
  #\indentrule#  #\indentrule#  // Create a new segment if needed.
  #\indentrule#  #\indentrule#  newSegm := Segment(id = cur.id + 1)
  #\indentrule#  #\indentrule#  if CAS(&cur.next, null, newSegm):
  #\indentrule#  #\indentrule#  #\indentrule#  // Is the previous tail removed?
  #\indentrule#  #\indentrule#  #\indentrule#  ##@if cur.removed(): cur.remove()@ #\label{line_apndx:r:oldTailRemove}#
  #\indentrule#  cur = cur.next
  return cur
}
fun moveForwardResume(to: Segment): Bool {
  while (true):
  #\indentrule#  cur := resumeSegm
  #\indentrule#  // Do we still need to update `resumeSegm`?
  #\indentrule#  if cur.id >= to.id: return true
  #\indentrule#  // Try to inc pointers to `to`.
  #\indentrule#  if ##@!to.tryIncPointers(): return false@ #\label{line_apndx:r:tryIncPointersMove}#
  #\indentrule#  // Try to update `resumeSegm`.
  #\indentrule#  if CAS(&resumeSegm, cur, to): #\label{line_apndx:r:moveUpdate}#
  #\indentrule#  #\indentrule#  // Dec pointers to cur.
  #\indentrule#  #\indentrule#  ##@if cur.decPointers(): cur.remove()@ #\label{line_apndx:r:removeOldTailMove}#
  #\indentrule#  #\indentrule#  return true
  #\indentrule#  // The `resumeSegm` update has failed,
  #\indentrule#  //  dec pointers to `to` back.
  #\indentrule#  if ##@to.decPointers(): to.remove()@
}
\end{lstlisting}
\end{minipage}
\hfill
\begin{minipage}[t]{0.45\textwidth}
\begin{lstlisting}[firstnumber=58,
basicstyle=\scriptsize\selectfont\ttfamily
]
class Segment {
 // Initialized with #\color{Mahogany}(2, 0)# for the first 
 // segment#\color{Mahogany};# stored in a single #\color{Mahogany}32-bit# Int.
 var (pointers, cancelled) = (0, 0) #\label{line_apndx:r:pc}#
 
 fun removed(): Bool = atomic { #\label{line_apndx:r:removed0}#
   return cancelled == SEGM_SIZE && 
          pointers = 0
 } #\label{line_apndx:r:removed1}#
 fun ##^onCancelledCell()^ = atomic {  #\label{line_apndx:r:onc0}#
   cancelled++; if removed(): remove()
 } #\label{line_apndx:r:onc1}#
 // Increments the number of pointers#\color{Mahogany};# 
 // Fails if the segment is removed.
 fun tryIncPointers(): Bool = atomic { #\label{line_apndx:r:tryInc0}#
   if removed():  return false
   pointers++; return true
 } #\label{line_apndx:r:tryInc1}#
 // Decrements the number of pointers and
 // returns `true` if the segment becomes 
 // logically removed, `false` otherwise.
 fun decPointers(): Bool = atomic { #\label{line_apndx:r:tryDec0}#
   pointers--; return removed()
 } #\label{line_apndx:r:tryDec1}#
 
 // Physically removes the current segment
 fun remove() = while(true) {
   // The tail segment cannot be removed.
   if next == null: return #\label{line_apndx:r:checkTail}#
   // Find the closest alive segments
   // on the left and on the right.
   prev := aliveSegmRight() #\label{line_apndx:r:findR}#
   next := aliveSegmLeft() #\label{line_apndx:r:findL}#
   // Link `next` and `prev`.
   next.prev = prev #\label{line_apndx:r:removeUpdPrev}#
   if prev != null: prev.next = next #\label{line_apndx:r:removeUpdNext}#
   // Are `prev` and `next` still alive?
   if next.removed() && next.next != null:#\label{line_apndx:r:restart0}#
   #\indentrule#  continue
   if prev != null && prev.removed(): 
   #\indentrule#  continue #\label{line_apndx:r:restart1}#
   return // this segment is removed.
 }
 
 fun aliveSegmLeft(): Segment {
   cur := prev
   while cur != null && cur.removed():
   #\indentrule# cur = cur.prev
   return cur // `null` if all are removed
 }
 fun aliveSegmRight(): Segment {
   cur := next
   while cur.removed() && cur.next#$\:$#!=#$\:$#null:
   #\indentrule# cur = cur.next
   return cur // tail if all are removed
 }
}
\end{lstlisting}
\end{minipage}
\caption{Pseudocode for the segment removal algorithm. On the left, the modified \texttt{resume(..)} is presented. In addition, \texttt{findAndModeForwardResume(..)} is split into two parts, the important changes of which are highlighted with yellow. The required changes to \texttt{Segment} are shown on the right. The \texttt{onCancelledCell()} function, highlighted with green, is called when a cell from this segment moves to \texttt{CANCELLED} state.}
\label{listing:remove_segment}
\end{figureAsListingWide}

\paragraph{The Removal Algorithm Overview.}
In order to remove a segment in $O(1)$ (under no contention), we add a \texttt{prev} pointer to the \texttt{Segment} structure, which references the nearest non-removed segment on the left, or equals \texttt{null} if all of them are removed or processed (e.g., when the segment being removed is the head of the list). By maintaining the \texttt{prev} pointer, we can perform physical removal by linking the previous and the next segments to each other. However, doing so correctly requires non-trivial tricks in a concurrent environment.

Once a segment is removed, we should guarantee that it is no longer reachable by the \texttt{next} and \texttt{prev} references starting from \texttt{suspendSegm} and \texttt{resumeSegm}. For this purpose, we split the removal procedure into two parts: \emph{logical} and \emph{physical}. 
We assume that the segment is logically removed if all the cells are in the \texttt{CANCELLED} state and neither \texttt{suspendSegm} nor \texttt{resumeSegm} references it (lines~\ref{line_apndx:r:removed0}--\ref{line_apndx:r:removed1}). At the same time, we need to guarantee that they cannot start referencing logically removed segments, making them ``alive'' again and causing memory leaks.

To solve this problem, we maintain the number of cancelled cells alongside the number of pointers that reference this segment in a single integer field (line~\ref{line_apndx:r:pc}). By storing these numbers in a single register, we are able to modify them atomically {---} to emphasize this, the corresponding code is wrapped with \texttt{atomic} block in the pseudocode. Thus, there are two ways for a segment to become cancelled. First, if neither \texttt{suspendSegm} nor \texttt{resumeSegm} references it, and the \texttt{cancelled} counter reaches \texttt{SEGM\_SIZE}, the segment becomes logically removed and the following \texttt{remove()} call should remove it physically.
In the second case, all the cells are already cancelled and the number of pointers reaches zero when \texttt{suspendSegm} or \texttt{resumeSegm} updates. In this case, the corresponding code must check whether the previously referenced segment became logically removed and invoke \texttt{remove()} if needed. 

Correspondingly, when \texttt{suspendSegm} or \texttt{resumeSegm} need to be updated, they should increment the number of pointers that reference the new segment. However, if this new segment is already logically removed, the increment fails and the update should be restarted. The corresponding logic is provided in the \texttt{tryIncPointers()} function (lines~\ref{line_apndx:r:tryInc0}--\ref{line_apndx:r:tryInc1}).
Similarly, when \texttt{suspendSegm} or \texttt{resumeSegm} stop referencing some segment, they decrement the number of pointers. The corresponding \texttt{decPointers()} function returns \texttt{true} if the segments becomes logically removed (lines~\ref{line_apndx:r:tryDec0}--\ref{line_apndx:r:tryDec1}).

The only exception from this is the attempt to remove the tail segment. We forbid removing the tail segment, as doing so would make it  more difficult to ensure that each segment has a unique id throughout the list's lifetime. Therefore, we ignore the attempts to remove the tail segment {---} see the first statement in \texttt{remove()} (line~\ref{line_apndx:r:checkTail}). Thus, if a segment was the tail of the list at the time of logical removal, the following \texttt{remove()} call does nothing, and the physical removal of this segment is postponed until it stops being the tail.

\paragraph{The \texttt{resume(..)} Operation.}
While the logic of \texttt{suspend()} stays the same, since the segment requested by it has non-\texttt{CANCELLED} cells and is not removed at that point, the \texttt{resume(..)} operation requires some modifications (lines~\ref{line_apndx:r:resume0}--\ref{line_apndx:r:resume1}). First, the semantics of \texttt{findAndMoveForward(..)} are slightly changed. Since the requested segment can be removed, it now returns the first \emph{non-removed} segment with id equal to or greater than the requested one, creating new segments if needed.
Thus, if the requested segment is not found (line~\ref{line_apndx:r:notFound}), we know that the required cell is in the \texttt{CANCELLED} state, and so can process it correspondingly: fail in the simple cancellation mode (line~\ref{line_apndx:r:simpleFail}) or efficiently skip a sequence of removed segments and restart the operation in the smart cancellation mode (lines~\ref{line_apndx:r:smartSkip0}--\ref{line_apndx:r:smartSkip1}). 
Otherwise, when the requested segment is successfully found, we process it as usual; see the Listing~\ref{listing:resume_and_cancellation} for details.
One more notable change is that we must clean the \texttt{prev} pointer in \texttt{findAndMoveForwardResume(..)} to avoid memory leaks (line~\ref{line_apndx:r:cleanPrevXXXXXX}) {---} all the previous segments are either processed or going to be processed by concurrent \texttt{resume(..)}-s.

\paragraph{The \texttt{findAndMoveForwardResume(..)} Operation.}
We split the operation into two parts. First, the \texttt{findSegm(..)} function finds the first non-removed segment with id equal to or greater than the requested one, creating new segments if needed (line~\ref{line_apndx:r:findSegm}). Once the segment is found, we try to make \texttt{resumeSegm} point to it {---} this part can fail if the found segment becomes logically removed in the meantime, and the procedure restarts in this case (line~\ref{line_apndx:r:move}).

Essentially, the \texttt{findSegm(..)} logic stays the same with two small modifications. First, it skips the logically removed segments in the search procedure (the highlighted query at line~\ref{line_apndx:r:highlighted}). Second, once the tail of the list is updated, it checks whether the old tail should be removed (line~\ref{line_apndx:r:oldTailRemove}).

As for the \texttt{moveForwardResume(..)} operation, we need to increment and decrement the numbers of pointers there. Thus, we first try to increment the number of pointers to the new segment (line~\ref{line_apndx:r:tryIncPointersMove}), returning \texttt{false} and causing \texttt{findAndMoveForwardResume(..)} to restart on failure. If the increment of the number of pointers succeeds, the operation tries to update \texttt{resumeSegm} to the new one (line~\ref{line_apndx:r:moveUpdate}). If the update succeeds, the number of pointers to the old segment (\texttt{cur} in the code) should be decremented, removing the segment physically if needed (line~\ref{line_apndx:r:removeOldTailMove}).
If the \texttt{resumeSegm} update fails, the operations decrements the number of pointers to the new segment back (removing it if needed) and restarts.

\paragraph{The \texttt{Segment.onCancelledCell()} Operation.}
The \texttt{onCancelledCell()} operation is called when the cell moves to the \texttt{CANCELLED} state, see Listing~\ref{listing:resume_and_cancellation}. It increments the number of cancelled cells and checks if this led to the segment becoming logically removed, in which case it invokes \texttt{remove()} (lines~\ref{line_apndx:r:onc0}--\ref{line_apndx:r:onc1}).

\paragraph{The \texttt{Segment.remove()} Operation.}
The last part is the \texttt{remove()} operation itself.
If the segment that is being removed is the tail, the removal is postponed and delegated to either \texttt{findSegmentResume(..)} or \texttt{findSegmentSuspend(..)} that will update the tail and check whether the old one should be removed (line~\ref{line_apndx:r:checkTail}). 

Otherwise, the algorithm finds the first non-removed segment to the right (line~\ref{line_apndx:r:findR}) by following \texttt{next} pointers, and the first non-removed segment on the left (line~\ref{line_apndx:r:findL}) by following \texttt{prev} pointers. 
After that, we link the segment on the right with the segment on the left by updating its \texttt{prev} pointer (line~\ref{line_apndx:r:removeUpdPrev}). If a non-removed segment on the left was not found, \texttt{prev} is updated to \texttt{null}. Otherwise, we link such a segment with the segment on the right by updating the \texttt{next} pointer (line~\ref{line_apndx:r:removeUpdNext}).
If all segments on the right are logically removed, we manipulate the tail one.

As a result, we successfully linked the segments found on the left and on the right with each other. However, they could have been removed in meantime. Therefore, we check if they became removed and re-start the removal if they did (lines~\ref{line_apndx:r:restart0}--\ref{line_apndx:r:restart1}). Otherwise, the removal procedure is completed. It is worth noting that it is possible that concurrent \texttt{remove()}-s keep the reference to our removed segment and can accidentally re-link it with some other segment(s). However, due to checks that the segments found on the left and on the right are non-removed after the linking procedure (lines~\ref{line_apndx:r:restart0}--\ref{line_apndx:r:restart1}), we can guarantee that even if such an accident occurs, the \texttt{remove()} that led to this error will fix the problem. Thus, we know that the segment will be removed eventually. 

\clearpage
\section{Semaphore and Blocking Pools: Implementation Details}\label{appendix:primitives}
In this section, we describe the semaphore and blocking pools algorithms in detail. Additionally, for semaphore, we also cover the version with \emph{synchronous} resumption mode, which is briefly introduced in Section~\ref{sec:basic} and discussed in Appendix~\ref{appendix:sync_async}. In essence, the synchronous resumption mode is needed to support non-blocking variants of blocking operations, such as \texttt{Mutex.tryLock()} and \texttt{Semaphore.tryAcquire()}. Please see Appendix~\ref{appendix:sync_async} for details.

\subsection{Semaphore}
The semaphore algorithm is similar to the mutex one discussed during the CQS details presentation; Listing~\ref{listing:mutex_basic} presents the basic mutex version without cancellation support, while Listing~\ref{listing:mutex_smart_cancellation} fills the gap presenting the cancellation handler for the \emph{smart} cancellation mode. When implementing semaphore, the only significant difference is that instead of a single ``unlocked'' state, the \texttt{state} counter stores the number of available permits. Listing~\ref{listing:semaphore_async_smart} shows the corresponding pseudocode.


\begin{wrapfigure}{r}{0.46\textwidth}
\vspace{-1.7em}
\begin{lstlisting}[
caption={Semaphore implementation on the top of CQS with asynchronous resumption mode and smart cancellation. The initial number of permits is $K$.},
label={listing:semaphore_async_smart}
]
val cqs = CQS<Unit>(
  resumptionMode = ASYNC, #\label{line:sema:async}#
  cancellationMode = SMART #\label{line:sema:smart}#
)
// Initialized#$\:$#with#$\:$#the#$\:$#number#$\:$#of#$\:$#permits
var state: Int = K #\label{line:sema:permits}#

fun acquire() {
  s := FAA(&state, -1) #\label{line:sema:acquire:dec}#
  // Is the permit acquired? 
  if s > 0: return #\label{line:sema:acquire:acquired}#
  // Suspend otherwise
  return cqs.suspend() #\label{line:sema:acquire:suspend}#
}
fun release() {
  s := FAA(&state, 1) #\label{line:sema:release:inc}#
  // Is there a waiter to be resumed?
  if s < 0: cqs.resume(Unit) #\label{line:sema:release:resume}#
}
fun onCancellation(): Bool {
  s := FAA(&state, 1) #\label{line:sema:onCancellation:inc}#
  // If the number of waiters was 
  // decremented, the cancellation 
  // successfully completes. Otherwise, 
  // there is a release() that is going 
  // to resume this waiter, refuse it.
  return s < 0
}
fun completeRefusedResume(permit:#$\:$#Unit)#$\:$#{#\label{line:sema:ref0}#
  // The permit is already been returned
  // to this semaphore#\color{Mahogany};# do nothing.
} #\label{line:sema:ref1}#
\end{lstlisting}
\vspace{-4em}
\end{wrapfigure}

In the presented implementation, we use the asynchronous resumption (line~\ref{line:sema:async}) and the smart cancellation (line~\ref{line:sema:smart}) modes; other variants are omitted and can be easily constructed based on the one we present. 
The \texttt{state} field (line~\ref{line:sema:permits}) stores either the number of permits when positive or $0$, or the number of waiters when negative. Similarly to the mutex algorithm, \texttt{acquire()} decrements this counter and suspends if needed, while \texttt{release()} increments it and resumes the first waiter if there is one. 

\paragraph{The \texttt{acquire()} Operation.}
First, the \texttt{state} counter is decremented (line~\ref{line:sema:acquire:dec}). If the counter was positive {---} thus, the number of available permits was positive, {---} the operation has successfully taken a permit and completes immediately (line~\ref{line:sema:acquire:acquired}). Otherwise, no permits are available, so it suspends in the \sqs{} (line~\ref{line:sema:acquire:suspend}).

\paragraph{The \texttt{release()} Operation.}
First, in decrements the \texttt{state} counter (line~\ref{line:sema:release:inc}). If the counter was non-negative, then no operation is waiting in the CQS, so the permit is successfully returned to the semaphore and the operation completes. Otherwise, the next waiter should be resumed (line~\ref{line:sema:release:resume}).

\paragraph{Cancellation.}
When a waiting \texttt{acquire()} is cancelled, it increments the \texttt{state} counter (line~\ref{line:sema:onCancellation:inc}). If the counter was negative, this increment successfully decremented the number of waiters, so the cancellation succeeds and \texttt{onCancellation()} returns \texttt{true}. Otherwise, if the counter was non-negative, there already is a concurrent \texttt{release()} that will resume this waiter eventually. In this case, the corresponding \texttt{resume()} should be refused, since and the permit is already returned back, the \texttt{completeRefusedResume(..)} operation does nothing  (lines~\ref{line:sema:ref0}--\ref{line:sema:ref1}).

\paragraph{Synchronous Resumption and \texttt{tryAcquire()}.}
In order to use the synchronous resumption mode, both \texttt{acquire()} and \texttt{release()} should restart when the \texttt{suspend()} and \texttt{resume(..)} invocations fail. Thus, it becomes possible to implement an additional \texttt{tryAcquire()} operation that attempts~to take a permit if it is available {---} it simply decrements the \texttt{state} counter if it is positive or fails~otherwise.



\subsection{Blocking Pools}\label{appendix:subsec:pools}
While the barrier, count-down-latch, and semaphore algorithms use \sqs{} only for synchronization, it is also possible to develop \emph{communication} primitives on top of it. In this section, we consider simple blocking pool implementations. 

When working with expensive resources such as database connections, sockets, etc., it is common to reuse them, which usually requires an efficient and accessible mechanism. The \emph{blocking pool} abstraction maintains a set of elements that can be retrieved to process some operation, after which the element is placed back in the pool. Operations \texttt{put(element)} and \texttt{take()} are provided:
\begin{itemize}
 \item \texttt{put(element)} either resumes the first waiting \texttt{take()} operation and passes the element to it, or puts the element into the pool;
 \item \texttt{take()} takes one of the elements from the pool (in an unspecified order), or suspends if it is empty: later 
 \texttt{put(e)} operations resume waiting \texttt{take()}-s in the first-in-first-out order.
\end{itemize}

\begin{wrapfigure}[41]{r}{0.51\textwidth}
\vspace{-1.9em}
\begin{lstlisting}[
caption={Abstract blocking pool implementation that maintains the \texttt{size} counter of available elements (if positive) or waiting retrievals (if negative). All the waiting \texttt{take()} operations are processed in first-in-first-out order, while the pool itself can use any concurrent data structure under the hood. In Listing~\ref{listing:pool_stack_and_queue}, we present two solutions: based on a concurrent stack, which returns the ``hottest'' element, and based on a queue, which is relatively more efficient.},
label={listing:pool_basic}
]
val cqs = CQS<E>(
  resumeMode = ASYNC, #\label{line:pools_basic:resumeMode}#
  cancellationMode = SMART #\label{line:pools_basic:cancellationMode}#
)
var size: Int = 0 #\label{line:pools_basic:size}#

fun put(element: E) = while (true) {
 s := FAA(&size, +1) #\label{line:pools_basic:put:inc}#
 if s < 0: // is there a waiting take()?
 #\indentrule#  // Resume#$\:$#the#$\:$#first#$\:$#waiter#$\:$#and#$\:$#complete.
 #\indentrule#  cqs.resume(element); return#\label{line:pools_basic:put:resume}#
 else:
 #\indentrule#  //#$\:$#Try#$\:$#to#$\:$#insert#$\:$#the#$\:$#element.#$\:$#Can#$\:$#fail#$\:$#due
 #\indentrule#  //#$\:$#to#$\:$#a#$\:$#race#$\:$#with#$\:$#a#$\:$#concurrent#$\:$#retrieve()
 #\indentrule#  if tryInsert(element): return #\label{line:pools_basic:put:tryInsert}#
}
fun take(): E = while(true) {
 s := FAA(&size, -1) #\label{line:pools_basic:take:dec}#
 if s > 0:
 #\indentrule#  //#$\:$#Try to retrieve an element. Can fail  
 #\indentrule#  //#$\:$#due to a race with a concurrent  put(e) 
 #\indentrule#  e := tryRetrieve() #\label{line:pools_basic:take:tryRetrieve0}#
 #\indentrule#  if e != null: return e #\label{line:pools_basic:take:tryRetrieve1}#
 else:
 #\indentrule#  return cqs.suspend() // no elements #\label{line:pools_basic:take:suspend}#
}
fun onCancellation(): Bool {
 // Similar to the semaphore algorithm.
 s := FAA(&size, 1) #\label{line:pools_basic:onCancellation:inc}#
 return s < 0 
}
fun completeRefusedResume(element: E) {  #\label{line:pools_basic:crr0}#
 if !tryInsert(e): put(e)
} #\label{line:pools_basic:crr1}#
//#$\:$#When#$\:$#tryInsert(e)#$\:$#fails,#$\:$#the#$\:$#conjunctive
//#$\:$#tryRetrieve()#$\:$#fails#$\:$#as#$\:$#well,#$\:$#and#$\:$#vice#$\:$#versa
fun tryInsert(element: E): Bool
fun tryRetrieve(): E?
\end{lstlisting}
\end{wrapfigure}

In this paper, we consider two pool implementations: queue-based and stack-based. Intuitively, the queue-based implementation is faster since it can be built on segments, similarly to the \sqs{}, and uses \texttt{Fetch-And-Add}-s on the contended path~\cite{LCRQ,YM16}. In contrast, the stack-based pool retrieves the last inserted, thus the ``hottest'', element.
Please note that both algorithms presented in this section are not linearizable and can retrieve elements out-of-order under some races. However, since pools do not guarantee that the stored elements are ordered, these queue and stack-based versions should be considered as bags with specific heuristics; these semantics matches practical applications.

We start with an abstract solution that does not rely on queues, stacks, or other containers. After that, we provide solutions for queue-based and stack-based pools on top of this abstract construct.

\paragraph{Abstract Blocking Pool.}
Intuitively, the blocking pool contract reminds of a semaphore. It, like the semaphore, transfers resources, with the only difference that semaphore transfers logical non-distinguishable permits while blocking pool works with actual elements. The rest, however, is almost the same. 
Listing~\ref{listing:pool_basic} presents the abstract blocking pool implementation on top of \sqs{} with asynchronous resumption and smart cancellation (lines~\ref{line:pools_basic:resumeMode} and~\ref{line:pools_basic:cancellationMode}). Like in the semaphore, the algorithm maintains the \texttt{size} counter (line~\ref{line:pools_basic:size}, in the semaphore this counter is called \texttt{state}), which represents the number of elements in the pool if it is non-negative, and the negated number of suspended \texttt{take()} requests otherwise. 

The \texttt{put(..)} operation increments \texttt{size} first (line~\ref{line:pools_basic:put:inc}), and either resumes the next waiter if the counter was negative (line~\ref{line:pools_basic:put:resume}) or adds the element to the pool structure via \texttt{tryInsert(..)} function if there was no waiter in the pool (line~\ref{line:pools_basic:put:tryInsert}). In our design, \texttt{tryInsert(..)} can fail if a concurrent \texttt{take()} comes between the counter increment and the \texttt{tryInsert(..)} call {---} both operations should restart in this case.

The \texttt{take()} operation decrements the \texttt{size} counter first (line~\ref{line:pools_basic:take:dec}), and either tries to retrieve an element from the pool structure via \texttt{tryRetrieve(..)} if the size was positive (lines~\ref{line:pools_basic:take:tryRetrieve0}--\ref{line:pools_basic:take:tryRetrieve1}) or suspends in the CQS if there is no element to retrieve (line~\ref{line:pools_basic:take:suspend}).

As for the cancellation logic, the \texttt{onCancellation()} implementation is identical to the one in semaphore: it increments the \texttt{size} counter (line~\ref{line:pools_basic:onCancellation:inc}) and returns \texttt{true} if the number of waiters was decremented (so the counter was negative) or returns \texttt{false} if there is an upcoming \texttt{resume(..)} that should be refused.
To complete the refused resume, the algorithm tries to insert the element back into the pool structure via \texttt{tryInstert(..)} and on its failure performs a full \texttt{put(..)} {---} see the \texttt{completeRefusedResume(..)} implementation (lines~\ref{line:pools_basic:crr0}--\ref{line:pools_basic:crr1})

\paragraph{Queue-Based Pool.} 
In order to complete the pool implementation, we need to specify the \texttt{tryInsert(..)} and \texttt{tryRetrieve()} functions.
The pool with a queue under the hood is shown on the left side of Listing~\ref{listing:pool_stack_and_queue}. Our implementation is based on an infinite array (line~\ref{line:poolq:infarr}), which can be emulated in a way similar to how it is done in the \sqs{} framework.

The \texttt{tryInsertQueue(..)} (we added \texttt{Queue} and \texttt{Stack} suffixes to distinguish the implementations) operation increments its \texttt{insertIdx} counter (line~\ref{line:poolq:insert:inc}) and tries to atomically change the corresponding slot in the infinite array from \texttt{null} to the given element via \texttt{CAS} (line~\ref{line:poolq:insert:cas}). If this \texttt{CAS} fails, it means that a concurrent \texttt{tryRetrieveQueue()}, which already discovered the preceding \texttt{size} increment, came to the same array slot and broke it (line~\ref{line:poolq:retrtieve:cas}) {---} \texttt{tryInsertQueue(..)} returns \texttt{false} in this case.

The \texttt{tryRetrieveQueue()} operation increments its \texttt{retrieveIdx} counter (line~\ref{line:poolq:retrtieve:inc}) and tries to retrieve an element from the corresponding infinite array slot (line~\ref{line:poolq:retrtieve:cas}). If the slot is empty, it breaks it by atomically replacing \texttt{null} with the \texttt{BROKEN} token and returns \texttt{false}, causing the paired \texttt{tryInsertQueue(..)} to fail as well (line~\ref{line:poolq:insert:cas}).


\begin{figureAsListingWide}
\begin{minipage}[t]{0.43\textwidth}
\begin{lstlisting}[
]
// The queue bases on an infinite 
// array with two counters.
val a = InfiniteArray() #\label{line:poolq:infarr}#
var insertIdx: Long = 0 #\label{line:poolq:insertIdx}#
var retrtieveIdx: Long = 0 #\label{line:poolq:retrtieveIdx}#

fun tryInsertQueue(element: E): Bool {
  // Get an index for this insertion
  i := FAA(&insertIdx, 1) #\label{line:poolq:insert:inc}#
  // Try to put the element into the
  // slot, failing if it has already  
  // been broken by a paired   
  // retrieval that came earlier.
  return CAS(&a[i], null, element) #\label{line:poolq:insert:cas}#
}
fun tryRetrieveQueue(): E? {
  // Get an index for this retrieval
  i := FAA(&retrtieveIdx, 1) #\label{line:poolq:retrtieve:inc}#
  // Replace the slot value with #\color{Mahogany}$\bot$#. 
  // If null, the slot becomes broken 
  // and this retrieval attempt fails.
  return GetAndSet(&a[i], BROKEN)#\label{line:poolq:retrtieve:cas}#
}
\end{lstlisting}
\end{minipage}
\hfill
\begin{minipage}[t]{0.52\textwidth}
\begin{lstlisting}[firstnumber=24
]
class Node(val element: E, next: Node)
var top: Node? = null

fun tryInsertStack(element: E): Bool {
 while (true) {
 #\indentrule#  t := top
 #\indentrule#  // Does this stack contain 
 #\indentrule#  // failed retrievals?
 #\indentrule#  if t != null && t.element == BROKEN: #\label{line:pools:insert:check}#
 #\indentrule#  #\indentrule#  // Remove the failed node and fail.
 #\indentrule#  #\indentrule#  if CAS(&top, t, t.next): return false #\label{line:pools:insert:failed}#
 #\indentrule#  else:
 #\indentrule#  #\indentrule#  // The stack is either empty 
 #\indentrule#  #\indentrule#  // or contains elements.
 #\indentrule#  #\indentrule#  if CAS(&top, t, Node(element, t)): return true #\label{line:pools:insert:normal}#
 }
}
fun tryRetrieveStack(): E? {
 while (true) {
 #\indentrule#  t := top
 #\indentrule#  // Does the stack have elements?
 #\indentrule#  if t == null || t.element == BROKEN:  #\label{line:pools:retrieve:check}#
 #\indentrule#  #\indentrule#  //#$\:$#The stack is either empty or  
 #\indentrule#  #\indentrule#  //#$\:$#contains failed retrievals;
 #\indentrule#  #\indentrule#  // add one more and fail
 #\indentrule#  #\indentrule#  if CAS(&top, t, Node(BROKEN, t): return null #\label{line:pools:retrieve:failed}#
 #\indentrule#  else:
 #\indentrule#  #\indentrule#  // The stack contains elements;
 #\indentrule#  #\indentrule#  // try to remove the top one.
 #\indentrule#  #\indentrule#  if CAS(&top, t, t.next): #\label{line:pools:retrieve:normal0}#
 #\indentrule#  #\indentrule#  #\indentrule#  return t.element #\label{line:pools:retrieve:normal1}#
 }
}
\end{lstlisting}
\end{minipage}
\caption{Blocking pool specializations built on top of the solution in Listing~\ref{listing:pool_basic} with a queue (on the left) and stack (on the right) under the hood. Intuitively, the queue-based implementation is faster since it is based on arrays and uses \texttt{Fetch-And-Add}-s on the contended path, while the stack-based pool retrieves the last inserted, thus the ``hottest'', element.}
\label{listing:pool_stack_and_queue}
\end{figureAsListingWide}

\paragraph{Stack-Based Pool.} 
The pool with a classic Treiber stack under the hood is presented on the right side of Listing~\ref{listing:pool_stack_and_queue}. Here we face a similar race when \texttt{put(..)}, which has already incremented the \texttt{size} counter but has not inserted the element yet, interferes with a concurrent \texttt{take()} that tries to retrieve an element. We use an approach similar to breaking slots in the previously discussed queue-based pool. The difference is that, instead of breaking slots, \texttt{tryRetrieveStack()} inserts a ``failed node'' if the stack is empty or contains other failed nodes (lines~\ref{line:pools:retrieve:check}--\ref{line:pools:retrieve:failed}); otherwise, it removes the top node with an element (lines~\ref{line:pools:retrieve:normal0}--\ref{line:pools:retrieve:normal1}). On the opposite side, \texttt{tryInsertStack()} checks that the stack does not have these ``failed nodes'', removing one and failing if they exist (lines~\ref{line:pools:insert:check}--\ref{line:pools:insert:failed}); otherwise, it inserts a node with the specified element (line~\ref{line:pools:insert:normal}). 
\section{Progress Guarantees}\label{appendix:progress_guarantee}
Here, we discuss the progress guarantees of both CQS \texttt{suspend(..)} and \texttt{resume(..)} operations and the primitives from Section~\ref{sec:primitives} built on top of the \sqs{} framework.
Notably, we consider both \emph{asynchronous} and \emph{synchronous} resumption modes, where the first is the default one described on the main body, and the \texttt{synchronous} resumption aims at supporting non-blocking operations, such as \texttt{Mutex.tryLock()} or \texttt{Semaphore.tryAcquire()} {---} it is briefly introduced in Section~\ref{sec:basic} and discussed in full detail in Appendix~\ref{appendix:sync_async}. 

Similarly to the dual data structures formalism~\cite{scherer2006scalable}, we reason about progress independently of whether the operation was suspended. Thus, when we say that some blocking operation is lock- or wait-free, we mean that it performs all the synchronization with this progress guarantee, either completing immediately or adding itself to the queue of waiters followed by suspension.
Specifically, we analyze the part of the operation prior to \texttt{Thread.park(..)} call, if one ever occurs.

\subsection{The CQS Operations}\label{subsec:progress:cqs}
First, we discuss the \texttt{suspend()} and \texttt{resume(..)} of the \sqs{} framework itself,  followed by the analysis of the barrier, the count-down-latch, the semaphore, and the blocking pools presented in Section~\ref{sec:primitives} and Appendix~\ref{appendix:primitives}.

\paragraph{The \texttt{suspend()} Operation.}
The \texttt{suspend()} operation obtains the id of the working cell by incrementing \texttt{suspendIdx}. It then finds the required segment in a bounded number of steps and either installs the currently running thread to the cell or returns the value already stored in it, failing if the cell is already broken by a concurrent \texttt{resume()}. In either case, it completes within a finite number of its own steps, and is, therefore, \emph{wait-free}.

\paragraph{The \texttt{resume(..)} Operation.}
The behaviour of \texttt{resume(..)} depends on the cancellation mode. If no cancellation happened during the execution, \texttt{resume(..)} obtains an id of the working cell by incrementing \texttt{resumeIdx}, finds the required segment in a bounded number of steps, and either places the element in the cell (optionally waiting in a bounded loop in the synchronous resumption mode) or resumes the stored waiter. In either case, \texttt{resume(..)} is wait-free.

With simple cancellation, \texttt{Thread.cancel()} moves the cell state to \texttt{CANCELLED}, and the \texttt{resume(..)} that processes this cell fails. Therefore, \texttt{resume(..)} remains \emph{wait-free}.

The situation is more complex in the smart cancellation mode. In this case, the progress guarantee of \texttt{resume(..)} depends on the resumption mode. In the synchronous resumption mode, \texttt{resume(..)} may wait in a spin-loop until the cell's state changes from $\mathtt{THREAD}_\mathtt{CANCELLED}$ to \texttt{CANCELLED} or \texttt{REFUSE}. Thus, \texttt{resume(..)} is \emph{blocking}. In the asynchronous mode, \texttt{resume(..)} is \emph{lock-free} due to a possibly infinite number of \texttt{suspend()}-s that place and immediately abort. However, the progress guarantee can degrade if the \texttt{completeRefusedResume(..)} implementation, which is specified by the user and invoked when a \texttt{resume(..)} detects that it was refused, ensures a weaker progress guarantee.

\paragraph{The \texttt{Thread.cancel(..)} Operation.}
With simple cancellation, \texttt{Thread.cancel()} moves the cell state to \texttt{CANCELLED} and potentially removes the segment if the last cell was cancelled. The segment removing procedure is lock-free, so cancellation obeys lock-freedom as well.

With smart cancellation, the handler invokes the \texttt{onCancellation()} function and can also invoke the \texttt{completeRefusedResume(..)} procedure {---} both of them are specified by the user. In addition, the handler can call \texttt{resume(..)} in the asynchronous (default) resumption mode. The \texttt{resume(..)} operation is at best lock-free, so the overall cancellation is lock-free if the functions specified by the user guarantee lock-freedom as well, and is bounded by their progress guarantees otherwise.

\subsection{Barrier} \label{subsec:progress:barrier}
Since our implementation does not support cancellation and the asynchronous resumption mode is used, it is guaranteed that both \texttt{suspend()} and \texttt{resume(..)} synchronizations are wait-free. The rest of the \texttt{arrive()} operation is also wait-free, which should be obvious from the code. Therefore, our implementation guarantees \emph{wait-freedom}.

\subsection{Count-Down-Latch} \label{subsec:progress:cdl}

Since \texttt{suspend()} is wait-free and does not fail, the \texttt{await()} operation is obviously \emph{wait-free} as well. 
The cancellation, however, is \emph{lock-free} due to possible segment removing.

As for the \texttt{countDown()} operation, it performs \texttt{Fetch-And-Add} at line~\ref{line:cdl_cd_inc} and invokes \texttt{resumeWaiters()} if the count has reached zero at line~\ref{line:cdl_cd_rw}. Thus, the progress guarantee for \texttt{countDown()} is completely dependent on the \texttt{resumeWaiters()} function. Surprisingly, even with the infinite loop wrapper, the number of failed \texttt{CAS}-s to set the \texttt{DONE\_BIT} at line~\ref{line:cdl_rw_setbit} is bounded by the number of concurrent \texttt{await()} invocations, and thus, by the parallelism level in general. If new \texttt{await()} invocations happen when \texttt{resumeWaiters()} is invoked, since the \texttt{count} is already zero, they complete immediately and neither change the \texttt{waiters} field nor suspend. As a consequence, \texttt{resume(..)} can skip a bounded number of cancelled cells and is wait-free. In sum, \texttt{resumeWaiters()} along with the \texttt{countDown()} operation are \emph{wait-free}.

\subsection{Semaphore} \label{subsec:progress:semaphore}
Consider the case where no cancellation happens during the execution. In this case, both \texttt{suspend()} and \texttt{resume(..)} are \emph{wait-free}, so \texttt{acquire()} and \texttt{release()} are also wait-free. However, when synchronous resumption is used, concurrent \texttt{suspend()} and \texttt{resume(..)} can lead to failing each other. Therefore, the operations may restart and interfere infinitely with synchronous resumption, so only \emph{obstruction-freedom} is guaranteed.

Cancellation weakens the progress guarantees. With asynchronous resumption, \texttt{resume(..)} is only \emph{lock-free} since there can be an infinite sequence of \texttt{suspend()}-s followed by successful cancellations, so any given \texttt{resume(..)} may not finish while the system makes progress. Since the cancellation handler can invoke \texttt{resume(..)}, it is, therefore, also \emph{lock-free}.
With the synchronous resumption, the \texttt{resume(..)} operation is \emph{blocking}, while the cancellation part is \emph{lock-free} due to a possible segment removal.

\subsection{Blocking Pools} \label{subsec:progress:pools}
The queue-based pool provides wait-free \texttt{tryInsert(e)} and \texttt{tryRetrieve()} functions, while in the stack-based version, they ensure lock-freedom.
However, additions and removals can interfere in an obstruction-free way due to the slot breaking in the queue-based version and publishing ``failed nodes'' in the stack-based one. Nonetheless, they always complete in a bounded number of steps when all other threads are paused. Therefore, all the operations, including the cancellation that can invoke \texttt{put(..)} as a part of \texttt{completeRefusedResume(..)}, are \emph{obstruction-free}. 
\newcommand\memLoc[1]{\texttt{#1}}%
\newcommand\methName[1]{\texttt{#1}}%
\newcommand\literal[1]{\texttt{#1}}%
\newcommand\field[1]{\texttt{#1}}%
\newcommand\defn[1]{\emph{#1}}%
\newcommand\stateName[1]{\texttt{#1}}%
\newcommand\futureState[1]{$\texttt{FUTURE}_\texttt{#1}$}%
\newcommand\suspendOp{\methName{suspend()}}%
\newcommand\onSlotCleaned{\methName{onCancelledCell()}}%
\newcommand\resumeOp{\methName{resume(v)}}%
\newcommand\marker[1]{\texttt{#1}}%
\newcommand\faa{\texttt{Fetch-And-Add}}
\newcommand\segId{\texttt{id}}
\newcommand\getAndSet{\texttt{GetAndSet}}
\newcommand\segmentSize{\literal{SEGM\_SIZE}}
\newcommand\fileInRepoNoSurround[1]{\url{https://github.com/Kotlin/kotlinx.coroutines/tree/cqs-proofs/theories/lib/#1}}
\newcommand\fileInRepo[1]{
(see file \fileInRepoNoSurround{#1})
}

\section{Formal Specification and Proofs for CQS}\label{sec:proofs}

This section outlines the formal proofs for the Coq formalization of \sqs{}; the specifications and proofs of the presented algorithms on top of CQS are discussed in Section~\ref{sec:proofs-primitives}.
{\bf The proofs themselves are available on GitHub \cite{proofRepo}}. The corresponding files are referenced throughout.

Providing formal proofs of correctness for concurrent data structures is
currently rare, and even more so for algorithms and data structures employed in a realistic production setting. 
(This is in spite of Iris certainly being powerful enough to express such proofs.) 
Notable exceptions include the verification of a concurrent queue used in the Dartino framework~\cite{krogh2016verifying}, proofs for algorithms used in real-world databases and
filesystems~\cite{krishna2020verifying}, the contextual refinement of a concurrent queue similar to one in the Java standard library~\cite{vindum2021contextual}, and the recent cases of the Meta company verifying several of its internally-used concurrent data structures~\cite{carbonneaux2022applying, vindum2022mechanized}. 
We also highlight the proofs for a wide range of libraries used throughout the Rust ecosystem~\cite{jung2017rustbelt}.

We suspect that the main reason for the dearth of such proofs is the high complexity barrier, preventing
users from using separation logic to encode the intuition behind the data structure design. 
Also, in our experience, obtaining formal proofs for complex data structures is not obvious: 
in total, the proofs of the claims of this paper span more than 10'000 lines
of Coq code, much of which required non-trivial reasoning.

We found the formalization process quite useful, as it identified subtle correctness issues in our implementation, especially in the case of the cancellation operation. We note that the proofs below do not attempt to show the FIFO property: proving such properties is known to be very challenging in our framework, and can be approached via classical proofs.

\paragraph{Reader Guide.}
This section outlines the basic ideas, definitions, and rationale behind our proofs in Coq, and functions essentially as ``liner notes'' for the formal proof. 
The experienced reader may wish to directly examine the proof text, perhaps in conjunction with Section~\ref{sec:cqs-proof}. 
Although we strive to justify our definitions and choices, we understand that some readers may find it difficult to internalize the fine details in this section. This is due to the fact that proofs of such massive algorithms as CQS are typically hard to follow and understand. That is the reason why we decided to prove the framework in Coq, which guarantees correctness of our proofs: manual proofs would provide too big of a surface for error for our liking.

\subsection{Structure of the Proofs}
\label{subsec:proof_structure}

\paragraph{Resources and Invariants.}
The discussion here is a high-level description of the notions on which Iris operates. Its purpose is to provide the reader with just enough intuition to be able to follow the outline provided in this paper. A more detailed and technical discussion can be found in the description of Iris itself~\cite{jung2018iris}.

There are two basic notions at the heart of the proofs: \defn{resources} and
\defn{invariants}.

A \emph{resource} is an entity that only exists in the logical
realm, does not affect the code execution in any way, and is used to keep track
of our knowledge about the state of the system. An example of a resource is an
exclusive right to write to a particular memory location. Each executing thread
keeps a collection of resources that it can use to perform various operations.

An \emph{invariant} is a collection of resources that is always owned by a data
structure itself, as opposed to some particular thread. This notion is not to be
confused with a loop invariant, which is a broadly similar, but meaningfully distinct
concept. The resources stored in an invariant can be used by any thread at
any time as long as no thread can ever observe the invariant not holding: in
particular, it is allowed for a thread to borrow the resources from an invariant
for the duration of an atomic operation, but not for longer. (Note that if the proofs
were performed in a weak memory model, this notion would have to be significantly more
elaborate.)

Resources can be allocated; some resources can be deallocated; some can be
duplicated, split into fractions, combined to form other resources, etc. In
this outline of the proof, the inner workings of these operations are omitted due to the sheer scope of the formal proof and the number of resources that needed to be defined;
instead, we postulate where needed the existence of resources with the required
properties or even imply it. We feel justified to focus on the general picture
due to the fact that Coq
has performed an automatic verification of the validity of our claims. For example,
we often say that a particular data structure ``knows'' that there only exists
a fixed number of copies of a particular resource; such knowledge is itself
represented as a resource that is stored in an invariant associated with that
structure, but presenting the proofs in accordance with this would, in our view,
obscure the general view in favor of minutiae.

\paragraph{Specifications of Methods.}
The proof of each method is provided in the form of specification of how its
behavior affects the available resources. Specifications have the following
form: ``If an expression $e$ is executed by a thread that owns $A$,
then the call does not break any invariants and, when it completes, it returns
a value $v$ and provides the calling thread with $B$'', where $A$ and $B$ are
(groups of) resources and $v$ can appear in the definition of $B$.

The specification can be parameterized with some values (usually the arguments
to the method), which can appear in definitions of $e$, $A$, and $B$.

As an example of a specification, we consider \getAndSet{}, also commonly known as \texttt{swap}(), which always successfully writes $v$ to
a memory location $\ell$ and returns the value that was stored there at the moment of the write, can be given as
follows: ``If $\getAndSet(v)$ is executed by a thread that owns the exclusive
knowledge
about the memory location $\ell$ containing $x$, then the call returns
$x$ and produces the exclusive knowledge about $\ell$ containing $v$''. The
``exclusivity'' here is mentioned because if some other parts of the system
knew that $\ell$ contained $x$, the method could not be correctly executed, as
it would violate the knowledge owned by the other parties.

There are some weaknesses to this form of specification: if a method
never finishes and instead hangs without breaking any invariants, then the
specification is still correct. In fact, a simple infinite loop that does not access
any state satisfies any specification. This is an important reason for why we
address the progress guarantees separately from correctness proofs.

\paragraph{Specifications of Logically Atomic Methods.}
An additional special case is that of methods that need to perform their
operations atomically in order to be correct. For example, consider the
specification of the \getAndSet{} operation given above. That specification is highly
impractical, as \getAndSet{} usually operates on shared state, so it is not
possible to provide only one thread with the exclusive knowledge of the contents
of a memory cell for the whole duration of \getAndSet{}: an attempt by any other thread to access the cell in the meantime would be invalid, as only one
thread has any knowledge about that memory.

To deal with this issue, a separate form of specifications exists: ``If $e$ is
executed and has access to $A$, then at some point in time it
atomically consumes $A$ and provides $B$; after it finishes, it returns a value
$v$'', where $v$ can appear in the definition of $B$ and $A$ can be
parameterized by some values that can also appear in the definition of $B$.
``Having access'' here means obtaining $A$ (possibly several times) and
immediately providing it back.

It is possible then to provide a useful specification of \getAndSet{}: ``If
$\getAndSet(v)$ is executed and has access to knowledge that $\ell$ contains
some value (we call the value it has at this moment in time $x$), then at some
point it replaces this knowledge with the fact that $\ell$ contains $v$; after
the operation finishes, it returns $x$''. To simplify the nomenclature, we
instead say that $\getAndSet(v)$ atomically replaces the value in $\ell$ with
$v$ and returns the initial value; this section is meant to define what
specifically we mean by $\getAndSet(v)$ being atomic even though, as usually
defined, it can access a memory location several times, which is atomic only
logically and not physically.

\paragraph{Specifications of Logical Operations.}
In addition to specifications of code, we recognize some operations that do not require any code to
execute and only operate on resources and invariants. For example, there could
exist an invariant that owned an instance of either an exclusive (that is,
one-of-a-kind) resource $C$ or a $D$; then a thread that owns the $C$ could
obtain an instance of $D$: given that $C$ is exclusive and the thread owns it,
the invariant must be holding $D$, so the thread can then swap the $D$ for its
$C$ without violating the invariant, and, most importantly for the point raised
here, without executing any code.

\subsection{Futures}
\label{subsec:proof_futures}

Throughout the paper, we used the notion of threads that allow parking, unparking, and canceling them. However, in HeapLang, the default language to describe computations in the Iris framework, threads are a very light concept that essentially just describes code running in parallel with arbitrary interleavings. It doesn't have a notion of parking, unparking, or cancellation.

Therefore, in order to perform formal proofs of operations that support cancellation, we need to introduce some model of blocking computations that is general enough to be adapted to any practical language or library, independently of whether they are built on threads or coroutines, while at the same time supporting all operations on threads that we used.

The model we chose is that of \texttt{Future}s, a structure that allows passing a unique value to it, checking whether the value is present, or canceling the computation that would provide the value. The following is a description of the model, including its pseudocode.

\paragraph{Example: the Mutex.}
Consider the \texttt{lock()} operation in mutex. Intuitively, it either takes the lock immediately or registers as a waiter and then is resumed by an \texttt{unlock()} operation. We can split \texttt{lock()} into two phases at the point of suspension. This idea is inspired by the dual data structures formalism~\cite{scherer2006scalable}, originally designed for synchronous queues, where these two phases are named ``registration'' and ``follow-up''. 

Unlike the dual data structures formalism, we make suspension \emph{explicit} by returning a special \texttt{Future} instance as a result of a blocking operation. With this change, \texttt{lock()} in mutex returns \texttt{Future<Unit>}. See Listing~\ref{lst:mutex_api_future} below.

%
%
\begin{lstlisting}[
label={lst:mutex_api_future},
caption={Mutex API via \texttt{Future}-s.}
]
interface Mutex {
 fun lock(): #\st{Unit}# @Future<Unit>@ { ... }
 fun release() { ... }
}
\end{lstlisting}
%

The \texttt{lock()} operation completes regardless of whether the lock has been successfully acquired or the request was put into the waiting queue. If the lock has not yet been acquired, calling \texttt{get()} on this \texttt{Future} returns \texttt{null} instead, but after the lock is transferred to the waiting \texttt{lock()} operation, \texttt{get()} starts returning \texttt{Unit}, indicating that the blocking part of the \texttt{lock()} operation has completed with the result \texttt{Unit}.

\paragraph{Implementation of \texttt{Future}s.}
Since it is possible for a potentially blocking operation to complete immediately, we have two \texttt{Future} implementations presented in Listing~\ref{listing:futures}: \texttt{ImmediateResult} is returned when the operation completes without suspension, while \texttt{Request} is returned when the operation suspends.

Though most synchronization primitives return \texttt{Unit} as a result of blocking request, there are plenty of data structures, such as blocking queues, where operations also manipulate some data. 
Thus, we make our \texttt{Future} generic in type parameter \texttt{R} (line~\ref{line:futures:future}).
In addition, we provide a way to cancel the waiting request via the \texttt{cancel()} operation (line~\ref{line:futures:future_cancel}). When the operation is not completed yet, \texttt{cancel()} succeeds and returns \texttt{true}, and \texttt{get()} starts returning $\bot$.
Also, the specified cancellation handler (line~\ref{line:futures:r_canchandler}) is invoked in this case.

\begin{lstlisting}[float,
caption={Implementations of \texttt{Future} for both suspending and immediate completing situations.},
label={listing:futures}
]
interface Future<R> { #\label{line:futures:future}#
 fun get(): R? or #$\bot$# // R - completed with R #\label{line:futures:future_get}#
                    // null - not completed 
                    // #\color{Mahogany}$\bot$# - cancelled
 fun cancel(): Bool // true - cancelled #\label{line:futures:future_cancel}#
                    // false - completed
}
// Use this Future without suspension.
class ImmediateResult<R>(
 val result: R // the operation result #\label{line:futures:ir_res}#
) : Future<R> {
 override fun get() = result #\label{line:futures:ir_get}#
 override fun cancel() = false #\label{line:futures:ir_cancel}#
}
// Use this Future when suspending.
class Request<R>( 
 val cancellationHandler: () -> Unit #\label{line:futures:r_canchandler}#
) : Future<R> {
 var result: R? or #$\bot$# = null // #\color{Mahogany}$\bot$ =># cancelled #\label{line:futures:r_result}#
  
 fun complete(r: R): Bool = CAS(&result, null, r) #\label{line:futures:r_complete}#
 
 override fun get() = result
 override fun cancel(): Bool {
  if CAS(&result, null, #$\bot$#): // mark as cancelled
  #\indentrule#  cancellationHandler() // invoke the handler #\label{line:futures:r_cancel}#
  #\indentrule#  return true // successfully cancelled
  return false // already completed
 }
}
\end{lstlisting}

The implementation in Listing~\ref{listing:futures} is certainly not the only one that will ensure the correct work of the provided data structures. All the proofs were performed against a generalized specification of the provided code, not against the actual code. As long as a set of fairly liberal requirements (listed below) is fulfilled, it's possible to implement such an interface in a wide variety of various programming languages and libraries, and the proofs will be immediately applicable.

For example, the blocking code throughout the paper relies on Java-like behavior of aborted threads throwing \texttt{InterruptedException}, which can be caught and processed by the user, which corresponds to canceling a future. Likewise, some coroutines libraries, such as Kotlin Coroutines~\cite{kotlincoroutines}, already support an API similar to the one in \texttt{Request}.

\clearpage

The actual requirements that are placed on the \texttt{Future} implementation are as follows:
\begin{itemize}
    \item A Future cannot be both cancelled and completed.
    \item Both cancellation and completion must happen in a logically atomic
    manner: there must be a single atomic operation that transfers a pending
    Future to one of the terminal states.
    \item At most one call to the cancellation handler may ever happen. If this property does not arise from the implementation of \texttt{Future}, it is easy to achieve this by replacing the cancellation handler with a version that checks whether the cancellation handler was already invoked and only invoking the original one if it was not.
    \item The right to complete a Future must be exclusive, that is, it must not ever be accessible by third parties.  Specifically, the
    implementation of the CQS would be certainly incorrect if a Future stored there could be completed
    by something other than a call to \resumeOp{}.
    \item There must exist an exclusive right to perform the acquisition of the
    logical resources stored in the completed Future. For example, in the case of a mutex, among the calls to \methName{future.get()} that
    return the unit value and not \literal{null} or $\bot$, only a single one of them
    actually has the right to enter the critical section.
    \item If the Future was ever completed or cancelled, it stays that way.
\end{itemize}

\paragraph{Specification.} There are several logical resources introduced for
the specification\fileInRepo{util/future.v}:
\begin{itemize}
    \item The \defn{completion permit}. This is a fractional resource: it can
    be split into several parts, but in order to perform some operations with
    the completion permit, the whole resource is required. There can not exist more than one completion permit for any given Future at any time.
    
    \item The \defn{cancellation permit}, with the same properties as the
    completion permit.

    Despite the name, the cancelling permit is also used as the exclusive right
    to acquire the logical resources stored in the Future. A separate permit
    could be introduced for this, but that would not affect the proof of the
    CQS but could complicate the specification of the Future.
    \item Knowledge that the Future was completed with some value $v$. This
    contradicts the knowledge that it was completed with some other value $v'$,
    with any fraction of a completion permit, or the knowledge that it was
    cancelled. Such knowledge is freely duplicable.
    \item Knowledge that the Future was cancelled; also freely duplicable and
    contradicting the existence of any fraction of the cancellation permit.
\end{itemize}

Additionally, each Future is parameterized with some logical resource;
we say that a Future is \defn{$R$-passing} if its parameter is $R$.
The following operations are supported on Futures:
\begin{itemize}
    \item Creation of a completed Future with \methName{ImmediateResult(..)}.
    When provided with an $R$, this method creates an $R$-passing Future,
    providing its cancellation permit and the knowledge that the Future was
    completed.
    \item Creation of an empty Future with \methName{Request()}. It creates
    an $R$-passing Future and provides its completion and cancellation permits.
    \item \methName{complete($\bot$)} atomically consumes the
    cancellation permit and returns either \literal{true} and the knowledge
    that the Future is cancelled or \literal{false} along with the untouched
    cancellation permit and the knowledge that the Future was already completed.
    \item The \methName{complete(v)} method obeys two specifications.

    It can be called as an atomic function that accepts the completion permit
    and an instance of $R$ and behaves symmetrically to
    \methName{complete($\bot$)},
    successfully providing the knowledge that the Future is completed, or failing,
    which shows that the Future was cancelled, and giving back the instance of $R$
    and the completion permit.

    Alternatively, if it is called with the completion permit and the knowledge
    that the Future is already cancelled, it always returns \literal{false} and
    gives back the completion permit.
    \item The \methName{get()} method accepts the cancellation permit as the
    exclusive right to perform acquisition of $R$. It atomically consumes the
    cancellation permit and either returns \literal{null} and provides back
    the cancellation permit or returns a value and provides an $R$ and half
    of the cancellation permit.
\end{itemize}

The specification of cancellation is not provided, as its effects heavily depend
on the behavior of the cancellation handler, and the proof as a whole would
become more difficult.

\paragraph{Invariants.} We register the following invariants:
\begin{enumerate}
\item Each Future is empty, completed, or cancelled.

\item If the Future is empty, then \memLoc{result} stores a \literal{null}, there
exist both the completion permit and the cancellation permit, and there does not exist
the knowledge that the Future was cancelled or completed.

\item If the Future is completed, then \memLoc{result} stores the value that the
Future was completed with, along with either a copy of $R$ or a half of the
cancellation permit. The cancellation permit exists, as does the knowledge that
the Future is completed.

\item  If the Future is cancelled, \memLoc{result} stores $\bot$, and there exist
both the completion permit and the knowledge that the Future was cancelled.
\end{enumerate}

\paragraph{Execution.}
The correctness of the methods can be verified by observing the effect
of the atomic operation underlying each of them and checking that the resources
entering and leaving the ownership of the Future are kept in balance.

\subsection{The Underlying Concurrent Linked-List}

The infinite array, a data structure that is key for defining the CQS,
is based on a concurrent linked list. Here, we discuss the part of the infinite
array that is dedicated to the management of segments.

For this proof, we abstract from
\methName{moveForwardSusp(..)} and
\methName{moveForwardResume(..)} to
just \methName{moveForward(..)} that works on any pointer to segments.

\subsubsection{Specification}\fileInRepo{concurrent_linked_list/list_spec.v} We introduce some additional logical resources
for describing the behavior of concurrent linked lists. First, we have 
the knowledge that a segment is \defn{logically removed}; this resource
is freely duplicable, which implies that a once-removed segment cannot stop being
removed, and is physically represented as \memLoc{cancelled} being equal to
\segmentSize{} and \memLoc{pointers} being $0$ simultaneously. The
second resource is the knowledge that a segment pointer $p$
(in this program, $p$ is either \memLoc{resumeSegm} or \memLoc{suspendSegm})
\defn{points to} $i$, which means two things: first, that $p$ contains a
reference to a segment whose \segId{} is $i$, and second, that it owns a piece
of the \memLoc{pointers} counter of that segment in the following
sense: as long as the piece exists, the counter stores at least $1$;
decreasing the counter by $1$ always requires the execution to relinquish a
piece of the counter. Note that the existence of a piece of a counter
contradicts the segment being logically removed, which follows directly from the
definitions.

If a segment $n$ was part of the linked list at some point, then segments
$[0..n-1]$ also were part of the list.

The \methName{findSegment(s, id)} operation takes as arguments a segment $s$ with identifier
$s.\segId{}$ and \segId{}, and returns some segment $t$ such that
$t.\segId{} \ge s.\segId{}$, $t.\segId{} \ge \segId{}$, and all the segments
in $[\max(s.\segId{}, \segId{}); t.\segId{})$ are cancelled.

The \methName{moveForward[$p$](to)} operation is a logically atomic operation that, if
$p$ points to \texttt{from}, returns \literal{true} if $p$ now points to
the maximum of \texttt{from} and $\texttt{to}.\segId{}$, and \literal{false}
if \texttt{to} is logically removed, in which case $p$ points to \texttt{from}.

The \methName{\onSlotCleaned{}} operation is a logically atomic operation that
is parameterized by some logical resources $\Phi$ and $\Psi$ such that the
existence of $\Phi$ implies that the \memLoc{cancelled} counter in the segment
is not yet equal to \segmentSize{} and it is possible to correctly
increase it by $1$ by relinquishing the ownership of $\Phi$, obtaining $\Psi$
in return. The operation atomically exchanges $\Phi$ for $\Psi$, which means
that the ability to increase \memLoc{cancelled} both existed and was utilized.

Last, setting the \memLoc{prev} of a segment to \literal{null} is always valid
and has no effect. This may seem like an incorrect statement, as it would mean
that not having a \memLoc{prev} field at all would not affect correctness even
though it would lead to \methName{remove()} not removing segments. This is
true and points to another limitation of the provided formal proofs: they do not
account for memory leaks and only concern themselves with invariants not being
violated and specifications being met. In fact, as will be shown later, the
specification for \methName{remove()} only claims that as long as it is only
called on logically removed segments, the invariants are not violated.

\subsubsection{Invariants}
For each segment $s$ of a concurrent linked list, the following holds:
\begin{itemize}
    \item If \memLoc{pointers} is $0$ and \memLoc{cancelled} is
    \segmentSize{}, the segment is logically removed; otherwise, it is not
    logically removed and either there exist some pieces of the
    \memLoc{cancelled} counter or \memLoc{pointers} is not yet \segmentSize{}.

    \item \memLoc{prev} contains either \literal{null} or a reference to a
    segment $s'$ such that $s'.\segId{} < s.\segId{}$ and all the segments
    between $s'$ and $s$ are cancelled.

    \item \memLoc{next} contains \literal{null} if it is the tail segment
    (that is, the rightmost segment that ever existed in the list);
    otherwise, it contains a reference to a
    segment $s'$ such that $s_\segId{} < s'_\segId{}$ and all the segments
    between $s$ and $s'$ are cancelled.

    Observe that this invariant implies that the current tail segment is always
    accessible from any other segment by following \memLoc{next} repeatedly:
    the chain of \memLoc{next} may only end with the tail.
\end{itemize}

\subsubsection{Execution}\fileInRepo{concurrent_linked_list/list_proof.v}

\paragraph{\methName{removed()}} This method checks whether the segment is
logically removed. If it is, the method returns \literal{true} and provides the
knowledge that the segment is logically removed. Otherwise, it just returns
\literal{false}.

\paragraph{\methName{remove()}} We claim that this operation does not violate any invariants if it
is called on a logically removed segment.

First, if the segment turns out to be the tail, nothing is done. Otherwise,
the segment is not the tail and can not become one anymore by definition.

Observe that \methName{aliveSegmLeft(s)} returns either \literal{null} or
a segment $s'$ such that $s'.\segId{} < s.\segId{}$ and all the segments between
$s'$ and $s$ are cancelled. This holds for the initial value of \memLoc{cur} due
to the invariant on \memLoc{prev}, and each loop iteration preserves this,
according to the same invariant on \memLoc{prev}, which can be easily observed.

Likewise, \methName{aliveSegmRight(s)}, when called on a non-tail segment,
returns a segment $s'$ such that
$s.\segId{} < s'.\segId{}$ and all the segments between $s$ and $s'$ are
cancelled. Note that \literal{null} can not be returned from this method.

Knowing this, we can easily observe that variables \memLoc{prev} and
\memLoc{next} defined in \methName{remove()} satisfy precisely the invariants
for when a value would be a valid content of fields \memLoc{prev} or \memLoc{next} respectively.

Then, the operation (which does not require any additional resources to be
initiated) is retried if it turns out that \memLoc{prev} and \memLoc{next} could
be validly pointing even further.

\paragraph{findSegment(start, id)}
We establish a loop invariant that \memLoc{cur} contains a reference to a segment $s$ whose
\segId{} is not less than that of \memLoc{start}, and also all segments from
$\max(\memLoc{start}.\segId{}, \memLoc{id})$ (inclusive) to $s.\segId{}$
(exclusive) are cancelled. This holds initially since there are no such
segments for $s = \memLoc{start}$.

If it turns out that $s.\segId{}$ is not less than \memLoc{id} and $s$ is not
removed, then, by the loop invariant, all the requirements for the return value
of the method are fulfilled.

Otherwise, it is known that either $s.\segId{}$ is less than \memLoc{id} or $s$
was removed. Two possibilities are then considered:
\begin{itemize}
    \item $s$ is not the current tail. Then, according to the invariant on
    \memLoc{next}, all the segments with $\segId{} \in (s.\segId{};
    s.\memLoc{next}.\segId{})$ are cancelled. To preserve the loop invariant,
    it is required to show that this fact implies that
    all the segments with $\segId{} \in [\max(\memLoc{start}.\segId{},
    \memLoc{id}); s.\memLoc{next}.\segId{})$ are cancelled.

    If $s.\segId{}$ is less than \memLoc{id}, then $s.\memLoc{next}.\segId{}
    \le \memLoc{id}$, so this is vacuously true. Otherwise, this loop iteration
    was performed because $s$ was removed. In this case, combining
    $[\max(\memLoc{start}.\segId{}, \memLoc{id}), s.\segId{}) \cup
    \{s.\segId{}\} \cup (s.\segId{}; s.\memLoc{next}.\segId{})$ provides the
    required interval of cancelled segments.
    \item $s$ is the current tail. In this case, a new segment must be appended.
    The only valid \segId{} for a segment to be appended is $s.\segId{} + 1$,
    according to the invariant on \memLoc{next}. An attempt is performed to
    append the
    new segment. Whether it succeeds or not, now it is known that $s$ is not
    anymore the current tail, as either \memLoc{next} already contained
    something or it does now as the result of a CAS. If the CAS succeeded, then,
    if the segment is logically removed, a call to \methName{remove()} is
    performed, which is valid. This is desirable since a call to
    \methName{remove()} that was performed after the segment became logically
    removed could have finished without performing any work, having observed
    that the segment was the tail.

    In any case, now that $s$ is not the current tail, the case reduces to the
    first one.
\end{itemize}

\paragraph{tryIncPointers()} This method atomically checks the current contents
of the fields \memLoc{pointers} and \memLoc{cancelled} in the current segment
and, if either of them is not in the terminal state, returns \literal{true} and increments
\memLoc{pointers}, providing the newly-created piece of the \memLoc{counter};
otherwise, it leaves everything unchanged and returns \literal{false} and
provides the knowledge that the segment is logically removed.

\paragraph{decPointers()} This method, when called with a piece of the
\methName{pointers} counter, consumes that piece, decrementing the counter and
returning \literal{true} if the segment became logically cancelled as a
result, or \literal{false} otherwise. This is fairly easy to observe from the
definitions. The tricky part is that the call does not violate the invariants,
but this follows from the fact that the existence of a piece of the counter
implies that the current value is nonzero.

\paragraph{moveForward(to)}
The $i$ that the segment pointer currently points to is read into \memLoc{cur}.
If its \segId{} is at least $\memLoc{to}.\segId{}$, then the moment of that
reading is the point where the method performs its atomic action, and
\literal{true} is returned.

Otherwise, an attempt is made to acquire a piece of the \methName{pointers}
counter of segment \memLoc{to}. If it fails, the segment must have been
cancelled, so the whole method returns \literal{false}. If it succeeds, this
means that the call obtains a piece of \methName{to.pointers}. \texttt{CAS} is
then attempted.

If the \texttt{CAS} fails, the piece of \methName{to.pointers} is used to
call \methName{to.decPointers()}, which may inform us that \methName{to} is
logically cancelled, in which case \methName{to.remove()} is called.

If the \texttt{CAS} succeeds, then the piece of \methName{to.pointers} is
transferred to the segment pointer so that it logically points to
$\memLoc{to}.\segId{}$. A piece of \methName{from.pointers} is acquired instead
and then used to call \methName{from.decPointers()}, which, likewise, can
show us that a call to \methName{from.remove()} could be valid.

\paragraph{\onSlotCleaned{}}
The difficult part of this proof was finding a suitable specification; the correctness of the execution follows from that.
Given a $\Phi$, in order to perform the \faa{}, this method atomically obtains
the right to increase \memLoc{cancelled} along with the knowledge that it was not
yet \segmentSize{}. Increasing the \memLoc{cancelled} yields a $\Psi$, which
is provided to the caller. Then, if the segment is logically removed,
\methName{remove()} is called.

\paragraph{Setting \memLoc{prev} to \literal{null}} No invariants are violated
by this: it is always valid for a \memLoc{prev} of any segment to contain
\literal{null}.

\subsection{Infinite Array Specification} 

The CQS does not actually need a true infinite array, which would use an unbounded
amount of memory. Instead, the data structure that is employed after all is able
to discard groups of cancelled cells and even lose access to cells from the
prefix of the ``array'' that are no longer needed. Thus, when the data structure
underlying the CQS is called an ``infinite array'', the term is used loosely and
for the lack of a more fitting name.

The code listings are defined (in order to avoid excessive abstractions
that would detract from the general idea) with the ad-hoc ``infinite array''
operations being interwoven with the logic of the CQS. However, it could just
as easily have been abstracted into a separate data structure by grouping
a segment and an index in that segment into an entity called a \defn{cell}:
for example, $(s, i)$ is the $i$'th cell in segment $s$, but its index in the
infinite array is $s.\segId \cdot \segmentSize + i$. This
is the approach taken in the Coq formalization to be able to prove the infinite
array operations independently from the CQS. The operations provided by the
ad-hoc infinite array are then the following\fileInRepo{concurrent_linked_list/infinite_array/array_interfaces.v}:

\begin{itemize}
    \item Array creation that allocates $n$ segment pointers (with $n = 2$ in
    case of the CQS, the pointers being \texttt{resumeSegm} and \texttt{suspendSegm})---performed by allocating the concurrent linked list.
    \item Reading a segment pointer.
    \item Moving a segment pointer forward---done with a call to
    \methName{moveForward(..)}.
    \item Finding a cell with the given \memLoc{id}---performed with a call to
    $s' = \methName{findSegment(\memLoc{s}, \memLoc{id} / \segmentSize{})}$ and
    later checking whether the requested segment identifier corresponds with the
    requested one and returning $(s', \memLoc{id} \bmod \segmentSize{})$ if so
    and $(s', 0)$ otherwise.
    \item Cancelling a cell---done with a call to \onSlotCleaned{} on
    the underlying segment;
    \item Setting \memLoc{prev} of the underlying segment of a cell to
    \literal{null};
    \item Checking the index of a cell---equal to $s.\memLoc{id} \cdot \segmentSize{} +
    i$, where the cell is $(s, i)$;
    \item Accessing the contents of a cell.
\end{itemize}

The code listings can easily be rewritten in terms of these operations (which would lead to \segmentSize{} not ever being mentioned outside of the infinite array abstraction) that, when inlined, would result in the code that is present
currently. The specifications of the listed operations mirror those of the
operations that they are wrapping.

Notable additions that the infinite array performs in the logical realm when
compared to the general concurrent linked list are the following\fileInRepo{concurrent_linked_list/infinite_array/array_spec.v}:
\begin{itemize}
    \item Introduction of a \defn{cancellation handle for the $i$'th cell}. This
    logical resource is provided for each cell and is the $\Phi$ that is passed
    to \onSlotCleaned{} when called on an infinite array segment. In accordance
    to the behavior required from $\Phi$, the existence of the cancellation handle for the cell $i$ from a segment $s$ prevents \memLoc{cancelled} from being equal to \segmentSize{}, and thus implies that $s$ is not yet
    logically removed.

    There can only exist a single cell cancellation handle for any given cell.
    \item The $\Psi$ acquired from \onSlotCleaned{} is another
    logical resource, the knowledge that the $i$'th cell was cancelled. This
    knowledge is mutually incompatible with the existence of the cancellation
    handle for the $i$'th cell.

    This resource is freely duplicable.
    \item A logical operation of accessing an infinite array cell for the first
    time is introduced. This operation provides each caller either with both the
    exclusive right for writing to the cell and the cell cancellation handle,
    or with the evidence that the caller was not in fact the first one to attempt
    this operation; this evidence is called the knowledge that the
    \defn{cell is owned}. The user of the infinite array themselves decide what
    logical resource is used to signify that the cell is owned.
\end{itemize}

The implementations of these logical entities\fileInRepo{concurrent_linked_list/infinite_array/array_proof.v} are purely technical and do not
provide the reader with a deeper understanding of the subject matter, so their
existence is simply postulated in this text.

\subsection{Infinite Array Iterators}
An additional construct is built on top of the infinite
arrays---\defn{iterators}. An iterator is a pair of a segment pointer and a
counter; for example, \memLoc{resumeSegm} and \memLoc{resumeIdx} form an
iterator, as do \memLoc{suspendSegm} and \memLoc{suspendIdx}.

Iterators provide a single operation \methName{step()}, an example of which is
shown in \ref{listing:array_iterator_step}. In the actual code, the operation is
inlined and simplified, but due to how crucial it is to the proof, it is
considered separately.

\begin{figureAsListing}
    \begin{lstlisting}
fun stepResume(): (Bool, Cell) {
  r := resumeSegm
  i := FAA(&resumeIdx, 1)
  s := findAndMoveForwardResume(r, i/SEGM_SIZE)
  if s.id == i/SEGM_SIZE: return (true, Cell(s, i))
  else: return (false, Cell(s, 0))
}
    \end{lstlisting}
    \caption{An implementation of an array step for the dequeue iterator.
    This is not used in the actual
    code and is only introduced as an abstraction to separate the parts of the formal proof.}
    \label{listing:array_iterator_step}
\end{figureAsListing}

\paragraph{Specification} (not in a separate file due to being seemingly non-generalizable)
Each iterator is parameterized by some logical resource $R$; we say that an
iterator is \defn{bounded by} $R$.

Each iterator introduces a logical resource that is parameterized by a
nonnegative number $i$. We call this resource the \defn{$i$'th permit from the
iterator}. This permit has two important properties:
\begin{itemize}
    \item It is exclusive: at most one permit exists for any given iterator for
    any number $i$.
    \item It implies that the iterator's counter is at least $i + 1$.
\end{itemize}

The \methName{step()} operation acquires an instance of an $R$ and returns one
of two possible results:
\begin{itemize}
    \item A pair of \literal{true} and a cell $c$ with index $i$. In this case,
    the $i$'th permit from this iterator is provided.
    \item A pair of \literal{false} and a cell $c$ with index $j$. In this case,
    the $i$'th permit from this iterator is provided for some $i < j$, and
    all the cells in $[i; j)$ are known to be cancelled.
\end{itemize}

Last, a logical operation of accessing the bounding resource is available. This
operation allows the caller to observe that there are at least $i + 1$ copies
of $R$ stored in the iterator if there exists the $i$'th permit. The correctness
of this operation follows directly from the invariants that follow.

\paragraph{Invariants}
The state of the invariant is described by the number $n$ currently stored in
its counter. Then the following is true:
\begin{itemize}
    \item The iterator stores $n$ copies of $R$.
    \item The segment pointer \defn{points to} (in the sense of containing the
    reference to a segment and owning a piece of its \memLoc{pointers} counter)
    some segment $i$ such that all the cells in $[n; i \cdot \segmentSize{})$
    are cancelled. Note that it is possible for $i \cdot \segmentSize{}$ to be
    $n$ or less, in which case no knowledge about cancelled cells is present.
    \item There exists the $i$'th permit for all $i \in [0; n)$.
\end{itemize}

Additionally, $n$ is nondecreasing with respect to time, that is, the iterator
can only go forward.

\paragraph{Proof of the Step Operation.} \fileInRepo{concurrent_linked_list/infinite_array/iterator/iterator_impl.v}
First, the segment pointer is read. The counter contained some number $n$ at
the moment, and, according to the invariant, the segment pointer contained a
reference to some segment $r$ such that all cells in $[n; r.\segId{} \cdot
\segmentSize{})$ are cancelled.

Next, the counter is incremented. If at that moment it contained $i$, then the
$i$'th permit from the iterator is created; it is also known that $i \ge n$,
due to monotonicity of the iterator.

Then \methName{findAndMoveForward(r, i/\segmentSize{})} is executed. According
to its specification, it returns some segment $s$ such that
$s.\segId{} \ge r.\segId{}$, $s.\segId{} \ge i / \segmentSize{}$, and all the
segments in $[\max(r.\segId{}, i / \segmentSize{}); s.\segId{})$ are cancelled.

If $s.\segId{}$ is $i / \segmentSize{}$, then \literal{true} is returned, and
the cell is formed correctly, obeying the provided specification. Otherwise,
\literal{false} is returned, along with a cell with index
$s.\segId{} \cdot \segmentSize{}$. The specification of \texttt{step()} requires that we show then
that $i < s.\segId{} \cdot \segmentSize{}$ (which easily follows from
$i / \segmentSize{} < s.\segId{}$) and that
all cells in $[i; s.\segId{} \cdot \segmentSize{})$ are cancelled, which
requires further elaboration.

We consider two possibilities in order to prove it.
\begin{itemize}
    \item $i / \segmentSize{} \ge r.\segId{}$. Then we know from the information
    provided by \methName{findAndMoveForward(..)} that all the segments in
    $[i / \segmentSize{}; s.\segId{})$ are logically removed, which means that
    all the cell in $[i; s.\segId{} \cdot \segmentSize{})$ are cancelled, which
    is what we needed to prove.
    \item $r.\segId{} > i / \segmentSize{}$. Then
    \methName{findAndMoveForward(..)} guarantees that segments in $[r.\segId{};
    s.\segId{})$ are logically removed, so cells in
    $[r.\segId{} \cdot \segmentSize{}; s.\segId{} \cdot \segmentSize{})$ are
    cancelled.
    It is also known that all cells in $[n; r.\segId{} \cdot \segmentSize{})$
    are cancelled and that $n \le i$; thus, all cells in $[i; r.\segId{} \cdot
    \segmentSize{})$ are cancelled. By combining the two intervals, we obtain
    the desired proposition.
\end{itemize}

\subsection{Proving Correctness of the CQS}
\label{sec:cqs-proof}

This is by far the most involved part of the proof, taking about a third of the lines of the proof code on its own\fileInRepo{thread_queue/thread_queue.v}.

\subsubsection{Specification} The general idea is that a CQS with the enqueue
resource $E$ and the
dequeue resource $R$ is a data structure that allows callers of
\suspendOp{} that provide an instance of $E$ to register a Future that, upon
being completed, receives an instance of $R$, and callers of \resumeOp{} that
provide an $R$ to receive an $E$ upon completion. In the mutex example, $E$ is
the empty resource {---} no special permissions are required to call
\methName{lock()} {---} and $R$ is the permission to enter the critical section.

More accurately, both \resumeOp{} and \suspendOp{} are split into a logical and a physical stage. Actually calling \resumeOp{} requires providing not an $R$ but an
\defn{awakening permit}, a logical resource introduced by this data structure;
it is the awakening permit that can be acquired in exchange for $R$, which is done on the logical level. Likewise,
a call to \suspendOp{} requires a \defn{suspension permit} that can be acquired
beforehand in exchange for $E$.

When correctly called with a suspension permit, \suspendOp{} returns an $R$-passing Future, but also provides the means of cancelling that Future: a
complete (but ineffectual) cancellation permit if the Future was constructed
with \methName{ImmediateResult}, and a half of its cancellation permit
otherwise. The other half is owned by the CQS, but is always
available, so owning a half of the cancellation permit is sufficient for
cancelling the Future. The complication that requires the protocol to require the
canceller to access the invariants of the CQS to cancel the Future is due to the
need for the CQS to reliably observe the effective state of each cell, which is
affected by cancellation.

If the resumption is synchronous, both \resumeOp{} and the
corresponding \suspendOp{} may fail due to the cell being broken by
the resumer; additionally, \resumeOp{} may fail if the cell was
``simply'' cancelled. In both failure modes for \resumeOp{}, instead
of receiving an instance of $E$ upon its completion, the caller instead gets an
$R$, allowing the operation to be restarted. Likewise, when \suspendOp{}
fails, it provides an instance of $E$ and doesn't return a Future.

The state of the CQS is represented as the number of threads that are
currently enqueued or are willing to be. We call this number the \emph{size} of
the CQS; a more specific definition will be provided later.

Users of the CQS are responsible for representing the size of the queue
as physical values. In the mutex example, the size is 0 if \field{state} is
nonnegative and $-\field{state}$ otherwise. The CQS supports the
following operations on its logical state:
\begin{description}
    \item[Enqueue registration] It is possible to provide an $E$, increasing the
    size and obtaining in exchange a suspension permit {---} a permission to call \suspendOp{}.
    \item[Dequeue registration] If it is known that the size is nonzero, it is
    possible to decrease the size and provide an $R$, obtaining in exchange an awakening permit {---} a permission to call \resumeOp{}.
    \item[Cancellation registration] Used in the ``smart'' cancellation
    mode and implemented for use in \methName{onCancellation()}, this
    operation can only be invoked
    once for each cell; this is ensured by introducing yet another set of logical resources, which don't influence the general idea and will be described later. If the CQS size at the time of this operation
    is $0$, then cancellation registration does not change the size, provides the caller with an
    instance of $R$ and logically performs the transition of the state of the
    cell to \stateName{REFUSED} (without physically writing anything to the
    cell). Otherwise, the operation decreases the size of the CQS,
    provides the caller with the cancellation permission for the cell in the
    infinite array, and performs the transition of the cell state to
    \stateName{CANCELLED}.
\end{description}

\subsubsection{Per-queue Invariants} Two logical values are maintained: the
\defn{dequeue front}, and the \defn{cell state list}. The \defn{deque front}
is the first cell about which it is not yet known that its state is going to
be observed by \resumeOp{} or already was. A cell is
\defn{inside the deque front} if its index is less than the deque front.
\defn{Cell state list} stores
the authoritative knowledge of the state of each \defn{relevant} cell in the
infinite array; a cell is considered relevant if it is known that its state is
going to be observed by a call to \suspendOp{} or already was.

The last cell inside the deque front must be a relevant cell, which effectively
means that we forbid calling \resumeOp{} unless it is known that the
corresponding \suspendOp{} is also eventually going to be called.

It is maintained that the last cell before the deque front can not be \defn{skippable}.
A cell is skippable if it was ``smartly'' cancelled and
\methName{onCancellation()} returned \literal{true}. Without this invariant,
the definition of the deque front would be violated: if it is known that the
last cell inside the deque front is skippable, it is going to be observed by a call to \resumeOp{}, but is going to be skipped (hence
the name), and thus some of the following cells are also known to be observed by
\resumeOp{} at a later time, thus fitting in the deque front, which contradicts the skippable cell being the last such cell.

For each \defn{relevant} cell, there exists a single instance of a logical
resource called the \defn{suspension permit}. For each cell before the deque
front, there exists a single instance of a logical resource called the
\defn{awakening permit}.

The deque front is nondecreasing, which is in line with its definition: once it
is known that a cell is about to be witnessed by a call to \resumeOp{},
this can not become false. Likewise, the length of the cell state list can not
decrease with time.

Physically, a CQS consists of an infinite array and two
iterators, one, the \defn{enqueue iterator}, bounded by the suspension permits,
and another, the \defn{dequeue iterator}, bounded by awakening permits.

Finally, the \emph{size} of a CQS is defined to be the number of
nonskippable relevant cells outside the deque front. This is the value in terms
of which the programmer-facing specifications are defined.

\subsubsection{Cell States}
The descriptions of cell states contained in the cell state list closely mirror
the state transition systems presented in Figure~\ref{fig:cell_simple_cancellation} and
Figure~\ref{fig:cell_smart_cancellation}, as in most cases the logical state of the cell is adequately described by its contents. Some superficial differences are that here, a cell stores a \texttt{Future} instead of a thread, and a \texttt{Future} filled with a value instead of just the value.

A change that is actually significant is that simple cancellation and smart cancellation are not
handled separately
    due to the amount of shared parts between them. Instead, a single state
    transition system is implemented, where there are two possible transitions
    from \futureState{WAITING} to a cancelled state, one for each
    cancellation mode. However, for each use of \sqs{} it must be true that
    either all the cancelled cells are ``smartly'' cancelled or all the cells
    are ``simply'' cancelled. Otherwise, it would be impossible for the
    \resumeOp{} operations to determine whether a cell in a removed
    segment is responsible for finalizing its own cancellation (as is done in
    ``smart'' cancellation).

A much more significant change is that
the transition system described in Figure~\ref{fig:cell_simple_cancellation}
    and Figure~\ref{fig:cell_smart_cancellation}
    is presented in terms of values that the cell contains at any given
    moment; however, the state transition in terms of observable behavior
    sometimes happens before anything is written to the cell or does not occur
    at all.
    For example, it does not matter for correctness whether a resumed Future or
    a \marker{RESUMED} is in the cell, the observable behavior is exactly the
    same in either case, so the two states are merged into one. Contrastingly,
    there are many more additional states required for describing smart
    cancellation due to the need to always distinguish cancelled cells where
    \methName{onCancellation()} returned \literal{true}, as such cells, being
    skippable, are important for defining the current state of the CQS.
    When a cell is smartly cancelled, it is initially \stateName{UNDECIDED}
    and stores the Future; ``undecidedness'' here is in regards to whether the
    cancellation registration will succeed. If cancellation registration fails,
    the cell is \stateName{REFUSED} even before the corresponding marker is
    written to it; otherwise, a race can happen between writing
    \marker{CANCELLED} and a resumer passing a value in the asynchronous mode.
    \stateName{SMARTLY-CANCELLED} is the state when the cancellation
    registration has already succeeded, and \stateName{CLOSED} is the state
    when \marker{CANCELLED} was written to the cell before the resumer managed
    to write its value.
    This race can be decided even before the cancellation registration attempt,
    if the resumer passes its value when the cell is still in the
    \stateName{UNDECIDED} state.

\begin{figure*}[!htbp]
    \centering
    \includegraphics[width=0.8\textwidth]{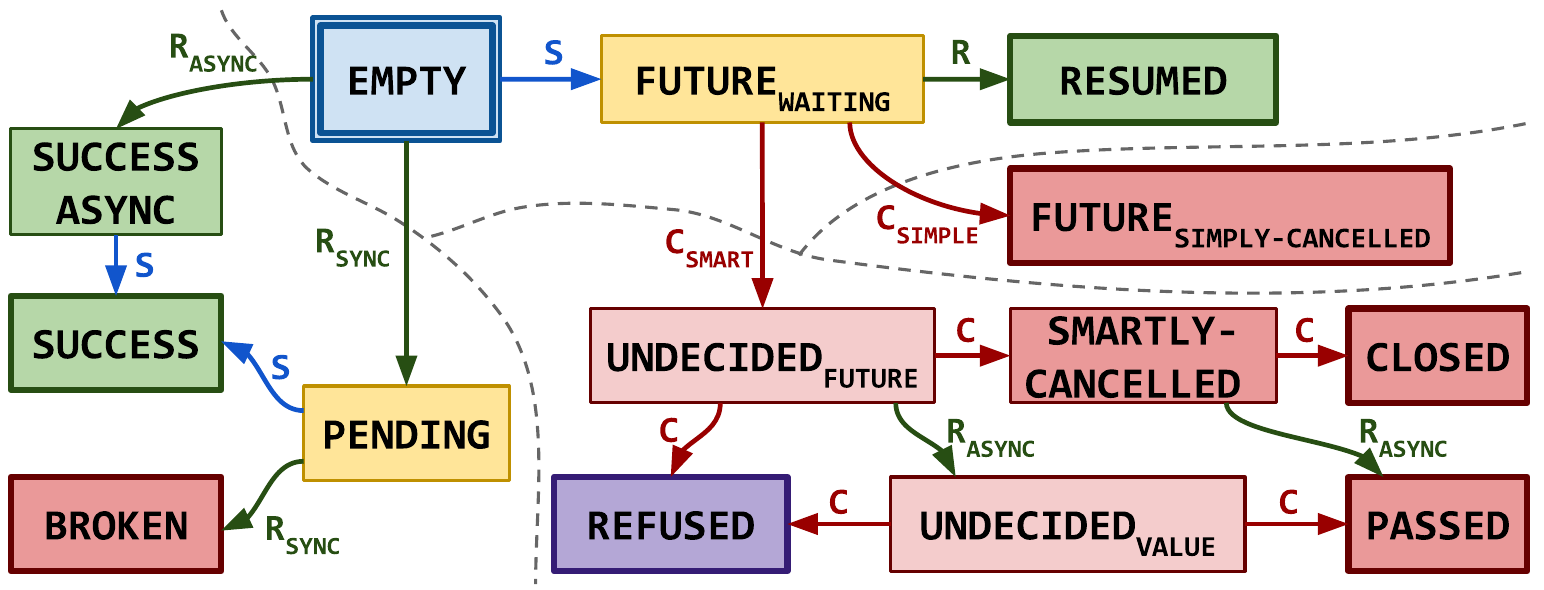}
    \caption{The state transition system for a single cell from the logical
    perspective.}
    \label{fig:proof_cell_full}
\end{figure*}

\subsubsection{Per-cell Tokens}
Some of the logical resources only exist when a cell is in a particular state;
each of the resources listed here is exclusive and parameterized by the cell
for which it is defined.
\begin{itemize}
    \item The \defn{cell breaking token} represents the exclusive right of the
    execution of \resumeOp{} to break the cell in case of elimination. This
    token exists in every state from the ``elimination'' execution path except
    for \stateName{BROKEN}, namely \stateName{SUCCESS-ASYNC},
    \stateName{SUCCESS}, and \stateName{PENDING}. This token is generated on a
    successful write of a value to an empty cell and is only destroyed if the
    cell is broken.
    \item The \defn{cancellation registration token} is generated when a Future
    stored in the CQS is successfully cancelled. It represents the exclusive
    right to call \methName{onCancellation()}, which, in term, must use it to
    attempt the logical operation of cancellation registration, which destroys the cancellation
    registration token and instead creates a
    cell cancelling token.
    \item The \defn{cell cancelling token}, provided by the cancellation
    registration, represents the right to write \marker{CANCELLED} or
    \marker{REFUSED} to the cell.
\end{itemize}

\subsubsection{Requirements for User-Defined Operations}
For the CQS to operate correctly, it is crucial that the operations
provided by the user behave in accordance with the established invariants.

\begin{description}
    \item[\methName{completeRefusedResume(v)}] must define some resource $X$ such that \methName{completeRefusedResume(v)} returns $E$ on completion when called with it.
    \item[\methName{onCancellation()}] This method is given a cancellation
    registration token and must invoke the operation of cancellation
    registration. If the registration is successful, it must return
    \literal{true}; otherwise, it must return \literal{false} and ensure
    the existence of an instance of $X$. The operation of cancellation
    registration also generates a cell cancelling token and, in case of a
    successful cancellation, provides a cancellation handle; these resources
    must be passed along, which can not be reflected in the code but
    is required in the formal specification.
\end{description}

\subsubsection{Per-cell Invariants and Transitions}
Each \defn{relevant} cell $i$ is said to own various collections of logical
resources depending on its state. This description heavily relies on Figure~\ref{fig:proof_cell_full}; so the reader is advised to consult with it periodically. In particular, the fact that some transitions are impossible (for example, the states in the elimination path are inaccessible once a Future is written to the cell) is only reflected in that figure. In the actual Coq proof, this, too, had to be specifically encoded using logical resources; we omit these details here, as they would greatly increase the size of the proof but would not provide additional clarity.

\begin{itemize}
    \item \textbf{\stateName{EMPTY}.} By definition of a \defn{relevant} cell, $E$ was
    passed to it
    during the enqueue registration, and no state transitions happened yet to
    extract that resource, so the cell owns a $E$. Additionally, it is possible
    that the cell is already inside the dequeue front, in which case the cell
    also owns an $R$.
    \item \textbf{\resumeOp{} arrived first.} If the call to \resumeOp{} was the first to
    modify the cell, we have the following:
    \begin{itemize}
        \item The cell owns the cancellation permit for the infinite array
        cell, meaning that the cell is never going to be cancelled;
        \item The cell owns the $i$'th permit from the dequeue iterator.
    \end{itemize}
    There are some additional resources that the cell owns, depending on its
    state:
    \begin{itemize}
        \item \textbf{\stateName{PENDING}.} The cell contains some value $v$, as well as
        both $E$ and $R$. This state can only be entered when in the synchronous
        resumption mode.

        Transition from \stateName{EMPTY} to \stateName{PENDING} happens when
        the resumer writes its value to the cell, providing the $i$'th permit
        from the dequeue iterator and receiving the cell breaking token. The
        transition happens as follows: given that the $i$'th permit from the
        dequeue iterator exists,
        $i$ is clearly inside the deque front, so initially the cell owned $E$
        and $R$. When the resumer first interacted with the cell, its
        cancellation handler was initialized. Therefore, we have all the
        resources that the cell needs to own. Additionally, the transition to
        \stateName{PENDING} creates the cell breaking token, which is taken by
        the resumer.
        \item \textbf{\stateName{BROKEN}.} The cell contains the \marker{BROKEN} marker.
        It also owns either the $i$'th permit from the enqueue iterator or an
        $E$. This state can only be entered when in the synchronous resumption
        mode.

        Transition from \stateName{PENDING} to \stateName{BROKEN} happens when
        the resumer stops waiting for the corresponding call to \suspendOp{} and
        writes \marker{BROKEN} to the cell, using up its cell breaking permit
        and getting back an $R$ in return. The structure of this transition is
        trivial.

        When \suspendOp{} observes \marker{BROKEN} in the cell, it gives up the
        $i$'th permit from the enqueue iterator in exchange for the $E$,
        performing no state transitions in the process.
        \item \textbf{\stateName{SUCCESS-ASYNC}.} The cell contains some value $v$, owns
            an $R$ and this cell's breaking token.

            Transition from \stateName{EMPTY} to this state happens in the
            asynchronous resumption mode when the resumer writes its value to
            the cell, relinquishing the $i$'th permit from the dequeue iterator
            in exchange for $E$. The cell was inside the deque front, so it
            owned an $R$; also, the transition creates the cell breaking token,
            which is not given to the resumer but is instead kept in the cell's
            ownership.
        \item \textbf{\stateName{SUCCESS}.} The cell contains the \marker{TAKEN} marker
            and owns the
            $i$'th permit from the enqueue iterator, as well as either $E$ or
            this cell's breaking token.

            Transition from \stateName{SUCCESS-ASYNC} happens when \suspendOp{}
            writes \marker{TAKEN}, having observed that the cell contains a
            value, providing its $i$'th permit from the enqueue iterator in
            exchange for $R$.

            Tranisition from \stateName{PENDING} happens under the same
            conditions.

            If the resumption is synchronous, the resumer can observe the state
            transition and give up its cell breaking token in exchange for $E$.
    \end{itemize}

    \item \textbf{\suspendOp{} arrived first.} If the call to \suspendOp{} was the first to
    modify the cell, writing a Future to it, the following is true:
    \begin{itemize}
        \item The cell owns the $i$'th permit from the enqueue iterator.
        \item There exists a unique $R$-passing Future $f$ associated with the
        cell.
    \end{itemize}

    To describe the resources the cell owns at various states, we first need to
    introduce a helpful ownership pattern. We say that $T$ is owned as a
    \defn{resource for resumer} if one of the following is true:
    \begin{itemize}
        \item The Future completion permit for $f$ and $T$ is owned;
        \item Half the Future completion permit, the $i$'th permit from the
        dequeue iterator, and $T$ is owned;
        \item The Future completion permit for $f$ and the $i$'th permit from
        the dequeue iterator is owned.
    \end{itemize}
    The logic behind this notion is that it would be unfeasible to use the state
    transition system to carefully track the progress that \resumeOp{} has made,
    as that would greatly multiply the number of states. Instead, in the
    interesting case, the resumer goes through the following stages when working with
    the cell, during which the cancellation process could advance the state
    several times:
    \begin{itemize}
        \item It owns the $i$'th permit from the dequeue iterator (and the
        system owns the Future completion permit and some $T$). After
        observing that the cell contains a Future, the resumer trades the dequeue iterator permit
        from the iterator for a half of the Future completion permit (it can not
        take the whole permit, as then the cell state would not be able to
        uniquely identify the corresponding state of the Future).
        \item The resumer tries to complete the Future. If it succeeds, then
        a state transition occurs. Otherwise, the completion attempt does not
        have an effect. In this case, the resumer trades back its half of the
        completion permit without performing a state transition and either
        receives the $i$'th permit from the dequeue iterator back or gets a $T$.
    \end{itemize}
    It should be noted that, with this ownership scheme, it is always possible
    for the resumer to determine exactly what the cell owns; this is due to the
    exclusivity of the iterator permits and Future completion permits.

    We also say that \emph{resumer has finished} if the cell owns a combination
    of the $i$'th permit from the deque iterator and the completion permit for
    $f$. The intuition behind this notion is that in the part of the state
    transition system where a Future was written to the cell it is impossible for
    the resumer to perform any operation without owning some of these two
    resources, so if they belong to the cell, the resumer can no longer perform
    meaningful operations. Due to the invariants of the system, it is only
    possible for the resumer to have finished if the cell is inside the deque
    front.

    \begin{itemize}
        \item \textbf{\futureState{WAITING}.} The cell contains $f$. It also owns the
        cancellation handle of this cell, as well as $E$. If the cell is inside
        the deque front, it owns $R$. It owns a half of a Future cancellation
        permit, and also owns unit (that is, an always-true statement that
        bears no information) as a resource for resumer.

        The transition from \stateName{EMPTY} to \futureState{WAITING} happens
        when \suspendOp{} writes a Future $f$ to the cell. It provides to the cell
        the $i$'th permit from the dequeue iterator, the full completion permit
        and half of the cancellation permit for $f$; in return, it receives the
        knowledge that $f$ is now part of the CQS. The non-obvious
        parts of the transition is that the cell owns the cancellation handle
        and owns the unit as a resource for resumer. The cancellation handle is
        allocated with the first access to the cell; the unit resource for
        resumer has the form where the cell owns the full completion permit and
        $T$, with the completion permit being provided by \suspendOp{} at the
        start of the transition, and $T$, being the unit, always trivially
        available.
        \item \textbf{\stateName{RESUMED}.} The cell contains either \marker{RESUMED} or
        $f$. It owns the $i$'th permit from the dequeue iterator, knowledge that
        the Future was completed, the cancellation handle, and half of a
        cancellation permit.

        The transition from \futureState{WAITING} to \stateName{RESUMED} happens
        when \resumeOp{} successfully completes $f$. For this to happen, it
        must have given up its $i$'th permit from the dequeue iterator in
        exchange for the half of a completion permit when the cell was still
        \stateName{WAITING} (utilizing the ownership of unit as a resource for
        resumer). Thus, during the transition, \resumeOp{} relinquishes its half
        of the completion permit in exchange for $E$. At the start, the cell
        owns a half of the completion permit, thus, the full completion permit
        is available for this operation and it is possible to safely try to
        complete the Future. However, in order to do that, it is also needed to
        provide an $R$. To obtain it, we observe that the interaction of
        \resumeOp{} with the cell implies that it is inside the deque front, so
        $R$ is owned by the cell at the start of the operation.
        \item \textbf{\futureState{SIMPLY-CANCELLED}.} The cell contains
        \marker{CANCELLED} or $f$ and owns the knowledge that the Future is
        cancelled. It also owns as a resource for resumer an $R$ if the cell is
        inside the deque front or unit if it is not. This state is only available
        with the simple cancellation mode.

        Transition to \futureState{SIMPLY-CANCELLED} from \futureState{WAITING}
        happens when the Future is successfully cancelled. The half of the
        Future cancellation permit owned by the cell and the half in possession
        of the end user of the CQS are combined to perform the
        cancellation, which, upon success, provides the knowledge that the cell
        was cancelled. The cancellation handler that is owned by the cell at the
        start of the operation is given to the canceller so that it is able to call \onSlotCleaned{}.
        \item \textbf{The Future is smartly cancelled.} The cell owns the knowledge that
        the Future was cancelled.
        \begin{itemize}
            \item \textbf{\stateName{UNDECIDED}.} The cell owns the cancellation handle
            and, if it is inside the deque front, an $R$. The cell may contain
            one of two things:
            \begin{enumerate}
                \item $f$. In this case, the cell owns $E$ and also owns the
                unit as a resource for resumer.

                Transition from \futureState{WAITING} to this happens when the
                Future is cancelled. The cancelling operation gives up its
                half of the cancellation permit which, combined with the half
                that was stored in the cell, allow cancelling the Future. In
                exchange, the cancellation registration token is created and
                given to the cancelling operation.
                \item The cell can contain some value $v$. In this case we know
                that the resumer has finished.

                There is no transition from \futureState{WAITING} directly to
                this form; instead, this is what happens in the ``smart async''
                cancellation mode when the state is already
                \stateName{UNDECIDED}, so the Future is cancelled, the resumer
                unsuccessfully attempts to resume it and writes its value to
                the cell, relinquishing its permits (and finishing) in exchange for the $E$ that was owned as a
                resource for resumer.
            \end{enumerate}
            \item \textbf{\stateName{REFUSED}.} The cell owns the cancellation handle and
            is known to be inside the deque front. There are three possibilities
            for what the cell can contain:
            \begin{enumerate}
                \item $f$. Similarly to the case for \stateName{UNDECIDED},
                the cell owns $E$ and owns the unit as a resumer resource.

                Transition from the first form of \stateName{UNDECIDED} happens
                when cancellation registration fails due to the CQS
                being logically empty. The cancelling operation receives the
                cell cancelling token and, since the registration fails only
                when the cell is inside the deque front, an $R$ in exchange for
                its cancellation registration token needed to invoke
                \methName{onCancellation()}.

                \item Some value $v$. Then the resumer has finished.

                Transition from the second form of \stateName{UNDECIDED} happens
                when cancellation registration fails due to the CQS
                being logically empty. Transition from the first form of
                \stateName{REFUSED} happens similarly to the transition from the
                first to the second form of \stateName{UNDECIDED}.

                \item \marker{REFUSED}. Then the cell owns the cancelling token
                and owns as a resource for resumer such a resource $X$ provided
                by \methName{onCancellation()} that
                \methName{completeRefusedResume(v)} is guaranteed to provide an $E$
                if given an $X$.

                Transition to this from the first form happens when the
                cancelling operation writes \marker{REFUSED} to the cell, giving
                up its cell cancelling token and the instance of $X$ received
                from \methName{onCancellation()} and thus finishing the
                cancellation process.

                Transition from the second form to this happens when writing
                \marker{REFUSED} and giving up its cell cancelling token, but
                keeping the instance of $X$ to be able to call
                \methName{completeRefusedResume(v)} from the cancellation handler.
                Keeping the $X$ is possible because the resumer is known to
                have finished, so owning $X$ as a resource for resumer has its
                final form where $X$ is not actually owned by the cell.
            \end{enumerate}
            \item \textbf{Cancellation was allowed.} When cancellation registration
            succeeds, a state from this group is entered; all of these states are skippable. For each of them, the awakening permit is mentioned, which raises a question of where would the awakening permit come from. The answer is that providing cells with
            awakening permits is the responsibility of either cancellation
            registration or dequeue registration.
            
            Recall that an awakening permit exists for each cell
            inside the deque front; observe also that if a skippable cell is
            inside the deque front, then it is known that it will be skipped by
            a resumer, so the deque front is increased until it encounters a
            nonskippable cell (which must exist, since both cancellation
            registration and dequeue registration only succeed for non-empty queues), possibly skipping
            many cancelled cells. Thus, for each skippable cell
            inside the deque front a new awakening permit is generated; this is
            the permit that is given to the cell as the result of the operation.

            \begin{description}
                \item[\stateName{SMARTLY-CANCELLED}] The cell stores $f$ and
                owns $E$. It also owns as a resource for resumer the resource
                that is the awakening permit if the cell is inside the deque
                front and the unit otherwise.

                The transition from the first form of \stateName{UNDECIDED}
                happens as follows. The cancellation handle that was owned by
                the cell is given to the cancelling operation.
                If this cell was inside the deque front and owned an $R$, then
                that $R$ is moved to the new end of the deque front, which must
                have been in the \stateName{EMPTY} or \futureState{WAITING}
                state, given that the resumer could not have yet interated with
                that cell and the end of the deque front can not be a cancelled
                cell. The awakening permit is then generated and provided to
                the cell as described above. If the cell was outside the deque
                front, then the only change to the resources owned by the cell
                is the cancellation handle provided to the cancelling operation.

                \item[\stateName{PASSED}] It is known that the resumer has
                finished. Additionally, the cell may store some value $v$ and
                own the awakening permit, or it may store \marker{CANCELLED} and
                own the cancelling token.

                The transition from the second form of \stateName{UNDECIDED}
                happens in almost the same way as the transition to
                \stateName{SMARTLY-CANCELLED}, with the notable difference being
                that since it is known that the resumer has
                finished, the cell is inside the deque front, so a new awakening
                permit was generated for it.

                A transition can also happen from \stateName{SMARTLY-CANCELLED}
                when the resumer finishes, getting an $E$ in return.

                Finally, when the cancelling operation writes
                \marker{CANCELLED}, it receives the awakening permit in exchange
                for its cancelling token.
                \item[\stateName{CLOSED}] The cell stores \marker{CANCELLED},
                owns the cancelling token, and owns as a resource for resumer
                a resource that is the awakening permit if the cell is inside
                the deque front and the unit otherwise.

                The transition from \stateName{SMARTLY-CANCELLED} here happens
                when the cancelling operation successfully replaces $f$ with
                \marker{CANCELLED}, providing its cancelling token in exchange
                for $E$.
            \end{description}
        \end{itemize}
    \end{itemize}
\end{itemize}

\subsubsection{Proofs of Logical Operations}

\paragraph{Enqueue Registration} This is the simplest operation to prove. Given
an $E$, we say that the first cell that we did not consider relevant before is
now \stateName{EMPTY}. $R$ does not need to be provided because the cell can not be
inside the deque front, as, by an invariant, before the registration started,
the last cell inside the deque front was relevant, so it could not have been the
cell in question, and enqueue registration does not advance the deque front.

That the size of the CQS is increased is obvious from the definition.

A suspension permit exists for each relevant cell, so it is valid to allocate
one.

\paragraph{Dequeue Registration} By definition of the CQS size, if it
is nonzero, then there exists a nonskippable relevant cell outside the deque
front. We say that the first such cell has an index of $d + i$, where $d$ is
the current deque front and $i \ge 0$. Then $i + 1$ more cells than before are
going to be observed by a call to \resumeOp{}: in order to access the cell
$d + i$, calls to \resumeOp{} will have to observe the $i$ skippable cells in
addition to the new nonskippable one. Therefore, the deque front is increased
by $i + 1$, and $i + 1$ awakening permits are generated.

We know that the $i$
skippable cells were not inside the deque front at the start of this operation;
observe also that a cell being skippable means that its state is, by definition,
from a group of states where the cancellation succeeded and can not be
\stateName{PASSED} (as a cell can only be in this state if it is inside the
deque front); therefore, each of these cells is either
\stateName{SMARTLY-CANCELLED} or \stateName{CLOSED} and owns the unit as a
resource for the resumer. Now that the deque front is advanced, these cells
must instead own an awakening permit as the resource for the resumer. Therefore,
$i$ of the $i+1$ allocated awakening permits are passed to individual skippable
cells to satisfy their invariants, and the last awakening permit is
provided to the caller.

That the size of the CQS is decremented follows directly from the
definition.

\paragraph{Cancellation Registration}
If the queue was empty, this means that every cell that is not yet cancelled is
inside the deque front, including the one we attept to cancel currently. Thus, a
transition is performed from \stateName{UNDECIDED} to \stateName{REFUSED},
providing an $R$ and a cell cancelling token.

Otherwise, the queue was not empty.
A transition is performed from \stateName{UNDECIDED} to
either \stateName{SMARTLY-CANCELLED} or \stateName{PASSED}, depending on
whether a value was already written to the cell, providing a cancellation
handle. If the cell was outside the deque front, this change is sufficient,
as the CQS size is obviously decremented due to this cell becoming
skippable. Otherwise, the transition additionally requires an awakening
permit and provides an $R$. This $R$ is then used to perform a deque
registration. The awakening permit that is obtained in such a way is used
to complete the transition.

\subsubsection{Execution}

\paragraph{\resumeOp{}} To reiterate what was said in discussion of the
specification of this method, it requires passing an awakening permit to it and
either finishes with \literal{true} and provides an $E$, or it finishes with
\literal{false} and provides an $R$, which is only possible in the synchronous
resumption mode or smart cancellation mode.

First, a single step of the dequeue iterator is performed. If it is successful,
the $i$'th cell is obtained along with the $i$'th permit from the dequeue
iterator. The correctness of the execution then follows from the described
transitions of the cell state. Otherwise, the step is unsuccessful, and
the $j$'th cell is obtained with the $i$'th permit from the dequeue iterator,
where cells in $[i; j)$ are known to be cancelled.

If the cancellation mode is
``simple'', then the $i$'th cell must be \futureState{SIMPLY-CANCELLED}. Given
that the $i$'th permit from the dequeue iterator exists, the $i$'th cell is
inside the deque front, and so stores $R$ as a resource for the resumer. The
resumer then obtains $R$ and finishes the call to \resumeOp{} with
\literal{false}.

Otherwise, the cancellation is ``smart'', in which case a CAS is performed to
set the counter of the dequeue iterator to be at least $j$. If the CAS fails,
it has no effect and so does not affect the correctness. If it succeeds, then
it is known that the counter in the dequeue iterator must still contain $i + 1$.
For this increase of the counter from $i + 1$ to $j$
to be valid, the invariant of iterators requires that $j - i - 1$ copies of the
awakening permit be provided. However, this is always possible. Observe that the
permits $[i + 1; j)$ from the dequeue iterator do not yet exist. Therefore,
the cells must be either \stateName{SMARTLY-CANCELLED} or \stateName{CLOSED},
so each of them must own as a resource for resumer an awakening permit if it
is inside the deque front. All these cells must be inside the deque front, as
for the current call to \resumeOp{} to finish, it must arrive at a cell $j$ or
later. Thus, the permits $[i + 1; j)$ from the dequeue iterator are generated
and exchanged for the awakening permits needed for the invariants of the
iterator. The $i$'th permit is also exchanged for an awakening permit, which is
then used to retry the whole \resumeOp{} by this caller.

\paragraph{\suspendOp{}} First, a step of the enqueue iterator is performed.
It can not fail: otherwise, the caller would obtain the $i$'th permit from the
enqueue iterator along with the knowledge that the cell $i$ is cancelled;
however, this can not be, as only inhabited cells can be cancelled, and each
inhabited cell owns its exclusive permit from the enqueue iterator. Therefore,
the execution always obtains the $i$'th cell along with the $i$'th permit from
the enqueue iterator.

A new Future is then created, along with its resumption and cancellation
permits. If writing the Future to the cell is successful, then the whole
resumption permit and half of the completion permit are passed to the thread
queue, as is described for the corresponding transition. The remaining half of
the permit is then passed alongside the Future to return to the caller of
\suspendOp{}. If the CAS was failed, but the CAS setting the contents of
the cell to \marker{TAKEN} succeeded, then an \methName{ImmediateResult} is
created with the acquired value, and the full cancellation permit is provided to
the caller. Last, if writing \marker{TAKEN} failed, then the execution acquires
a copy of $E$ in exchange for the $i$'th permit from the enqueue iterator.

\paragraph{Cancellation Handler} If the cancellation mode is ``simple'', then
writing \marker{CANCELLED} does not affect the execution or break any invariants;
also, the canceller acquires the cancellation handle, which allows it to
cancel the infinite array cell.

Otherwise, the cancellation mode is ``smart'', the initial state is
\stateName{UNDECIDED}, and the execution owns the cancellation registration
token. A call to \methName{onCancellation()} is then made, and we consider two
possibilities:
\begin{itemize}
    \item The call was successful. The execution is provided with a cell
    cancelling token and a cancellation handle.
    An attempt is made to write \marker{CANCELLED} to the cell in place of the Future; on success,
    the state transitions to \stateName{CLOSED}, and otherwise, the state was
    \stateName{PASSED} and stays this way. In the latter case, the execution
    obtains an awakening permit that is then used to call \resumeOp{}.
    \item The call was unsuccessful. The execution is provided with a cell
    cancelling token and a copy of $X$. If an attempt to write \marker{REFUSED} in place of the Future
    succeeds, $X$ is passed to the CQS so that the resumer obtains it later;
    otherwise, $X$ is kept so that it can be used to call
    \methName{completeRefusedResume(..)}.
\end{itemize}

\subsubsection{Observable Behavior} There are a few useful properties that can
be derived from the specification of the CQS.

Our main claim is that each successful call to \resumeOp{}
either completes exactly one Future or performs one call to
\methName{completeRefusedResume(..)} if there are no Futures left in the queue, and
each unsuccessful call completes no Futures and does not perform a call to
\methName{completeRefusedResume(..)}.

To see this, observe that the only cases when a cell does not
own an $E$ are the following:
\begin{itemize}
    \item \resumeOp{} finished its execution by
    writing a value to the cell (in which case the corresponding call to
    \suspendOp{} observes it and completes its Future)

    \item The Future was successfully resumed by \resumeOp{};

    \item The Future was cancelled in the simple cancellation mode (and the call
    to \resumeOp{} fails);

    \item In smart cancellation mode, the cancellation handler succeeded in
    rewriting the Future contained in the cell with a marker and took the $E$
    (in which case either the marker is \marker{CANCELLED} and
    \resumeOp{} attempts to work with another cell, or the marker is
    \marker{REFUSED} and \methName{completeRefusedResume(..)} is called with the
    knowledge that \methName{onCancellation(..)} returned \literal{false}, which
    means that cancellation registration failed, which can only happen due to
    the CQS being empty);

    \item In smart async cancellation mode, the resumer managed to write its
    value to the cell before the cancellation handler wrote a marker; this
    case is identical to the previous one, except that the cancellation handler
    does the described actions on behalf of the \resumeOp{}.
\end{itemize}

Since each cell only has a single $E$ associated with it, the fact that a
successful call to \resumeOp{} obtains an $E$ means that one of the
aforementioned situations must have happened.

The converse is also true: both each call to \methName{completeRefusedResume(..)} and each completion of a Future was due to a call to \resumeOp{}. Observe that each call to \resumeOp{} only provides
a single $R$. Each Future requires an $R$ to complete, and obtaining the
resource $X$ also requires some cancellation registration to have failed, in
which case the caller is provided with an instance of $R$. For these operations
to be balanced, it is required that there are no more total calls to
\methName{completeRefusedResume(..)} and completed Future than there were
successful calls to \resumeOp{}.

\clearpage
\section{Formal Specifications and Proofs for the Presented Primitives}\label{sec:proofs-primitives}

In this section we outline the proofs for the presented primitives on top of \sqs{} in a way similar to Section~\ref{sec:proofs}.
The proofs are presented in the same model as the one described in Subsection \ref{subsec:proof_structure}, and are defined in terms of the logical model of futures described in Subsection \ref{subsec:proof_futures}.

\subsection{The Barrier Correctness}\label{subsec:barrier_proof}

\newcommand{\partiesV}{\texttt{parties}}

See file \fileInRepoNoSurround{barrier/barrier.v}.

\paragraph{Specification.} 
When the barrier is initialized to wait for
$k$ parties, it produces $k$ logical resources called \defn{entry permits}.
The \methName{arrive()} operation consumes an entry permit and returns a
\defn{brokenness-evidence}-passing Future. \emph{Brokenness evidence} is knowledge of
the fact that the barrier is broken; it can be freely duplicated and is
mutually exclusive with entry permits, meaning that it is impossible for an
entry permit to exist when it is already known that the barrier is broken.

\paragraph{Invariants.}
The \memLoc{remaining} counter stores some
value $r$. We define $n = \min(\partiesV{} - r, \partiesV{})$, roughly corresponding to the number of parties yet to arrive.

If $n$ is less than $\partiesV{}$, then $r > 0$, and we say that the barrier has
not yet been broken. The following statements hold in this case:
\begin{itemize}
    \item at most $r$ entry permits still exist, and the brokenness evidence does not
    exist yet;
    \item the size of the CQS is $n$.
\end{itemize}

If $n$ equals to $\partiesV{}$, then $r \le 0$ {---} we say that the barrier has been
broken in this case, and the following is true:
\begin{itemize}
    \item no entry permits exist, and the brokenness evidence if free to create;
    \item the size of the CQS is $0$.
\end{itemize}

The barrier stores a CQS with the unit resource as the enqueue resource $E$ the brokenness
evidence as the dequeue resource $R$.


\paragraph{Initialization.} Initially, all the invariants are satisfied: either \partiesV{}
is positive, in which case \partiesV{} entry permits are allocated and given to
the user, or it is non-positive, meaning that the barrier was created broken.

\paragraph{The \methName{arrive()} Operation.} The method is called with an entry permit.
This means that when the \faa{} at line \ref{line:barrier_faa} is performed, the barrier cannot not broken, so
$n < \partiesV{}$ and $r > 0$.
This way, we can consider the following two cases:
\begin{description}
    \item[$r > 1$:] This means that we are not the last party to be arrived.
    The invariant is preserved with $n' = n + 1 < \partiesV{}$. The entry permit
    with which the call to \methName{arrive()} was performed is destroyed,
    and the enqueue registration is performed, appending to the CQS and
    providing the suspension permit to the caller.
    After the \faa{} at line~\ref{line:barrier_faa}, the suspension permit is used to suspend in the CQS at line \ref{line:barrier_susp}.
    
    \item[$r = 1$:] We are the last party. Thus, $n = \partiesV{} - 1$
    and there are $n$ waiters in the CQS.
    The invariant is preserved with $n' = \partiesV{}$. Now that all the entry
    permits are destroyed, it is possible to construct the brokenness evidence.
    Since it is freely duplicable, we create \partiesV{} copies of it and use
    them to perform $\partiesV{} - 1$ dequeue registrations, acquiring
    $\partiesV{} - 1$ awakening permits.

    The awakening permits are then used to perform $\partiesV{} - 1$ calls to
    \methName{resume(..)} at line \ref{line:barrier_resume}, which is guaranteed not to fail with the chosen modes of the CQS.
\end{description}

\paragraph{Observable Behavior.}
The provided specification ensures that the implemented barrier is correct: as long as at least one entry permit exists, it is impossible for any of the \texttt{Future}s to complete. On the other hand, if any of the \texttt{Future}s are completed, no entry permits exist anymore.
\newcommand{\latchInitCount}{\texttt{initCount}}
\newcommand{\latchCount}{\memLoc{count}}
\newcommand{\latchWaiters}{\memLoc{waiters}}
\newcommand{\latchDoneBit}{\literal{DONE\_BIT}}
\newcommand{\latchOpen}{\stateName{Open Latch}}
\newcommand{\latchClosed}{\stateName{Closed Latch}}
\newcommand{\latchFinish}{\stateName{Finish}}

\subsection{The Count-Down-Latch Correctness}

See file \fileInRepoNoSurround{countdownlatch/countdownlatch.v}.

\paragraph{Specification.}
The state of the latch is represented as a
natural number logically equal to the number of
\methName{countDown()} invocations until the count reaches zero.
When the latch is initialized, the user obtains control of this state. Once the state reaches zero, it no longer changes.

The \methName{await()} operation returns an \defn{openness-evidence}-passing \texttt{Future}. The
openness evidence is a logical resource that signifies that the state of the
latch has become zero. It contradicts any non-zero latch state and can be
freely acquired from the zero state.

The \methName{countDown()} operation atomically decrements the latch state if it was positive, and keeps it $0$ otherwise. 


\paragraph{Invariants.} 
The state of the latch is
$\max(\latchCount, 0)$.
The latch operates in three phases, each with its own set of invariants:
\begin{itemize}
    \item \textbf{\texttt{Closed Latch.}} When the state is non-zero, the \latchWaiters{} counter does not have
    \latchDoneBit{} set and stores the current
    size of the CQS.
    
    \item \textbf{\texttt{Open Latch.}} The state is $0$, but the \latchWaiters{} counter does not have
    the \latchDoneBit{} yet, and still stores the size of the CQS. The notable
    difference from the previous state is that the latch is already logically
    open, so new calls to \methName{await()} will immediatelly return a completed \texttt{Future}.
    However, both concurrent requests can  add themselves to the CQS, and the waiters are eligible to abort.
    
    \item \textbf{\texttt{Finish}}. The state is $0$, and \latchWaiters{} has \latchDoneBit{} set. In this case, the CQS is empty.
\end{itemize}

It is impossible to go from \latchFinish{} to any other phase, and from
\latchOpen{} to \latchClosed{}.
The latch stores a CQS with the unit resource as the enqueue resource $E$ and the openness
evidence as the dequeue resource $R$.


\paragraph{Initialization} The invariants hold initially. \latchWaiters{} is
initialized with $0$ (see line~\ref{line:cdl_waiters}). If \latchInitCount{} is positive, then the latch is closed, and its
initial state equals \latchInitCount{}; otherwise, it is already open and its state
is $0$. In both cases, the CQS is empty.

\paragraph{The \methName{resumeWaiters()} Operation.} This method is called only when the
openness evidence exists, so the latch is not closed.

First, the \latchWaiters{} counter is checked to see whether the current phase is already
\latchFinish{}, by checking for \latchDoneBit{} at line~\ref{line:cdl_rw_bitcheck}. In this case, the CQS is empty and there is nothing to do.

Otherwise, if the phase is not \latchFinish{}, it must be \stateName{Open},
given that the openness evidence exists. The operation then attempts to set \latchDoneBit{}, 
performing the phase transition to \latchFinish{} (line~\ref{line:cdl_rw_setbit}). The corresponding CAS may fail due to a concurrent \latchDoneBit{} setting or the \latchWaiters{} counter increment or decrement; in this case, the operation restarts. 
Otherwise, if the CAS succeeds, then enough copies of the openness
evidence are created to perform dequeue registration for each waiter in the CQS (line~\ref{line:cdl_rw_resume}).

\paragraph{The \methName{countDown()} Operation.}
First, a decrement via \faa{} is performed (line~\ref{line:cdl_cd_inc}). If the latch was already in \latchOpen{} or \latchFinish{} phase, this action does not have any effect, as the state was zero and stays zero when further decrementing \latchCount{}; in this case we simply obtain the
freely-duplicable evidence that the latch is open.

However, if the latch was closed at the point of decrement, there are two possibilities:
\begin{itemize}
    \item The state was larger than $1$. This subtraction logically decrements the state of the latch, but has no further effect, as all the invariants still hold.
    \item The state equaled $1$. This means that when the \faa{} was
    performed, the latch entered the \latchOpen{} phase. Thus, we perform the
    corresponding phase transition and obtain the knowledge that the latch is closed.
\end{itemize}

Observe that if the latch is closed at the end of the operation, it means that
the state was either $0$ or $1$, which, in turn, shows that \latchCount{} was $\le1$. In this case, it is valid to call
\methName{\methName{resumeWaiters()}} (line~\ref{line:cdl_cd_rw}).

\paragraph{The \methName{await()} Operation.} 
Initially, \latchCount{} is checked; if the state
turns out to be $0$, the Future is completed immediately. This is not just done
as an optimization but has an observable effect in cases where the latch was
initialized with its state being $0$: if no call to \methName{countDown()}
occurs, the Futures will not ever be completed despite the latch being open.

If the state was not $0$, this means that the latch was initialized with a
non-zero state, which ensures that it is safe to suspend in its CQS:
when the latch is opened, some of the calls to \methName{countDown()} will
eventually complete the Futures.
After that, \faa{} is invoked (line~\ref{line:cdl_a_winc}). Its effect depends on whether the current phase is
\latchFinish{}. If it is, then the \faa{} has no effect: since \latchDoneBit{} is
already set, the CQS can only be empty from now on; in this case,
the call observes that the bit was set and completes its Future. Otherwise,
the \faa{} performs enqueue registration; \methName{suspend()} is then called
with the resulting awakening permit (line~\ref{line:cdl_a_susp}).

\paragraph{Cancellation.} There are two modes of cancellation described. Here
we show the correctness of both of them:
\begin{itemize}
    \item \textbf{Simple.} Cancellation of a Future does not change the size of the
    CQS in this mode. Thus, when a call to
    \methName{resumeWaiters()} succeeds in setting the \latchDoneBit{},
    the number of waiters that it receives includes both alive and cancelled
    ones. \methName{resume()} is repeatedly invoked, and it may fail for
    cancelled cells, but this does not matter: the goal here is to complete all
    the existing Futures, not a set number of them.
    
    \item \textbf{Smart}. Given the modes of operation of the CQS, it suffices
    to show that \methName{onCancellation()} and the trivial version of
    \methName{completeRefusedResume(..)} are correct.

    \methName{onCancellation()} performs a \faa{} (line~\ref{line:cdl_onc_decw}). If the phase was already
    \latchFinish{}, then this operation has no effect, as the CQS is
    empty and the \latchDoneBit{} is set, in which case cancellation
    registration fails and the call returns \literal{false} accordingly.
    Otherwise, the \faa{} decrements the size of the CQS, and the
    cancellation registration succeeds.

    \methName{completeRefusedResume(..)} is valid since if the CQS is
    empty, it is safe to simply quit the resumption operation: no valuable
    resources are being passed that could be lost otherwise.
\end{itemize}

\paragraph{Observable Behavior.}
The specification ensures that at least \latchInitCount{} calls are needed for any
of the Futures to complete.

\subsection{The Semaphore Correctness}

See files in \fileInRepoNoSurround{interruptible_semaphore}, depending on the CQS mode used.

\newcommand{\semaphorePermits}{\memLoc{permits}}
\paragraph{Specification.} The semaphore does not introduce any logical
resources on its own; instead, it is parameterized by the type of the resource
$T$ that is needed to perform operations in the critical section guarded by the
semaphore. Thus, in order to call \methName{release()} in a valid manner, the
caller must provide an instance of $T$, and \methName{acquire()} returns a
$T$-passing Future. Additionally, in order to create a semaphore that initially stores $K$, then $K$ copies of $T$ must be provided during initialization.
This specification is inspired by the way to express the mutex property for
locks that is commonly used in the example code for Iris.

\paragraph{Invariants.}
There are two possibilities for the state of the semaphore:
\begin{itemize}
    \item The CQS is empty. In this case, \semaphorePermits{} stores
    some number $n$, and the semaphore owns $n$ copies of $T$.
    \item The size of the CQS is $m > 0$. Then \semaphorePermits{}
    stores $-m$.
\end{itemize}

The semaphore stores a CQS with the unit resource as the enqueue resource $E$ and $T$ as the dequeue resource $R$.


\paragraph{Initialization.} The invariants hold initially. The CQS is empty, and the semaphore owns $K$ copies of $T$ (which are required in order to initialize it), which is reflected in
\semaphorePermits{} storing $K$.

\paragraph{The \methName{acquire()} Operation.} 
First, a \faa{} is performed (line~\ref{line:sema:acquire:dec}). We consider two
possibilities for the initial state of \semaphorePermits{}: if it contained a
number greater than zero, then the semaphore owned at least one copy of $T$,
which is taken by the call and put into an immediately complete Future (line~\ref{line:sema:acquire:acquired}).
Otherwise, the semaphore did not own a single copy of $T$, and the call to
\faa{} is used to perform the enqueue registration needed to call
\methName{suspend()} (line~\ref{line:sema:acquire:suspend}).

\paragraph{The \methName{release()} Operation.} 
First, a \faa{} is performed (line~\ref{line:sema:release:inc}). If the CQS was
empty, then \semaphorePermits{} contained a nonnegative number, in which case
the operation simply passes its copy of $T$ to the semaphore and finishes.
Otherwise, \semaphorePermits{} contained a negative number, and the \faa{}
performs a dequeue registration, the awakening permit from which is used to call
\methName{resume()} at line~\ref{line:sema:release:resume}, which never fails given the chosen mode of the CQS.

\paragraph{Cancellation.} 
With the chosen modes of the CQS, it suffices
to show that \methName{onCancellation()} and the trivial version of
\methName{completeRefusedResume(..)} are correct.
The \methName{onCancellation()} function performs a \faa{} at line~\ref{line:sema:onCancellation:inc}, attempting to perform a
cancellation registration. If the CQS was empty
(and the value in \semaphorePermits{} is negative), then the $T$ that is
obtained as part of the cancellation registration is passed to the semaphore,
like it is done in \methName{release()}. If the CQS did contain other
Futures, the size of the CQS is decremented, successfully registering
the cancellation.
The \methName{completeRefusedResume(..)} invocation is valid since the call to
\methName{onCancellation()} has already provided the semaphore with $T$.

\paragraph{Observable Behavior.}
Choosing $T$ to be some resource that only exists in $n$ copies, we can ensure
that the provided implementation is indeed a semaphore: if a block of code is
guarded by awaiting on a Future returned from \methName{acquire()}, it would
be impossible for this code to be executed concurrently by more than $n$
threads.

\subsection{The Blocking Pools Correctness}

See file \fileInRepoNoSurround{blocking_pool/pool.v}.

\newcommand{\poolSize}{\memLoc{size}}
The idea of the blocking pool algorithm is similar to the one for semaphore, with the difference that pools transfer actual values instead of logical permits.

In this proof, the concurrent data structure chosen for passing values outside
of the CQS is called the \defn{outer storage} of the pool (``outer''
refers to the activity happening outside of the CQS).

\paragraph{Specification.} The pool does not use any user-visible logical
resources and is parametrized by a predicate $U(..)$ that describes the values
passed in the pool.
A call to \methName{put(e)} requires providing an instance of $U(e)$, and
\methName{take()} returns a $U(v)$-passing Future if it completes with a $v$.

Note the similarities between the semaphore specification and this one; this is
explained by the fact that a semaphore can be thought of as a pool of unit
values with an optimization that allows the implementation to only keep a
counter of available permits instead of actually storing them in some
concurrent data structure.

\paragraph{Requirements for the Outer Storage.}\fileInRepo{blocking_pool/outer_storage_spec.v} The outer storage must expose
a state that is represented by a multiset of values that it contains and
the number of failed retrieval attempts that are yet to be balanced by the
corresponding insertions.
Then, \methName{tryInsert(..)} must atomically add its value to the multiset, returning \literal{true}, or decrease the number of failed retrievals, returning
\literal{false}.
\methName{tryRetrieve()}, likewise, must atomically extract some value from the
multiset and return it, or increase the number of failed retrievals, returning
\literal{null}.

It can be easily shown that the proposed data structures fulfill these
requirements (see files \fileInRepoNoSurround{blocking_pool/stack_outer_storage.v}, \fileInRepoNoSurround{blocking_pool/queue_outer_storage.v}). For example, if the head of the stack is $\bot$, then it contains only failed retrievals; otherwise, the contents of the stack form the multiset of the contained elements.
\methName{tryInsertStack(..)} performs a single successful \texttt{CAS} during
its operation, either observing that there are no failed retrievals and placing
its value in the stack or deregistering one of the retrievals.
\methName{tryRetrieveStack()} behaves in a symmetrical manner.
We say that the \defn{size} of the outer storage is $n - m$ if $n$ elements are
stored in the multiset and there are $m$ failed retrievals.

\paragraph{Invariants.}
The implementation uses internally some logical resources that are not exposed
to the user of the pool, namely \defn{insertion permits} and \defn{retrieval
permits}. We say that $k_i$ is the number of the insertion permits in existence,
and $k_r$ is the number of retrieval permits. If the current size of the outer storage is $s_o$, it is known that $s_o + k_i \ge k_r$;  in simple words, there can not be more retrieval permits than
there would be elements in the outer storage if all insertion attempts
succeeded.
\poolSize{} stores $s_o - s_t + k_i - k_r$, where $s_t$ is the current CQS size.
We know that either $s_t$ is zero or $s_o + k_i - k_r$ is, so either the CQS or the outer storage is effectively empty.

For each instance of a value $v$ stored in the multiset representation of the
outer storage, the pool owns a copy of $U(v)$.
The pool stores an CQS with the unit resource as the enqueue resource $E$, $U(v)$ for
element each element $v$ passed through it as the dequeue resource $R$, and a pair of the insertion permit
and $U(v)$ as $X$, the resource required to call \methName{completeRefusedResume(..)}.


\paragraph{Initialization.} Initially the invariants hold. Both the CQS
and the outer storage are empty, no insertion or retrieval permits exist, and
\poolSize{} is zero (line~\ref{line:pools_basic:size}).

\paragraph{The \methName{take()} Operation.} First, \faa{} is performed at line~\ref{line:pools_basic:take:dec}. We consider two
possibilities for the initial state of \poolSize{}.
\begin{itemize}
    \item If it contained a number greater than zero, then $s_o + k_i - k_r$ is
    positive. A new retrieval permit is created, incrementing $k_r$. With this
    permit, a call to \methName{tryRetrieve()} is attempted (line~\ref{line:pools_basic:take:tryRetrieve0}); regardless of
    whether it succeeds, both $s_o$ and $k_r$ are decremented: the retrieval
    permit is destroyed in the attempt, and either a value is taken from the
    logical multiset or a new failed retrieval is registered.
    \item If it contained a nonpositive number, then $s_o + k_i - k_r$ is
    zero. $s_t$ is incremented via enqueue registration, and the resulting
    suspension permit is used to place a Future in the CQS (line~\ref{line:pools_basic:take:suspend}).
\end{itemize}

\paragraph{The \methName{put()} Operation.} \faa{} is performed (line~\ref{line:pools_basic:put:inc}). Like with \methName{take()},
we consider two possibilities.
\begin{itemize}
    \item If \poolSize{} contained a nonnegative number, the CQS is
    empty. In this case, a new insertion permit is generated, which increments
    $k_i$. A call to \methName{tryInsert()} is performed at line~\ref{line:pools_basic:put:tryInsert}. The permit is
    destroyed, and either $s_o$ increases because a new value was successfully
    placed in the multiset, or it increases because a failed retrieval was
    removed.
    \item Otherwise, \poolSize{} contained a negative number, which means that
    $s_t > 0$. A dequeue registration is performed, and the awakening permit
    is then used to call \methName{resume(..)} (line~\ref{line:pools_basic:put:resume}).
\end{itemize}

\paragraph{Cancellation.} With the given mode of the CQS, we show that \methName{onCancellation()} and \methName{completeRefusedResume(..)} are correct.
The \methName{onCancellation()} operation performs a \faa{}, attempting to perform a
cancellation registration, at line~\ref{line:pools_basic:onCancellation:inc}. If the CQS was empty
(and the value in \poolSize{} is negative), then a new insertion permit is
generated, and the cancellation procedure obtains an $X$ using the $U(v)$
provided to it, where $v$ is the value of the resumer of this cell (possibly
undecided at the moment).
If the CQS did contain other
Futures, the size of the CQS is decremented, successfully registering
the cancellation.

The \methName{completeRefusedResume(..)} implementation is valid since $U(v)$ and the
insertion permit, both of which are needed to perform \methName{tryInsert(..)},
are present as part of $X$. If \methName{tryInsert(..)} fails, $U(v)$ is taken back
and can be used in a call to \methName{put(..)}.
%


\end{document}